\documentclass[11pt,a4paper]{article}

\usepackage[utf8]{inputenc}
\usepackage[T1]{fontenc}
\usepackage{amsmath,amssymb,bm}
\usepackage{geometry}
\usepackage{booktabs}
\usepackage{graphicx}
\usepackage{xcolor}
\usepackage{float}
\usepackage{placeins}
\usepackage{tabularx,array}
\usepackage{enumitem}
\usepackage{needspace}
\usepackage{microtype}
\usepackage[numbers,sort&compress]{natbib}
\usepackage{hyperref}
\hypersetup{colorlinks=true,linkcolor=blue,citecolor=blue,urlcolor=blue,
 pdftitle={A Quantitative Framework for Testing the Hubble Tension in a Bianchi Type I Cosmological Background},
 pdfauthor={Luigi Tedesco},
 pdfkeywords={Hubble tension; Bianchi type I; anisotropic cosmology; Lambda CDM}}
\newcommand{\rev}[1]{#1}
\newenvironment{revision}{}{}
\newcommand{\sectionlead}[1]{%
\Needspace{5\baselineskip}%
\medskip
\noindent\textbf{#1.}\par\nopagebreak\smallskip\noindent%
}

\title{\rev{A Quantitative Framework for Testing the Hubble Tension in a\\
Bianchi Type I Cosmological Background}}
\author{Luigi Tedesco\\\small Dipartimento di Fisica di Bari and INFN di Bari, 
\\
via Amendola 173, 70126 Bari\\\small e-mail: luigi.tedesco@ba.infn.it}
\date{}

\begin{document}
\maketitle

\begin{revision}
\begin{abstract}
The Hubble tension is usually formulated as a disagreement between two determinations of a single scalar parameter, $H_0$, within an exactly isotropic FLRW model. We develop a quantitative framework in which the tension is instead treated as a consistency test of the scalar FLRW compression of cosmological data in a homogeneous but anisotropically expanding Bianchi type I background. Beyond synthesizing established results on Bianchi I kinematics, null geodesics and optical propagation, the original contribution is a worked weak-shear, axisymmetric calculation that maps a specified shear history into a low-redshift luminosity-distance quadrupole. The calculation explicitly separates the direction-dependent redshift--affine-parameter mapping from the Jacobi-focusing contribution and then propagates the resulting distance quadrupole through an analytic polar-cap toy window. For freely decaying shear we obtain
$A_D(z)=-B_{H0}+(2q_0-1)B_{H0}z/2+(5-q_0-18q_0^2+6j_0)B_{H0}z^2/12+\mathcal O(z^3,B_{H0}^2)$,
where $B_{H0}=(H_{\parallel0}-H_{\perp0})/H_0$ and $j_0$ is the mean jerk parameter. A representative BBN limit, $\Omega_{\sigma0}\lesssim10^{-23}$, implies $|B_{H0}|\lesssim9.5\times10^{-12}$ and a distance-modulus quadrupole below approximately $2.4\times10^{-11}$ mag at $z=0.15$. \rev{The early-Universe bound used in this comparison is adopted from previous work and is not itself a new result of the present analysis; the novelty is its propagation through the derived direction-dependent redshift and Sachs--Jacobi mapping into quantitative limits on the luminosity-distance quadrupole and on the catalogue-window bias of an isotropic $H_0$ fit.} By contrast, even a maximally aligned one-percent directional shift requires $\Omega_{\sigma0}\simeq2.5\times10^{-5}$, while a shift comparable with the \rev{Planck 2018--SH0ES 2022 benchmark separation} requires $\Omega_{\sigma0}\simeq1.8\times10^{-3}$. Thus the minimal shear-only model cannot resolve the tension, although the framework supplies a falsifiable programme for testing sustained late-time anisotropy with supernovae, BAO, distance-ladder measurements and future standard sirens.
\end{abstract}
\end{revision}


\section{Introduction}

\begin{revision}
The standard cosmological model, usually denoted by $\Lambda$CDM, rests on a small number of physical and geometrical assumptions. The geometrical part is especially important: on sufficiently large scales the Universe is assumed to be described by a spatially homogeneous and isotropic Friedmann--Lemaitre--Robertson--Walker (FLRW) spacetime, as in the standard relativistic-cosmology framework \cite{Ellis1969Bianchi,EllisVanElst1999}. This assumption enters the interpretation of the cosmic microwave background (CMB), the calibration of baryon acoustic oscillations (BAO), the construction of distance--redshift relations, and the inference of the present expansion rate. The success of $\Lambda$CDM indicates that the FLRW approximation is extremely accurate. At the same time, the increasing precision of modern cosmological observations makes it legitimate to ask whether small departures from exact isotropy can bias, mimic, or partly relieve some of the tensions that have emerged within the concordance model.

Among these tensions, the discrepancy between the value of the Hubble constant inferred from early-Universe observations and the value obtained from local distance-ladder measurements is the most persistent. Planck CMB data, interpreted within base $\Lambda$CDM, give a value close to $H_0\simeq 67.4\,\mathrm{km\,s^{-1}\,Mpc^{-1}}$ \cite{Aghanim2020PlanckParams}, whereas the SH0ES distance ladder gives a value close to $73\,\mathrm{km\,s^{-1}\,Mpc^{-1}}$ \cite{Riess2022SH0ES}. Recent JWST observations have sharpened the debate by providing higher-resolution tests of Cepheid crowding and by enabling independent TRGB and JAGB calibrations \cite{Riess2024JWSTCrowding,Freedman2024CCHP,Lee2024JAGB,Hoyt2025TRGB,Riess2025PerfectHost,Li2025TRGBComplete,Scolnic2025Coma}. The situation is therefore not simply a disagreement between two numbers, but a stress test of the assumptions that connect observations at very different epochs and scales.
Most proposed solutions modify the matter sector, the dark-energy sector, the pre-recombination expansion history, neutrino physics, or gravity itself \cite{Verde2019Tension,Knox2020Hunter,Shah2021Buyer,DiValentino2021Review,Abdalla2022Intertwined,Efstathiou2025Challenges,Hu2023Evidence}. Recent analyses based on sound-horizon-free measurements, late-time phenomenology, early dark energy and modified recombination sharpen the same conclusion: a viable solution must fit CMB, BAO, supernovae and local calibration data simultaneously, not merely shift one inferred number \cite{Pantos2026Dissecting,Bansal2026LateTime,Poulin2026ACTDESIEDE,LeeZhou2026Recombination}. A complementary route is to examine the background geometry. The question addressed in this paper is whether relaxing exact FLRW isotropy in favor of a homogeneous but anisotropically expanding Bianchi type I geometry can affect the inference of $H_0$ in a controlled and observationally testable way.

Bianchi type I is the simplest anisotropic generalization of the spatially flat FLRW spacetime. It preserves spatial homogeneity but allows three independent directional scale factors. It is therefore a natural laboratory for testing the robustness of the isotropy assumption and is closely connected with the classical relativistic literature on anisotropic cosmologies, isotropization and cosmic no-hair behavior \cite{Misner1968Mixmaster,Wald1983CosmicNoHair}. The Bianchi I metric introduces shear, directional Hubble rates, and, in the absence of anisotropic stress, an effective contribution to the mean expansion rate that scales as $a^{-6}$. This rapid decay is a crucial fact: it implies that the minimal shear-only extension of $\Lambda$CDM is strongly constrained by early-Universe physics and is unlikely to produce a large late-time shift in $H_0$ unless additional physical ingredients are introduced. A previous work has shown that the shear contribution is strongly constrained by late-time data and becomes extremely small when BAO, CMB or big-bang nucleosynthesis constraints are included \cite{Akarsu2019Bianchi}. On the other hand, more recent studies have revisited anisotropic cosmologies, anisotropic Hubble diagrams and isotropy tests in the context of the Hubble tension, finding that anisotropy remains phenomenologically interesting even when it does not by itself solve the tension \cite{Anton2024HubbleDiagrams,Deliyergiyev2025MNRAS,Gron2024Symmetry,PalaciosCordoba2026Anisotropic,Zhou2025HubbleIsotropy}. The ellipsoidal-universe programme is especially relevant here because it discusses how a small anisotropy of the large-scale spatial geometry could affect CMB anomalies and the Hubble tension \cite{Campanelli2006Ellipsoidal,Campanelli2007CMBQuadrupole,Cea2007Polarization,Cea2010Polarization,Cea2014Planck,Cea2022EllipsoidalH0}.

Previous studies have established complementary parts of the problem. Constraints on the mean shear contribution in the minimal Bianchi I extension were obtained in Ref.~\cite{Akarsu2019Bianchi}; directional Hubble diagrams and statistical anisotropy tests were developed in Refs.~\cite{Anton2024HubbleDiagrams,Zhou2025HubbleIsotropy}; and exact or covariant optical constructions in anisotropic spacetimes were given in Refs.~\cite{Gasperini2011LightCone,Fanizza2013Jacobi,Fanizza2015LightSignals,Fleury2016GLC}. These ingredients are not claimed as new here. The original contribution is their organization into a single inference chain together with three explicit quantitative steps. First, we derive the weak-shear axisymmetric luminosity-distance quadrupole through relative order $z^2$, separating the redshift--affine-parameter contribution from the Jacobi-focusing term. Second, we compare representative BBN and BAO+CMB limits with the anisotropy required for a one-percent shift and for the \rev{Planck 2018--SH0ES 2022 benchmark separation}. Third, we propagate the finite-redshift distance quadrupole through a specified analytic polar-cap toy window to quantify quadrupole-to-monopole leakage. The polar cap is used only as a transparent theoretical selection function; it is not presented as the angular or redshift distribution of a particular survey.

\rev{In particular, we do not claim as new the conclusion that freely decaying Bianchi I shear is severely restricted by BBN, CMB and BAO constraints; this was established in previous analyses, including Ref.~\cite{Akarsu2019Bianchi}. The new element is the explicit propagation of those independently obtained shear bounds through a direction-dependent redshift map and the Sachs--Jacobi optical response, yielding quantitative limits on $B_{H0}$, $A_D(z)$, $A_\mu(z)$ and the leakage into the scalar $H_0$ value recovered by an isotropic catalogue fit.}

We do not assume that Bianchi I anisotropy solves the Hubble tension. Instead, we formulate and evaluate a framework in which this possibility can be tested. The central questions are: how does anisotropic expansion enter the distance--redshift relation; what is the difference between decaying shear and sustained late-time anisotropy; how should one build a likelihood for supernovae, BAO, CMB priors, cosmic chronometers and local $H_0$ measurements; and what level of anisotropy would be required to produce an observationally relevant shift in the inferred Hubble constant?

The paper is organized as follows. Section~\ref{sec:h0status} summarizes the observational status of the Hubble tension and the role of geometrical assumptions. Section~\ref{sec:bianchi} introduces the Bianchi type I background, the directional expansion rates and the shear scalar. Section~\ref{sec:observables} derives the main consequences for directional observables, including luminosity distance and distance modulus, and gives an explicit weak-shear low-redshift mapping from $\Delta H_{\rm ax}(t)$ to the distance quadrupole. Section~\ref{sec:minimal} discusses the minimal shear-only model, corrects the flat-reference comparison, and provides a numerical hierarchy between BBN/CMB bounds and the amplitudes required to affect $H_0$. Section~\ref{sec:phenomenology} introduces a phenomenological late-time anisotropy parameterization suitable for observational tests. Section~\ref{sec:data} describes a possible likelihood strategy for real data. Section~\ref{sec:diagnostics} discusses diagnostic outcomes. Section~\ref{sec:discussion} discusses interpretation, limitations and extensions. Section~\ref{sec:conclusions} presents the conclusions. Appendix~\ref{app:shear} gives an expanded derivation of the Bianchi I kinematics and shear evolution, while Appendix~\ref{app:pipeline} gives an expanded practical likelihood pipeline.
\end{revision}

\section{The Hubble Tension and the Role of Geometry}\label{sec:h0status}

\rev{The Hubble constant $H_0$ is one of the fundamental parameters of observational cosmology, but its operational meaning depends on the spacetime model used to interpret the data. In an exactly FLRW background there is a unique scale factor. The expansion rate is the scalar quantity $H(t)=\dot a/a$, and its present-day value is identified with $H_0$. Observationally, however, $H_0$ is never measured as an isolated local field variable. It is inferred from a chain of distance, redshift, calibration and model assumptions. This distinction is particularly important in anisotropic cosmology because a data set compressed into a single scalar Hubble constant may contain information about the angular structure of the expansion.}

\rev{It is useful to separate three meanings of $H_0$ that coincide in FLRW but need not coincide exactly once isotropy is relaxed. The first is the \emph{mean kinematical} Hubble rate, defined by the volume expansion of the homogeneous spatial hypersurfaces. The second is a \emph{directional} Hubble rate, measured along a given line of sight and affected by the projection of the shear tensor. The third is the \emph{effective scalar} value obtained when an isotropic FLRW template is fitted to a finite and angularly non-uniform data set. The present paper is mainly concerned with the relation among these three quantities. Even when the underlying anisotropy is small, the fitted scalar value may depend on sky coverage, redshift distribution, covariance modelling and the particular combination of observables used in the likelihood.}

\rev{The early-Universe determination is most commonly associated with CMB anisotropies and with the inference of the sound horizon within a specified pre-recombination model \cite{Aghanim2020PlanckParams,Knox2020Hunter}. The CMB does not directly measure $H_0$. It measures angular scales, relative acoustic peak heights, polarization spectra and damping effects; these observables are translated into cosmological parameters only after a background model and a perturbation model have been specified. Within base $\Lambda$CDM, the angular acoustic scale, the physical baryon density, the physical cold-dark-matter density and the assumption of a standard sound horizon tightly determine the late-time expansion history. Recent small-scale CMB analyses and forecasts, including ACT DR6 (Atacama Cosmology Telescope - Data Release 6) and SPT-3G (South Pole Telescope - 3rd Generation), add information on the same parameter space and on extensions that try to alter the early-Universe inference of $H_0$ \cite{Louis2025ACTDR6,Calabrese2025ACTExtended,Farren2025ACTUnWISE,Vitrier2025SPT3G}. In a Bianchi I setting this point is restrictive: the model must not only shift the background distance to last scattering, but must also preserve the high degree of observed CMB isotropy and the consistency of the acoustic pattern.}

\rev{The late-Universe determination follows a different logic. The local distance ladder calibrates absolute distances using nearby anchors and propagates this calibration to Type Ia supernovae in the Hubble flow. Cepheids, water megamasers, detached eclipsing binaries, TRGB stars and JAGB stars provide different routes to the absolute calibration of supernova luminosities \cite{Riess2022SH0ES,Freedman2024CCHP,Lee2024JAGB,Hoyt2025TRGB,Riess2025PerfectHost,Li2025TRGBComplete,Scolnic2025Coma} and the analysis is correspondingly sensitive to calibration, photometry, metallicity corrections, crowding, peculiar velocities, sample selection and covariance modelling. Modern supernova compilations and recalibrations, including DES-SN5YR (Dark Energy Survey - Supernovae 5-Year), its recalibrated analyses and Union3, are therefore important cross-checks of any anisotropic Hubble-diagram signal \cite{Abbott2024DESY5,Popovic2025DESReanalysis,Rubin2025Union3}. For the present purpose, the relevant fact is that a local distance ladder is not an all-sky measurement of an ideal scalar field. It samples a finite angular window and a finite redshift range, and therefore it can be tested for dipolar or quadrupolar residuals relative to an isotropic best fit.}

\rev{BAO measurements provide a third element. In fact, BAO are standard-ruler measurements usually expressed through quantities such as $D_M(z)/r_d$, $D_H(z)/r_d$ and $D_V(z)/r_d$, where $r_d$ is the sound horizon at the baryon drag epoch. Surveys such as SDSS/eBOSS and DESI have made BAO one of the most precise probes of late-time geometry \cite{Eisenstein2005BAO,Alam2021Eboss,DESI2024DR1BAO,DESI2025DR2BAO,DESI2025Cosmology}. Recent DESI-DR2-based analyses and updated 2D BAO compilations further test whether the inferred expansion rate is redshift dependent or connected with evolving dark energy \cite{Jia2025DESIH0,Liu2025DESIR2H0,Sabogal2025BAO2D,ForeroSanchez2026DESIFS}. In the standard interpretation, BAO analyses assume that the large-scale background is well described by an isotropic FLRW metric, even when the observables are split into radial and transverse components. A Bianchi I interpretation must therefore specify whether BAO are treated as constraints on the volume-averaged expansion, on direction-dependent distances, or merely as compressed FLRW priors used to reject anisotropic models that are already too far from the standard solution.}

\rev{Table~\ref{tab:h0routes} summarizes the observational routes used in the remainder of the paper. It is not intended as a numerical data compilation; rather, it clarifies which aspect of each probe becomes relevant when the scalar FLRW compression is replaced by a directional Bianchi I analysis.}

\begin{table}[!htbp]
\centering
\caption{Schematic role of the main observational routes to $H_0$ in a Bianchi type I analysis. The entries identify the quantity usually constrained in an FLRW analysis and the aspect that becomes relevant when directional dependence is allowed.}
\label{tab:h0routes}
\footnotesize
\setlength{\tabcolsep}{3.5pt}
\renewcommand{\arraystretch}{1.05}
\begin{tabularx}{\textwidth}{@{}>{\raggedright\arraybackslash}p{0.22\textwidth}>{\raggedright\arraybackslash}p{0.31\textwidth}>{\raggedright\arraybackslash}X@{}}
\toprule
Route & Main quantity & Relevance for anisotropy \\
\midrule
CMB & Acoustic scale and physical densities & Early anchor; severe constraint on shear, quadrupolar geometry and any departure from near-isotropy \\
BAO & $D_M/r_d$, $D_H/r_d$, $D_V/r_d$ & Standard ruler; can be used either as FLRW-compressed information or as a test of directional distance consistency \\
Cepheid--SNe ladder & Calibrated Hubble-flow intercept & Sensitive to sky coverage, peculiar velocities, local flows and angular systematics in the Hubble diagram \\
TRGB/JAGB routes & Independent stellar calibrations & Cross-checks of calibration systematics and of whether a local value of $H_0$ depends on the adopted anchors \\
Cosmic chronometers & $H(z)$ from differential ages & Probe of the mean late-time expansion rate, with less direct sensitivity to angular anisotropy \\
Strong lenses & Time-delay distances & Intermediate-redshift geometrical probe; potentially sensitive to line-of-sight environment and anisotropic distance conversion \\
Standard sirens & Gravitational-wave luminosity distance & Future direct directional-distance test, especially when electromagnetic counterparts provide redshifts \\
\bottomrule
\end{tabularx}
\end{table}

\rev{The geometrical point can now be stated more formally. In FLRW cosmology, the luminosity distance is a function only of redshift and cosmological parameters:}
\begin{equation}
 D_L=D_L(z;H_0,\Omega_{m0},\Omega_{\Lambda0},\ldots),
 \label{eq:DLFLRWsec2}
\end{equation}
\rev{In a homogeneous but anisotropic spacetime, the same observable can also depend on the direction of observation:}
\begin{equation}
 D_L=D_L(z,\hat n;H_0,\Omega_{m0},\Omega_{\Lambda0},\sigma_{ij},\ldots),
 \label{eq:DLanisSec2}
\end{equation}
where $\hat n$ is the line-of-sight direction and $\sigma_{ij}$ is the shear tensor. The difference between Eqs.~\eqref{eq:DLFLRWsec2} and \eqref{eq:DLanisSec2} is not a formal detail: it changes the statistical object that is being fitted. In an isotropic analysis the residuals of supernovae, BAO or time-delay distances are treated as functions of redshift and nuisance parameters, whereas in an anisotropic analysis they must also be tested against angular templates, and the best-fit monopole can be correlated with the quadrupole amplitude.

At low redshift the distinction can be understood kinematically. A directional expansion rate may be written schematically as
\begin{equation}
 H_{\rm dir}(z,\hat n)=H(z)+\sigma_{ij}(z)\hat n^i\hat n^j+\mathcal{O}(\sigma^2),
 \label{eq:HdirSec2}
\end{equation}
\rev{Up to convention-dependent signs and higher-order corrections, the angular average of the shear projection vanishes for uniform sky coverage. A realistic catalogue, however, has a mask, calibration inhomogeneities and redshift-dependent selection. Consequently, a small quadrupolar distance residual can leak into the monopole inferred by an FLRW fit. This leakage is the mechanism that makes anisotropy relevant to the Hubble tension even when the mean expansion history is close to $\Lambda$CDM.}

\rev{This possibility should not be confused with the claim that the Universe is observationally anisotropic at a level sufficient to solve the tension. The CMB quadrupole, the isotropy of large-scale structure, the success of BAO compression and the smallness of observed Hubble-diagram anisotropies all impose strong constraints. In particular, a pure shear contribution in the Einstein equations behaves as a stiff component and typically scales as $a^{-6}$, which makes it dangerous at early times and inefficient at late times. The role of Section~\ref{sec:minimal} is precisely to show why this minimal shear-only scenario is too constrained to produce a large shift in $H_0$. The more phenomenological alternatives discussed later require a source of anisotropic stress, an anisotropic dark-energy sector, or an effective ellipsoidal geometry that changes the late-time distance relation without violating early-Universe bounds.}

\rev{The observational question is therefore not whether one can choose an arbitrary anisotropic correction to reproduce a desired value of $H_0$. A viable model must pass a hierarchy of tests. First, it must fit the mean distance--redshift relation at least as well as the corresponding FLRW baseline. Second, the inferred anisotropic amplitude must be stable under changes of sky mask, redshift cuts and calibration subset. Third, the preferred axis, if present, should not be a survey artefact. Fourth, the model must not spoil CMB, BAO, BBN or structure-growth constraints. Finally, the improvement in goodness of fit must be large enough to justify the additional parameters. Together, these requirements turn the Hubble tension into a well-defined geometrical consistency test rather than a purely phenomenological adjustment.}

\begin{figure}[H]
\centering
\includegraphics[width=0.70\textwidth]{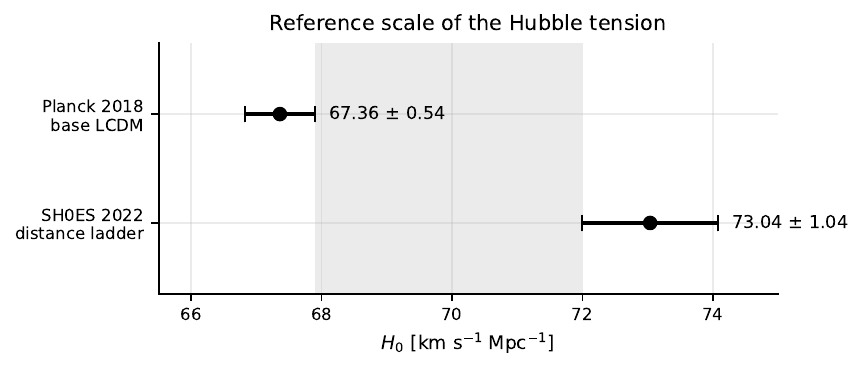}
\caption{Reference scale of the Hubble tension. The two points show the Planck 2018 base-$\Lambda$CDM value and the SH0ES 2022 local distance-ladder value quoted in the literature \cite{Aghanim2020PlanckParams,Riess2022SH0ES}. The shaded region between the one-sigma intervals emphasizes that the issue is not a small graphical offset, but a statistically significant mismatch between early- and late-Universe routes to $H_0$.}
\label{fig:h0comparison}
\end{figure}

Figure~\ref{fig:h0comparison} displays the reference scale of the tension used throughout the paper. It compares the Planck 2018 CMB inference in base $\Lambda$CDM with the SH0ES local distance-ladder determination. The plot is not meant as a complete compilation of all existing $H_0$ measurements; it is included to make visually clear the gap that any geometrical or physical extension must address.

\rev{The central question can now be formulated precisely: we do not ask whether an arbitrary anisotropy can be introduced to move one value of $H_0$ toward another. Rather, we ask whether a controlled Bianchi I deformation of the background geometry can produce a direction-dependent distance--redshift relation large enough to affect the inferred Hubble constant, while remaining compatible with CMB, BAO, nucleosynthesis and the observed near-isotropy of the Universe. This turns the Hubble tension into a quantitative test of the geometrical assumptions underlying precision cosmology.}

\section{Bianchi Type I Background}\label{sec:bianchi}

The Bianchi type I spacetime is the minimal homogeneous departure from the spatially flat FLRW geometry. It keeps the spatial hypersurfaces Euclidean and translationally invariant, but it relaxes invariance under arbitrary spatial rotations; for the problem considered here, this is an important compromise. The model is more general than FLRW because it admits directional expansion, but it remains simple enough to relate redshift, shear and distance observables to a small number of geometrical quantities. It is therefore well suited for asking whether the scalar compression implicit in an FLRW fit can hide a small anisotropic component without introducing the additional complications of a fully inhomogeneous spacetime.

Recent Bianchi-I and anisotropic-expansion analyses also make clear that the same metric can be used in two logically distinct ways: as a strict dynamical extension of $\Lambda$CDM, constrained by early-Universe physics, or as a phenomenological diagnostic of directional residuals in late-time distance data. This distinction is useful here because recent work has derived Bianchi-I Hubble diagrams, early-Universe anisotropy bounds, and self-consistent recombination or acoustic-scale observables in anisotropic backgrounds \cite{LeDelliou2020AnisotropicModel,HertzbergLoeb2024Anisotropic,NgChu2025LRSBianchi,NgChu2025CMBScale}.

In a principal-axis coordinate system the metric can be written as
\begin{equation}
 ds^2=-dt^2+a_1^2(t)dx^2+a_2^2(t)dy^2+a_3^2(t)dz^2,
 \label{eq:metric}
\end{equation}
where $a_i(t)$ are the three directional scale factors. The average scale factor is defined by the volume expansion,
\begin{equation}
 a(t)=\left[a_1(t)a_2(t)a_3(t)\right]^{1/3},
 \label{eq:averagescale}
\end{equation}
so that a comoving volume evolves as $V\propto a^3$. The directional Hubble rates are
\begin{equation}
 H_i(t)=\frac{\dot a_i}{a_i}, \qquad i=1,2,3,
 \label{eq:Hi}
\end{equation}
and their average defines the volume Hubble rate,
\begin{equation}
 H(t)=\frac{1}{3}\left(H_1+H_2+H_3\right)=\frac{\dot a}{a}.
 \label{eq:Hmean}
\end{equation}
The FLRW limit is obtained when $a_1=a_2=a_3=a$ and hence $H_1=H_2=H_3=H$. In the present paper $H$ always denotes this volume expansion rate, not a directional or catalogue-averaged estimator. This distinction is essential because the question of the Hubble tension concerns an inferred scalar parameter, whereas the Bianchi I geometry contains a tensorial expansion field.

A first useful step is to factor out the isotropic volume expansion by writing
\begin{equation}
 a_i(t)=a(t)e^{\beta_i(t)}, \qquad \sum_{i=1}^{3}\beta_i(t)=0,
 \label{eq:betai}
\end{equation}
\rev{The trace-free condition ensures that $a(t)$ is exactly the geometric mean of the three scale factors and the functions $\beta_i$ describe the shape distortion of an initially cubic comoving cell into a rectangular parallelepiped of the same volume. The directional Hubble rates become}
\begin{equation}
 H_i=H+\dot\beta_i,
 \label{eq:Hibeta}
\end{equation}
\rev{The condition $\sum_i\dot\beta_i=0$ leaves two independent anisotropic components, as expected for a symmetric trace-free tensor in the diagonal principal frame.}

The expansion tensor of the comoving congruence can be decomposed covariantly as
\begin{equation}
 \nabla_{\nu}u_{\mu}=\frac{1}{3}\theta h_{\mu\nu}+\sigma_{\mu\nu},
 \qquad \theta=3H,
 \label{eq:kinematicdecomp}
\end{equation}
where $u^\mu=\delta^\mu_0$, $h_{\mu\nu}=g_{\mu\nu}+u_\mu u_\nu$ is the spatial projector and $\sigma_{\mu\nu}$ is the shear tensor; vorticity and four-acceleration vanish for the comoving observers of the metric \eqref{eq:metric}. In the orthonormal principal-axis frame the nonzero shear eigenvalues are simply
\begin{equation}
 \sigma_i=H_i-H=\dot\beta_i, \qquad \sum_i\sigma_i=0,
 \label{eq:sheareigenvalues}
\end{equation}
The shear scalar is therefore
\begin{equation}
 \sigma^2=\frac{1}{2}\sigma_{\mu\nu}\sigma^{\mu\nu}
     =\frac{1}{2}\sum_{i=1}^{3}\left(H_i-H\right)^2
     =\frac{1}{2}\sum_{i=1}^{3}\dot\beta_i^2,
 \label{eq:shearscalar}
\end{equation}
\rev{This normalization is convenient because the shear contribution enters the average Friedmann equation as a positive term. A dimensionless measure of the anisotropic contribution is}
\begin{equation}
 \Omega_{\sigma}(z)=\frac{\sigma^2(z)}{3H^2(z)},
 \label{eq:omegasigma}
\end{equation}
\rev{The quantity $\Omega_\sigma$ is a geometrical analogue of a density parameter and should not be confused with the energy density of a new material component. Although it behaves like an effective stiff source in the mean expansion equation, this distinction remains useful when comparing shear-only models with models in which anisotropic stress is generated by a physical sector.}

The dynamics of these variables follow from the Einstein equations. Let the matter stress tensor be written as
\begin{equation}
 T^{\mu}{}_{\nu}=\mathrm{diag}\left[-\rho,p_1,p_2,p_3\right],
 \label{eq:Tdiag}
\end{equation}
with mean pressure $p=(p_1+p_2+p_3)/3$ and trace-free anisotropic pressure
\begin{equation}
 \pi^i{}_{j}=\mathrm{diag}\left[p_1-p,p_2-p,p_3-p\right],
 \qquad \pi^i{}_{i}=0.
 \label{eq:anisstressdef}
\end{equation}
The average Einstein equations may then be written schematically as
\begin{align}
 3H^2 &= 8\pi G\rho + \sigma^2,\label{eq:meanfriedmann}\\
 \dot H &= -4\pi G(\rho+p)-\sigma^2,\label{eq:meanraychaudhuri}\\
 \dot\sigma^i{}_{j}+3H\sigma^i{}_{j} &= 8\pi G\pi^i{}_{j},\label{eq:shearsource}
\end{align}
\rev{The last equation shows explicitly why the physical status of the anisotropy matters. If $\pi^i{}_{j}=0$, shear is a freely decaying geometrical mode. If $\pi^i{}_{j}\neq0$, a late-time source can maintain or regenerate anisotropy. This difference separates the minimal model considered in Section~\ref{sec:minimal} from the phenomenological scenarios of Section~\ref{sec:phenomenology}. It is also the reason why anisotropic dark energy and ellipsoidal-universe models cannot be reduced to a simple $a^{-6}$ correction without loss of physical content \cite{Koivisto2008AnisotropicDE,Rodrigues2008AnisotropicLambda,Appleby2013ProbingADE,Verma2025AnisotropicDE,Cea2022EllipsoidalH0,Tedesco2024CosmicShear}.}

For a matter source with isotropic pressure and no anisotropic stress, Eq.~\eqref{eq:shearsource} reduces to
\begin{equation}
 \dot\sigma_{ij}+3H\sigma_{ij}=0,
 \label{eq:shearevolution}
\end{equation}
which gives
\begin{equation}
 \sigma_{ij}=\frac{\Sigma_{ij}}{a^3}, \qquad \sigma^2=\frac{\Sigma^2}{a^6},
 \label{eq:sigmaa6}
\end{equation}
where $\Sigma_{ij}$ is a constant traceless tensor. Assuming radiation, pressureless matter, a cosmological constant and freely decaying shear, the mean Friedmann equation becomes
\begin{equation}
 H^2(a)=H_0^2\left[\Omega_{r0}a^{-4}+\Omega_{m0}a^{-3}+\Omega_{\Lambda0}+\Omega_{\sigma0}a^{-6}\right],
 \label{eq:Hsheronly}
\end{equation}
with
\begin{equation}
 \Omega_{\sigma0}=\frac{\sigma_0^2}{3H_0^2}
 \label{eq:omegasigma0}
\end{equation}
and the closure relation
\begin{equation}
 \Omega_{r0}+\Omega_{m0}+\Omega_{\Lambda0}+\Omega_{\sigma0}=1
 \label{eq:closure}
\end{equation}
\rev{for the spatially flat Bianchi I case. Equation~\eqref{eq:Hsheronly} is the simplest Bianchi I extension of $\Lambda$CDM considered in several phenomenological analyses \cite{Akarsu2019Bianchi,Akarsu2023CurvatureAnisotropy,Gron2024Symmetry}. It is mathematically simple and observationally useful, but its interpretation is restrictive: because the shear term scales as $a^{-6}$, a value large enough to influence the local Hubble diagram would generally be unacceptable at nucleosynthesis or recombination unless the shear is not freely decaying at all epochs.}

A particularly useful special case is the axisymmetric limit, in which two directional scale factors are equal. One may set
\begin{equation}
 a_1=a_2=a_{\perp}, \qquad a_3=a_{\parallel},
 \label{eq:axisym}
\end{equation}
with
\begin{equation}
 H_{\perp}=\frac{\dot a_{\perp}}{a_{\perp}}, \qquad
 H_{\parallel}=\frac{\dot a_{\parallel}}{a_{\parallel}}, \qquad
 H=\frac{2H_{\perp}+H_{\parallel}}{3}.
 \label{eq:axisymH}
\end{equation}
The shear eigenvalues are then
\begin{equation}
 \sigma_{\perp}=H_{\perp}-H=-\frac{1}{3}\Delta H_{\rm ax},
 \qquad
 \sigma_{\parallel}=H_{\parallel}-H=\frac{2}{3}\Delta H_{\rm ax},
 \qquad
 \Delta H_{\rm ax}\equiv H_{\parallel}-H_{\perp},
 \label{eq:axisymshear}
\end{equation}
and the shear scalar is
\begin{equation}
 \sigma^2=\frac{1}{3}\left(\Delta H_{\rm ax}\right)^2.
 \label{eq:axisymsigmascalar}
\end{equation}
This one-parameter form is especially transparent observationally because the leading angular dependence is an axial quadrupole and it also connects directly with ellipsoidal-universe descriptions, in which the ratio $a_{\parallel}/a_{\perp}$ or an associated eccentricity parameter encodes the departure from isotropic expansion \cite{Campanelli2006Ellipsoidal,Campanelli2007CMBQuadrupole,Cea2007Polarization,Cea2010Polarization,Cea2014Planck,Cea2022EllipsoidalH0}. In such a language, the Bianchi I model does not introduce a preferred centre, as in radial inhomogeneous models, but a preferred axis of expansion.

For small anisotropy one may define an eccentricity-like variable by
\begin{equation}
 e_B(t)\equiv \ln\left(\frac{a_{\parallel}}{a_{\perp}}\right)
   =\beta_{\parallel}-\beta_{\perp},
 \qquad \dot e_B=\Delta H_{\rm ax},
 \label{eq:eccentricityB}
\end{equation}
\rev{The integrated anisotropy $e_B$ affects distances through the accumulated photon propagation, whereas $\Delta H_{\rm ax}$ controls the instantaneous directional expansion. This distinction is relevant for data analysis: a low-redshift Hubble-flow measurement is mainly sensitive to the present value of the directional expansion, while CMB and high-redshift distance measures constrain the integrated anisotropic history. A model that appears harmless in the local Hubble diagram may therefore still be excluded if its integrated shear produces an excessively large CMB quadrupole or an inconsistent acoustic geometry.}

\rev{This section defines the background variables used throughout the paper. Section~\ref{sec:observables} translates the same geometry into photon propagation and distance observables. Section~\ref{sec:minimal} then explains why the freely decaying solution \eqref{eq:sigmaa6} is too tightly constrained to provide a generic resolution of the Hubble tension.}

\section{Directional Observables in a Bianchi I Spacetime}\label{sec:observables}

\rev{The observational effect of Bianchi I geometry is not exhausted by the mean expansion rate. In an FLRW spacetime, redshift, angular-diameter distance and luminosity distance depend only on emission time, or equivalently on redshift. The standard relation between luminosity and angular-diameter distances is tied to Etherington's reciprocity theorem and its modern relativistic-cosmology formulation \cite{Etherington1933,Ellis2007Etherington}. In a Bianchi I spacetime, the same quantities also depend on the direction of propagation relative to the principal axes. Consequently, Type Ia supernovae, standard sirens, cosmic parallax, angular BAO and residual maps of the Hubble diagram can test anisotropic expansion even when the mean expansion history remains close to flat $\Lambda$CDM \cite{KristianSachs1966,Sachs1961Optics,Fleury2016GLC,Gasperini2011LightCone,Fanizza2013Jacobi,Fanizza2015LightSignals,Quercellini2009CosmicParallax,Fontanini2009CosmicParallax,Campanelli2011Parallax,Anton2024HubbleDiagrams,Zhou2025HubbleIsotropy}.}

\rev{The observational problem must be formulated carefully because the sky position of a source is measured in the observer's orthonormal tetrad, whereas the geodesic constants are naturally expressed in the comoving coordinates of Eq.~\eqref{eq:metric}. The two descriptions coincide only in the isotropic limit. For weak anisotropy their difference is perturbatively small, but that difference generates the quadrupolar corrections to redshift and distance. A consistent analysis should therefore distinguish the physical direction $\hat n$ measured on the sky, the conserved coordinate momenta $p_i$, and the redshift-dependent photon direction relative to the principal axes.}

\subsection{Null geodesics and anisotropic redshift}

For the diagonal Bianchi I metric introduced in Section~\ref{sec:bianchi}, the photon trajectory is described by a wave vector $k^{\mu}=dx^{\mu}/d\lambda$, where $\lambda$ is an affine parameter and the null condition and the geodesic equation are
\begin{equation}
 ds^2=0, \qquad k^{\nu}\nabla_{\nu}k^{\mu}=0.
 \label{eq:nullgeodesic}
\end{equation}
The spatial homogeneity of the metric gives three Killing vectors, $\partial_i$, and therefore three conserved covariant momenta along the ray,
\begin{equation}
 p_i=g_{i\mu}k^{\mu}=a_i^2(t)\frac{dx^i}{d\lambda}=\mathrm{constant}, \qquad i=1,2,3.
 \label{eq:conservedpi}
\end{equation}
\rev{These constants are not themselves the observed direction on the sky; they are coordinate momenta. Their physical meaning becomes clear only after projecting the photon momentum on the orthonormal tetrad carried by the comoving observer. The temporal component is fixed by the null condition:}
\begin{equation}
 k^0(t)=\frac{dt}{d\lambda}=\left[\sum_{i=1}^{3}\frac{p_i^2}{a_i^2(t)}\right]^{1/2}.
 \label{eq:k0bianchi}
\end{equation}
For a comoving observer with four-velocity $u^{\mu}=\delta^{\mu}_{0}$, the photon energy is
\begin{equation}
 E(t)=-u_{\mu}k^{\mu}=k^0(t).
 \label{eq:energy_def}
\end{equation}
The observed redshift between emission at $t_s$ and observation at $t_o$ is therefore
\begin{equation}
 1+z=\frac{E_s}{E_o}
 =\frac{\left[\sum_i p_i^2/a_i^2(t_s)\right]^{1/2}}
    {\left[\sum_i p_i^2/a_i^2(t_o)\right]^{1/2}}.
 \label{eq:redshiftbianchi}
\end{equation}
Eq.~\eqref{eq:redshiftbianchi} is the exact redshift relation for null rays in the diagonal Bianchi I background and it shows already the essential physical point: redshift is not a function of the emission time alone, but of the orientation of the photon momentum relative to the three scale factors.

It is useful to rewrite the same result in terms of quantities measured by the observer. In fact, in the tetrad frame $e_{\hat 0}=u$ and $e_{\hat i}=a_i^{-1}\partial_i$, the instantaneous direction cosines of the photon are defined by
\begin{equation}
 q_i(t)=\frac{k^{\hat i}}{E}=\frac{p_i}{a_i(t)E(t)},
 \qquad
 \sum_i q_i^2(t)=1,
 \label{eq:qidef}
\end{equation}
\rev{At the observer one may set $q_i(t_o)=n_i$, where $n_i$ are the components of the measured line-of-sight direction in the principal-axis tetrad. Using the arbitrary normalization of the affine parameter to set $E_o=1$, Eq.~\eqref{eq:redshiftbianchi} can then be written in the observational form}
\begin{equation}
 1+z=\left[\sum_i n_i^2\frac{a_i^2(t_o)}{a_i^2(t_s)}\right]^{1/2}.
 \label{eq:redshift_observed_direction}
\end{equation}
This form is invariant under constant rescalings of the spatial coordinates and makes the FLRW limit immediate: if $a_1=a_2=a_3=a$, then $1+z=a_o/a_s$ independently of direction.

\rev{The photon direction in the local tetrad is generally not constant along the ray. Differentiating Eq.~\eqref{eq:qidef} with respect to cosmic time gives}
\begin{equation}
 \frac{dq_i}{dt}=q_i\left(H_{\hat q}-H_i\right),
 \qquad
 H_{\hat q}\equiv \sum_i H_i q_i^2.
 \label{eq:qievolution}
\end{equation}
\rev{Thus the principal axes focus or defocus the direction cosines according to the difference between the directional expansion rate $H_i$ and the expansion rate sampled by the photon. The quantity $H_{\hat q}$ can be written covariantly as}
\begin{equation}
 H_{\hat q}=H+\sigma_{ij}q^i q^j,
 \label{eq:Hq_shear}
\end{equation}
\rev{Using Eq.~\eqref{eq:k0bianchi}, the energy-loss equation takes the compact form}
\begin{equation}
 \frac{d\ln E}{dt}=-H_{\hat q}
 =-H-\sigma_{ij}q^i q^j.
 \label{eq:energy_loss_bianchi}
\end{equation}
Therefore the exact redshift can equivalently be expressed as a path integral,
\begin{equation}
 \ln(1+z)=\int_{t_s}^{t_o}\left[H(t)+\sigma_{ij}(t)q^i(t)q^j(t)\right]dt,
 \label{eq:redshift_integral}
\end{equation}
This equation is conceptually important because it separates the isotropic volume expansion from the shear projection along the ray. In FLRW the second term vanishes and the integral reduces to $\ln(a_o/a_s)$, whereas in Bianchi I the observed redshift contains an accumulated quadrupolar correction.

For weak anisotropy we write, as in Section~\ref{sec:bianchi},
\begin{equation}
 a_i(t)=a(t)e^{\beta_i(t)},
 \qquad
 \sum_i\beta_i(t)=0, 
 \label{eq:ai_beta_41}
\end{equation}
\rev{Expanding Eq.~\eqref{eq:redshift_observed_direction} to first order gives}
\begin{equation}
 1+z\simeq \frac{a_o}{a_s}
 \left[1+\sum_i n_i^2\beta_i(t_o)-\sum_i n_i^2\beta_i(t_s)\right].
 \label{eq:redshiftlinear}
\end{equation}
\rev{Equivalently, since $\dot\beta_i=\sigma_i$ in the principal-axis frame,}
\begin{equation}
 \frac{1+z}{1+z_{\rm FLRW}}-1
 \simeq \sum_i n_i^2\int_{t_s}^{t_o}\sigma_i(t)\,dt
 =\int_{t_s}^{t_o}\sigma_{ij}(t)n^i n^j\,dt,
 \label{eq:redshift_shear_integral_linear}
\end{equation}
\rev{where the replacement $q_i(t)\rightarrow n_i$ inside the shear term is sufficient at first order. Equation~\eqref{eq:redshift_shear_integral_linear} shows explicitly that the leading anisotropic correction is trace-free and quadrupolar. A constant value of $\beta_i$ at the observer can be changed by a constant rescaling of the spatial coordinates; the physical information is contained in the relative anisotropic deformation between source and observer, or, equivalently, in the integrated shear.}

In the axisymmetric limit introduced in Eq.~\eqref{eq:axisym}, with $a_1=a_2=a_\perp$ and $a_3=a_{\parallel}$, let $\mu\equiv \hat n\cdot\hat e$ be the cosine of the angle between the observed direction and the symmetry axis. From Eqs.~\eqref{eq:axisymH} and~\eqref{eq:axisymshear} one obtains
\begin{equation}
 H_{\hat n}=H+\Delta H_{\rm ax}\left(\mu^2-\frac{1}{3}\right),
 \qquad
 \Delta H_{\rm ax}\equiv H_{\parallel}-H_{\perp}.
 \label{eq:Hn_axisymmetric_41}
\end{equation}
and the corresponding first-order redshift correction is
\begin{equation}
 \frac{1+z}{1+z_{\rm FLRW}}-1
 \simeq \int_{t_s}^{t_o}\Delta H_{\rm ax}(t)
 \left[\mu^2(t)-\frac{1}{3}\right]dt.
 \label{eq:redshift_axisym_integral}
\end{equation}
If the anisotropy is very small, the change of $\mu$ along the ray is a second-order effect and one may replace $\mu(t)$ by the observed value $\mu_o$. This is the origin of the quadrupolar templates used later for luminosity-distance and distance-modulus residuals.

Several consequences follow. 

First, the redshift--time conversion itself becomes direction dependent. A source observed at fixed $z$ along the symmetry axis and a source observed at the same $z$ in the orthogonal plane need not correspond to exactly the same emission time in the underlying homogeneous foliation. 

Second, the anisotropic redshift is not a dipole generated by the observer's peculiar velocity. A velocity correction is odd under $\hat n\rightarrow -\hat n$, whereas the Bianchi I shear projection is even and has the angular structure of a quadrupole. 

Third, the low-redshift limit of Eq.~\eqref{eq:Hn_axisymmetric_41} gives the directional Hubble law used in Section~\ref{sec:observables}, while the finite-redshift expression in Eq.~\eqref{eq:redshift_axisym_integral} explains why the distance residual is controlled by the integrated history of the shear and not only by its present value.

This distinction is central for the Hubble-tension problem. In fact, an isotropic analysis assigns the same background relation $t_s=t_s(z)$ to all sky directions and then fits a single scalar expansion rate, whereas in a Bianchi I analysis the mapping between redshift, emission time and distance carries a quadrupolar dependence from the start. A finite and anisotropically distributed catalogue can therefore mix the quadrupole into the fitted monopole if the analysis imposes an FLRW template. The effect is nevertheless strongly constrained, because the same shear that perturbs the local Hubble law also contributes to the accumulated redshift and to the optical distortions discussed in the next subsection.

Finally, this homogeneous anisotropic mechanism should be kept distinct from radially inhomogeneous alternatives. Bianchi I introduces a preferred axis but no preferred spatial centre, whereas Lema\^{\i}tre--Tolman--Bondi or local-void descriptions introduce radial dependence and depend on the observer's position relative to a centre. Both classes of models can alter the interpretation of an isotropically fitted Hubble diagram, but their angular signatures and consistency requirements are different.

\subsection{Optical distances, Hubble residuals and the Jacobi map}

\rev{The connection between this optical subsection and the Hubble tension is direct. The local value of $H_0$ is not measured from redshift alone; it is inferred from a calibrated distance--redshift relation, most importantly from Type Ia supernova luminosity distances after the distance ladder has fixed the absolute magnitude. A Bianchi I model can therefore be relevant to the Hubble tension only if it specifies both the directional redshift of Section~\ref{sec:observables} and the directional luminosity distance that enters the Hubble-flow intercept. At low redshift, $D_L\simeq cz/H_0$, so a direction-dependent correction to $D_L$ is equivalent, when analysed with an isotropic FLRW template, to a direction-dependent effective value of $H_0$. The Jacobi map is introduced because it is the relativistic object that connects the anisotropic expansion of the metric to the observed area of a light beam and hence to the distance modulus used in Hubble-diagram analyses.}

In the geometric-optics limit the angular-diameter distance is obtained from the Jacobi map, whose evolution is fixed by the screen-projected geodesic-deviation equation. This is the standard Sachs-optical description of light propagation, but in the present context it has a specific role: it tells us whether the shear that changes the photon redshift also changes the beam area in a way that can mimic, bias or constrain a scalar FLRW estimate of $H_0$ \cite{Sachs1961Optics,KristianSachs1966,Gasperini2011LightCone,Fanizza2013Jacobi,Fanizza2015LightSignals,Fleury2016GLC}. The works of Gasperini, Marozzi, Nugier and Veneziano on geodesic-light-cone coordinates, and the exact Jacobi-map construction of Fanizza, Gasperini, Marozzi and Veneziano, are used here in this limited and precise sense: they provide a covariant optical language in which luminosity and area distances can be defined without first forcing the light cone into an FLRW form.

Let $k^\mu$ be the photon wave vector and let $s_A^{\mu}$, with $A=1,2$, be a Sachs screen basis orthogonal to both $k^\mu$ and the observer four-velocity. If $\xi^A$ is the screen-projected separation between two neighbouring light rays, the Jacobi map ${\cal D}^{A}{}_{B}$ is defined by
\begin{equation}
 \xi^A(\lambda_s)={\cal D}^{A}{}_{B}(\lambda_s,\lambda_o)\left.\frac{d\xi^B}{d\lambda}\right|_{o},
 \label{eq:jacobi_definition}
\end{equation}
and satisfies
\begin{equation}
 \frac{d^2}{d\lambda^2}{\cal D}^{A}{}_{B}
 ={\cal R}^{A}{}_{C}{\cal D}^{C}{}_{B},
 \qquad
 {\cal R}_{AB}=R_{\mu\nu\alpha\beta}k^\nu k^\alpha s_A^{\mu}s_B^{\beta},
 \label{eq:jacobi_equation}
\end{equation}
with observer boundary conditions
\begin{equation}
 {\cal D}^{A}{}_{B}(\lambda_o)=0,
 \qquad
 \left.\frac{d}{d\lambda}{\cal D}^{A}{}_{B}\right|_{o}=\delta^{A}{}_{B}.
 \label{eq:jacobi_boundary}
\end{equation}
The angular-diameter distance is then
\begin{equation}
 D_A^2=\det {\cal D}^{A}{}_{B},
 \qquad
 D_L=(1+z)^2D_A,
 \label{eq:jacobi_distance}
\end{equation}
where the second equality is Etherington reciprocity. Equations~\eqref{eq:jacobi_definition}--\eqref{eq:jacobi_distance} are the optical bridge to the Hubble tension: supernovae and standard sirens constrain $H_0$ through $D_L$, while BAO and strong-lensing time delays involve $D_A$ or combinations of angular-diameter distances and, if the optical tidal matrix is direction dependent, the data being compressed into one FLRW number actually contain a monopole plus possible angular residuals.

\rev{For the homogeneous Bianchi I background, the Weyl part of the optical tidal matrix is tied to the shear of the expansion rather than to localized lenses. Consequently, even in the absence of clumpy matter, a light bundle can experience direction-dependent focusing. The geodesic-light-cone approach is useful because it formulates this statement directly on the observer's past light cone and shows that angular averages of distance observables need not reduce to the FLRW result when a global anisotropy is present \cite{Gasperini2011LightCone,Fanizza2013Jacobi,Fanizza2015LightSignals,Fleury2016GLC}. The aim here is not to develop a full non-perturbative optical solution, but to obtain a controlled diagnostic form for the distance residual that can be confronted with the Hubble diagram.}

\begin{revision}
The local expansion of the Jacobi map also makes the separation between redshift mapping and beam focusing explicit. Orient an affine parameter from the observer toward the source and denote its separation by $\ell$. Integrating Eq.~\eqref{eq:jacobi_equation} twice with the boundary conditions in Eq.~\eqref{eq:jacobi_boundary} gives the exact Volterra form
\begin{equation}
 {\cal D}^{A}{}_{B}(\ell)=\ell\,\delta^{A}{}_{B}
 +\int_{0}^{\ell}{\rm d}\ell'\,(\ell-\ell')
 {\cal R}^{A}{}_{C}(\ell')\,{\cal D}^{C}{}_{B}(\ell').
 \label{eq:jacobi_volterra}
\end{equation}
This expression displays explicitly the cumulative character of optical focusing: the Jacobi map at the source depends on the optical tidal matrix along the entire null ray. Expanding the integrand about the observer gives
\begin{equation}
 {\cal D}^{A}{}_{B}(\ell)=\ell\,\delta^{A}{}_{B}
 +\frac{\ell^3}{6}{\cal R}^{A}{}_{B}(o)
 +\frac{\ell^4}{12}\left.\frac{{\rm d}{\cal R}^{A}{}_{B}}{{\rm d}\ell}\right|_{o}
 +{\cal O}(\ell^5),
 \label{eq:jacobi_local_series}
\end{equation}
so that
\begin{equation}
 D_A(\ell)=\ell\left[1+\frac{\ell^2}{12}\,\mathrm{tr}\,{\cal R}(o)
 +\frac{\ell^3}{24}\left.\frac{{\rm d}}{{\rm d}\ell}\mathrm{tr}\,{\cal R}\right|_{o}
 +{\cal O}(\ell^4)\right].
 \label{eq:DA_local_series}
\end{equation}
The appearance of the observer-side matrix ${\cal R}(o)$ is therefore not an assumption that the curvature acts only at the observer or remains constant along the ray. It is the leading coefficient in the observer-centred expansion of the exact line-of-sight integral in Eq.~\eqref{eq:jacobi_volterra}. With the curvature-sign convention of Eq.~\eqref{eq:jacobi_equation}, the sign is carried by ${\cal R}_{AB}$. Optical focusing first modifies the distance at absolute order $\ell^3$, or relative order $z^2$, and this coefficient is fixed by ${\cal R}(o)$. Derivatives of the curvature, and hence the first correction associated with its variation along the ray, enter at absolute order $\ell^4$, or relative order $z^3$. The leading and linear-in-$z$ distance quadrupole is instead fixed by the directional redshift--affine-parameter mapping. The calculation in Section~\ref{sec:lowzworked} retains the complete relative-$z^2$ term. It is thus the correct local limit of the cumulative focusing problem, but it is not a substitute for exact line-of-sight integration at finite redshift.

For a general finite-redshift analysis we define the fractional luminosity-distance quadrupole through a multiplicative correction to the isotropic luminosity distance,
\begin{equation}
 D_L(z,\hat n)=D_L^{\rm FLRW}(z)\,[1+\Delta(z,\hat n)],
 \label{eq:DLpert}
\end{equation}
where $D_L^{\rm FLRW}$ is evaluated for the corresponding mean expansion history and $\Delta$ contains the directional correction. Unless a shear or anisotropic-stress history has been specified and the optical equations have been solved, $A_D(z)$ below must be regarded as a phenomenological fractional-distance amplitude rather than as an independently derived shear observable. The weak-shear calculation in Section~\ref{sec:lowzworked} supplies an explicit dynamical relation at low redshift. At fixed redshift, a positive $\Delta$ means that the source appears farther than in the isotropic template. If such a correction is ignored, part of it can be absorbed into the supernova intercept, the absolute-magnitude calibration, or the scalar value of $H_0$. Conversely, the absence of a stable quadrupole in well-controlled data would rule out this geometrical explanation of the tension.

For an axisymmetric model the leading correction is naturally written as
\begin{equation}
 \Delta(z,\hat n)=A_D(z)\left[(\hat n\cdot \hat e)^2-\frac{1}{3}\right],
 \label{eq:quadDL}
\end{equation}
\rev{Here $\hat e$ is the preferred axis and $A_D(z)$ is the fractional luminosity-distance quadrupole.} The subtraction of $1/3$ is important for the interpretation of $H_0$: in an ideal full-sky and perfectly uniform catalogue the quadrupole has no monopole, whereas in a real distance-ladder sample the sky mask, the calibrator distribution and the Hubble-flow selection can allow part of this quadrupole to leak into the fitted scalar intercept. More generally, the quadrupole may be expanded as
\begin{equation}
 \Delta(z,\hat n)=\sum_{m=-2}^{2}A_{2m}(z)Y_{2m}(\hat n),
 \label{eq:harmDL}
\end{equation}
and this harmonic representation is useful because it separates a physical axial Bianchi I signal from a generic survey residual. A genuine axisymmetric background predicts a stable axis and a coherent redshift dependence of the five $A_{2m}$ coefficients; a calibration or selection effect may instead produce a quadrupole with no stable axis, with additional dipole or higher-multipole contamination, or with a strong dependence on the adopted sample cuts.

The distance modulus is
\begin{equation}
 \mu(z,\hat n)=5\log_{10}\left[\frac{D_L(z,\hat n)}{\rm Mpc}\right]+25,
 \label{eq:mu_def}
\end{equation}
for $|\Delta|\ll 1$,
\begin{equation}
 \delta\mu(z,\hat n)\simeq \frac{5}{\ln 10}\Delta(z,\hat n),
 \label{eq:deltamu}
\end{equation}
thus a percent-level distance modulation corresponds to a distance-modulus residual of order $2\times10^{-2}$ mag. This number is small compared with the intrinsic dispersion of a single Type Ia supernova, but it is not negligible for a large and well-calibrated catalogue because the inferred $H_0$ is controlled by the collective Hubble-flow intercept. Figure~\ref{fig:quadrupoleresidual} should therefore be read as a Hubble-tension diagnostic rather than as a merely illustrative optical effect: it shows the angular residual that must be fitted, marginalized over, or constrained before one can claim that an anisotropic geometry has shifted the scalar FLRW value of $H_0$.

\begin{figure}[!t]
\centering
\includegraphics[width=0.74\textwidth]{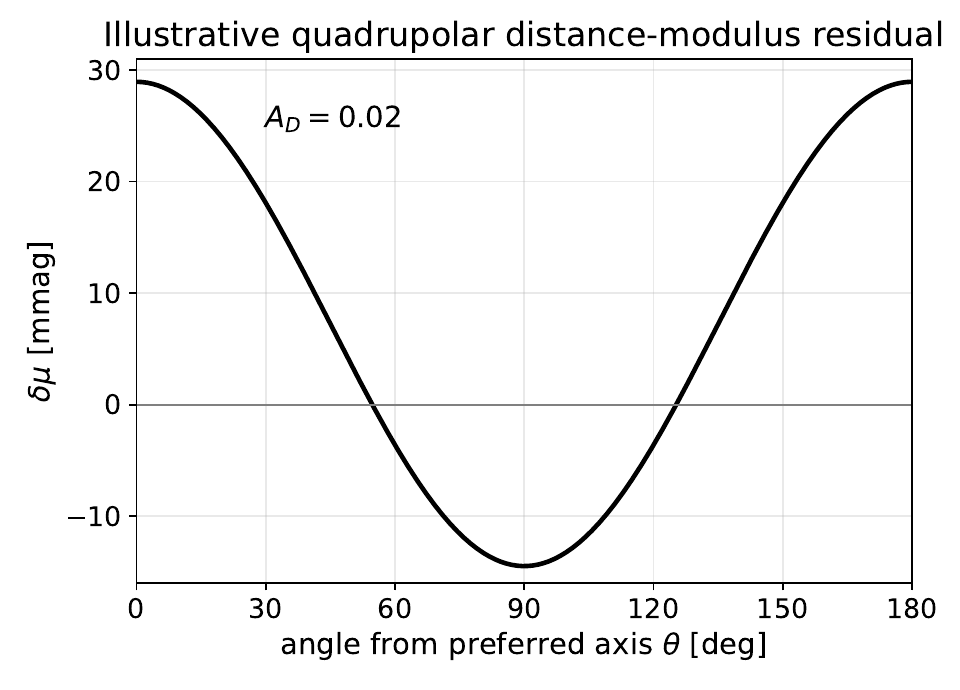}
\caption{Illustrative quadrupolar distance-modulus residual generated by Eq.~\eqref{eq:deltamu} with $A_D=0.02$ in Eq.~\eqref{eq:quadDL}. The vertical scale is in millimagnitudes. The figure is not a fit to supernova data; it is a diagnostic template showing the angular pattern that should be searched for in Hubble-diagram residuals before interpreting a scalar shift in $H_0$ as evidence for or against Bianchi-I-like anisotropy.}
\label{fig:quadrupoleresidual}
\end{figure}

\end{revision}
\begin{revision}
\subsection{Low-redshift limit and effective \texorpdfstring{$H_0$}{H0} bias}

The low-redshift limit is the place where the connection between anisotropic expansion and the operational definition of $H_0$ is most transparent. In an exactly isotropic spacetime the first term of the distance--redshift relation is universal: every sufficiently nearby source obeys the same linear Hubble law after correcting for peculiar velocities, while in a Bianchi I spacetime the same statement is no longer true. The local expansion measured by a comoving observer depends on the direction of the incoming photon through the shear projection. Neglecting, for the moment, peculiar velocities and the quadratic terms of the cosmographic series, one may write
\begin{equation}
 cz\simeq H_{\rm loc}(\hat n)D,
 \label{eq:directionalHlaw}
\end{equation}
where $D$ is the physical distance assigned in the local tetrad and
\begin{equation}
 H_{\rm loc}(\hat n)=H_0+\sigma_{ij,0}\hat n^i\hat n^j.
 \label{eq:Hloc_shear_general}
\end{equation}
This equation is only the first term of the Kristian--Sachs expansion, but it already contains the essential point: the linear Hubble law measures a scalar only if the shear term is absent or if the angular distribution of the sample averages it away \cite{KristianSachs1966,EllisVanElst1999}.

For the axisymmetric case, with $H_\parallel-H_\perp\equiv \Delta H_{\rm ax}$, Eq.~\eqref{eq:Hloc_shear_general} becomes
\begin{equation}
 H_{\rm loc}(\hat n)=H_0\left\{1+B_{H0}\left[(\hat n\cdot\hat e)^2-\frac{1}{3}\right]\right\},
 \qquad
 B_{H0}\equiv \frac{\Delta H_{\rm ax,0}}{H_0},
 \label{eq:Hlocdirectional}
\end{equation}
\rev{Here $\hat e$ is the preferred axis and $B_{H0}$ measures the present-day directional expansion contrast. The subtraction of $1/3$ removes the monopole of the quadrupolar template, so an ideal full-sky average measures the mean expansion rate $H_0$. The coefficient $B_{H0}$ should not be identified with the finite-redshift distance amplitude $A_D(z)$ introduced above. The former is a local, instantaneous quantity controlled by the present shear projection, whereas $A_D(z)$ contains the accumulated effects of redshift evolution, photon propagation and optical focusing. In a shear-only model, $B_{H0}$ and $A_D(z)$ are dynamically linked; in a late-time phenomenological model, they should be treated as related but non-identical observables.}

\subsubsection*{Worked weak-shear mapping}\label{sec:lowzworked}

The weak-shear mapping can be completed analytically through relative order $z^2$. Define
\begin{equation}
 H_{\hat n}=H\left[1+B_H(t)Q(\hat n)\right],
 \qquad B_H(t)\equiv\frac{\Delta H_{\rm ax}(t)}{H(t)},
 \qquad Q(\hat n)\equiv(\hat n\cdot\hat e)^2-\frac{1}{3}.
 \label{eq:BH_definition}
\end{equation}
At first order in the shear, the change of the observed direction along the ray contributes only at ${\cal O}(B_H^2)$. The directional deceleration and jerk parameters may therefore be written as
\begin{align}
 q_{\hat n}&\equiv-1-\frac{\dot H_{\hat n}}{H_{\hat n}^2}
 =q+\delta q_H Q(\hat n)+{\cal O}(B_H^2),\nonumber\\
 \delta q_H&=-\left[\frac{\dot{\Delta H}_{\rm ax}}{H^2}+2(1+q)B_H\right],
 \label{eq:q_directional_explicit}\\
 j_{\hat n}&\equiv1+3\frac{\dot H_{\hat n}}{H_{\hat n}^2}
 +\frac{\ddot H_{\hat n}}{H_{\hat n}^3}
 =j+\delta j_H Q(\hat n)+{\cal O}(B_H^2),\nonumber\\
 \delta j_H&=\frac{\ddot{\Delta H}_{\rm ax}}{H^3}
 +3\frac{\dot{\Delta H}_{\rm ax}}{H^2}-3(j+q)B_H.
 \label{eq:j_directional_explicit}
\end{align}
Here $q=-1-\dot H/H^2$ and $j=1+3\dot H/H^2+\ddot H/H^3$ refer to the mean background.

To display the optical contribution separately, decompose the trace of the observer-side optical tidal matrix as
\begin{equation}
 \mathrm{tr}\,{\cal R}(o,\hat n)={\cal R}_0+{\cal R}_2 Q(\hat n),
 \qquad s_0\equiv\frac{{\cal R}_0}{12H_0^2}.
 \label{eq:optical_trace_decomposition}
\end{equation}
Here ${\cal R}_0$ and ${\cal R}_2$ are, respectively, the monopolar and quadrupolar coefficients of the optical tidal matrix evaluated at the observer. They are local cosmographic coefficients, not line-of-sight averages of the curvature. Their use is sufficient for the relative-$z^2$ Kristian--Sachs expansion because Eq.~\eqref{eq:DA_local_series} shows that derivatives of ${\cal R}_{AB}$ first enter one relative order later. A finite-redshift prediction, however, requires the evolution of the full optical tidal matrix along the complete null ray through Eq.~\eqref{eq:jacobi_volterra}.
Combining the redshift--affine-parameter expansion with Eqs.~\eqref{eq:jacobi_local_series} and \eqref{eq:DA_local_series} gives the directional Kristian--Sachs series, with $c=1$ inside the dimensionless coefficients,
\begin{align}
 D_L(z,\hat n)&=\frac{cz}{H_{\hat n,0}}
 \left[1+\frac{1-q_{\hat n,0}}{2}z+C_{\hat n,0}z^2+{\cal O}(z^3)\right],
 \label{eq:DL_directional_cosmography}\\
 C_{\hat n,0}&=\frac{-j_{\hat n,0}+3q_{\hat n,0}^2+2q_{\hat n,0}}{6}
 +\frac{\mathrm{tr}\,{\cal R}(o,\hat n)}{12H_{\hat n,0}^2}.
 \label{eq:C_directional_cosmography}
\end{align}
The first term in Eq.~\eqref{eq:C_directional_cosmography} is fixed by the direction-dependent redshift--affine-parameter mapping. The second is the beam-area contribution obtained from the Jacobi map.

Dividing by the corresponding mean-background series and retaining terms linear in the anisotropy yields
\begin{equation}
 \frac{D_L(z,\hat n)}{D_L^{\rm FLRW}(z)}-1
 =A_D(z)Q(\hat n)+{\cal O}(z^3,B_H^2),
 \qquad A_D=A_{\rm map}+A_{\rm foc},
 \label{eq:AD_map_foc_decomposition}
\end{equation}
where
\begin{align}
 A_{\rm map}(z)&=-B_{H0}-\frac{1}{2}\delta q_{H0}z
 +\left[-\frac{\delta j_{H0}}{6}
 +\frac{9q_0+7}{12}\delta q_{H0}\right]z^2,
 \label{eq:AD_map_general}\\
 A_{\rm foc}(z)&=\left[\frac{{\cal R}_2}{12H_0^2}-2B_{H0}s_0\right]z^2.
 \label{eq:AD_foc_general}
\end{align}
Equations~\eqref{eq:AD_map_general} and \eqref{eq:AD_foc_general} explicitly identify the two contributions requested by the optical problem. The coefficient ${\cal R}_2$ is the quadrupolar part of the Ricci focusing. The Weyl part of ${\cal R}_{AB}$ is trace-free on the screen; consequently, at first order in the background anisotropy and at the observer-centred order retained here, it changes the beam shape but not the determinant of the Jacobi map. This statement is local and perturbative, not a claim that Weyl focusing is irrelevant at finite distance. In the full Sachs system, the Weyl-generated optical shear evolves cumulatively along the ray and contributes to the beam area through terms quadratic in the optical shear; those terms must be retained in an exact finite-redshift integration.

For the minimal shear-only model, the matter pressure is isotropic. The observer-side Ricci contraction is then direction independent, so ${\cal R}_2=0$. The isotropic focusing coefficient satisfies $s_0=-(1+q_0)/6$ with the sign convention of Eq.~\eqref{eq:jacobi_equation}. Hence
\begin{equation}
 A_{\rm foc}(z)=\frac{1+q_0}{3}B_{H0}z^2
 \qquad(\pi^i{}_j=0).
 \label{eq:AD_foc_isotropic_pressure}
\end{equation}
Although the direct quadrupolar Ricci term vanishes, the isotropic Ricci focusing still contributes to the distance quadrupole through the direction-dependent normalization by $H_{\hat n,0}$.

For freely decaying axisymmetric shear, $\dot{\Delta H}_{\rm ax}+3H\Delta H_{\rm ax}=0$. Equations~\eqref{eq:q_directional_explicit} and \eqref{eq:j_directional_explicit} then give
\begin{equation}
 \delta q_{H0}=(1-2q_0)B_{H0},
 \qquad \delta j_{H0}=3(1-j_0)B_{H0}.
 \label{eq:dq_dj_shearonly}
\end{equation}
The two pieces become
\begin{align}
 A_{\rm map}(z)&=-B_{H0}+\frac{2q_0-1}{2}B_{H0}z
 +\frac{1-5q_0-18q_0^2+6j_0}{12}B_{H0}z^2,
 \label{eq:AD_map_shearonly}\\
 A_{\rm foc}(z)&=\frac{1+q_0}{3}B_{H0}z^2,
 \label{eq:AD_foc_shearonly}
\end{align}
and therefore
\begin{equation}
 \boxed{A_D(z)=-B_{H0}+\frac{2q_0-1}{2}B_{H0}z
 +\frac{5-q_0-18q_0^2+6j_0}{12}B_{H0}z^2}
 +{\cal O}(z^3,B_{H0}^2).
 \label{eq:AD_lowz_shearonly}
\end{equation}
This result is the completed low-redshift map from the specified shear history to the luminosity-distance quadrupole. It includes both the redshift correction and the first non-vanishing Jacobi-focusing contribution.

The domain of validity of Eq.~\eqref{eq:AD_lowz_shearonly} must be emphasized. It is a local Kristian--Sachs cosmographic expansion about $z=0$, truncated at relative order $z^2$, and is used here only as a controlled low-redshift benchmark, in particular for the numerical scale comparison at $z_{\rm eff}=0.15$. It should not be extrapolated over the full redshift range of modern supernova or BAO catalogues. At intermediate and high redshift, the neglected $\mathcal O(z^3)$ and higher-order terms need not remain subdominant, and the accuracy of the truncated series becomes dependent on the background expansion and on the adopted shear history. A catalogue-level prediction must instead be obtained by integrating the exact direction-dependent null-geodesic equations together with the Sachs--Jacobi system for the specified shear or anisotropic-stress history, and by constructing $D_L(z,\hat n)$ without a cosmographic truncation in redshift. The phenomenological functions $A_D(z)$ introduced below provide flexible residual-level templates, but they should not be identified with the exact finite-redshift Bianchi I prediction unless the full optical system has been solved.

For $\Omega_{m0}=0.315$, $\Omega_{r0}=9\times10^{-5}$ and a flat cosmological-constant background, $q_0\simeq-0.527$ and $j_0\simeq1.000$. Thus
\begin{equation}
 A_D(z)\simeq-B_{H0}\left(1+1.027z-0.544z^2\right),
 \qquad A_\mu(z)\equiv\frac{5}{\ln10}A_D(z),
 \label{eq:AD_numeric_lowz}
\end{equation}
where $A_\mu$ is the quadrupole amplitude in magnitudes. This expression is used in Section~\ref{sec:minimal} for the numerical scale comparison.

The immediate consequence of Eq.~\eqref{eq:Hlocdirectional} is that a scalar value of $H_0$ inferred from a real catalogue is not necessarily the mathematical monopole of the spacetime. It is a weighted estimator over the angular and redshift distribution of the objects that enter the fit. If the catalogue window is denoted by $W(\hat n,z)$, the strictly local Hubble-law limit is
\begin{equation}
 H_0^{\rm eff}\simeq H_0\left[1+\left\langle B_H(z)Q(\hat n)\right\rangle_W\right],
 \label{eq:H0windowbias}
\end{equation}
with
\begin{equation}
 \left\langle X\right\rangle_W=
 \frac{\int {\rm d}z\,{\rm d}\Omega\,W(\hat n,z)X(\hat n,z)}
 {\int {\rm d}z\,{\rm d}\Omega\,W(\hat n,z)}.
 \label{eq:windowaverage}
\end{equation}
The finite-redshift form follows directly from the luminosity-distance quadrupole:
\begin{equation}
 \frac{\delta H_0^{\rm fit}}{H_0}
 \simeq-\left\langle A_D(z)Q(\hat n)\right\rangle_W.
 \label{eq:H0windowbias_finitez}
\end{equation}
Equation~\eqref{eq:H0windowbias_finitez} reduces to Eq.~\eqref{eq:H0windowbias} as $z\rightarrow0$, because $A_D(0)=-B_{H0}$. For a uniform full-sky catalogue, $\langle Q\rangle_W=0$. For a finite survey with non-uniform sky coverage, redshift-dependent selection, calibration subregions or masked zones, the average need not vanish; a genuine quadrupole can then leak into the fitted monopole.

The calibration step makes the same point more sharply. If the calibrator and Hubble-flow samples have different windows, the leading finite-redshift shift is
\begin{equation}
 \frac{\delta H_0}{H_0}\simeq
 -\left\langle A_DQ\right\rangle_{\rm HF}
 +\left\langle A_DQ\right\rangle_{\rm cal}.
 \label{eq:ladder_window_bias}
\end{equation}
A quadrupole can therefore be absorbed partly into the inferred absolute magnitude, partly into the Hubble-flow intercept and partly into peculiar-velocity corrections. Its observable effect depends on the relative angular and redshift distributions of anchors, calibrators and Hubble-flow supernovae.

In distance-modulus language, since $D_L\propto H_0^{-1}$ at fixed redshift to leading order, a small directional change of the locally inferred Hubble rate produces
\begin{equation}
 \delta\mu(\hat n)\simeq -\frac{5}{\ln 10}\,B_{H0} Q(\hat n),
 \label{eq:mubiaslowz}
\end{equation}
up to the sign convention used for the distance residual. More generally,
\begin{equation}
 \delta\mu(z,\hat n)\simeq -\frac{5}{\ln 10}\,B_H(z)Q(\hat n)+\delta\mu_{\rm int}(z,\hat n),
 \label{eq:mubiaslowz_general}
\end{equation}
where $\delta\mu_{\rm int}$ denotes the additional integrated contribution associated with the propagation between source and observer. This term is absent from a purely local Hubble-law argument but is present in the Jacobi-map description and it is one reason why a low-redshift quadrupole and an intermediate-redshift quadrupole should not automatically be forced to have the same amplitude.

The effect is difficult to isolate observationally because the same angular templates can be contaminated by other low-redshift phenomena. Moreover, the observed redshift may be written schematically as
\begin{equation}
 z_{\rm obs}=z_{\rm cos}+(1+z_{\rm cos})\frac{\bm v\cdot\hat n}{c}+\delta z_{\rm grav}+\delta z_{\rm sys},
 \label{eq:zobs_peculiar}
\end{equation}
where $\bm v$ is the peculiar velocity of the source relative to the chosen frame. The velocity term is dominantly dipolar for a coherent bulk flow, but survey geometry, Malmquist selection, extinction corrections and calibration inhomogeneities can mix dipolar and quadrupolar patterns in a finite sample. A few-millimagnitude quadrupole is already interesting in a large supernova catalogue, but it is easily confused with calibration gradients, peculiar-velocity modelling, host-galaxy population corrections or uneven angular sampling \cite{Zhao2013Anisotropy,Lin2016JLAIsotropy,Javanmardi2015Probing,Colin2019Evidence,Soltis2019Percent}.

A useful order-of-magnitude estimate shows the severity of the requirement. The \rev{Planck 2018--SH0ES 2022 benchmark separation} shown in Fig.~\ref{fig:h0comparison} corresponds to a fractional difference of order several percent; if this were produced only by the local leakage term in Eq.~\eqref{eq:H0windowbias}, one would need
\begin{equation}
 B_{H0}\simeq \frac{\delta H_0^{\rm fit}/H_0}{\langle Q\rangle_W}.
 \label{eq:requiredB0}
\end{equation}
Since $|Q|\leq 2/3$ and realistic window averages are usually much smaller than the maximum value, a large scalar shift would require a substantial directional expansion contrast. Such a contrast is precisely what CMB isotropy, BAO consistency, BBN and Hubble-diagram isotropy tests strongly restrict. The low-redshift formula is therefore best regarded as a diagnostic relation, not as evidence that anisotropy automatically resolves the tension.

A professional analysis should consequently fit the monopole, dipole and quadrupole simultaneously. The absolute-magnitude calibration, the velocity field and the quadrupolar amplitude should not be optimized in separate steps, because a change in one can mimic a change in the others. A minimal phenomenological residual model is
\begin{equation}
 \Delta\mu_i = \Delta\mu_{\rm mono}(z_i)+D(z_i)\,\hat n_i\cdot\hat p
 +A_\mu(z_i)\left[(\hat n_i\cdot\hat e)^2-\frac{1}{3}\right]+\epsilon_i,
 \label{eq:lowzresidualmodel}
\end{equation}
where the dipole term accounts for bulk-flow and velocity-frame effects, while the quadrupole term represents the Bianchi-I-like contribution. In matrix notation the corresponding Gaussian likelihood is
\begin{equation}
 \chi^2_{\rm SN}=\left[\Delta\bm\mu-\bm T(\Theta)\right]^{\rm T}
 {\bf C}^{-1}
 \left[\Delta\bm\mu-\bm T(\Theta)\right],
 \label{eq:snquadrupole_chi2}
\end{equation}
where ${\bf C}$ must include both statistical and systematic covariance contributions and $\bm T$ contains the monopole, dipole and quadrupole templates. If the covariance matrix is ignored, or if the sky mask is not propagated into the fit, the apparent significance of an angular residual can be badly misestimated.

The preferred axis $\hat e$ should be inferred from the data and then tested for stability under changes of sky mask, redshift cuts, calibrator subset, light-curve standardization, host-mass correction and peculiar-velocity model. If the preferred axis moves when these choices are changed, the signal is more naturally interpreted as a catalogue or calibration effect than as homogeneous anisotropic expansion. Eq.~\eqref{eq:H0windowbias} therefore gives the operational meaning of a possible effective $H_0$ bias: one must ask whether the same data, analysed without imposing exact isotropy, prefer a statistically stable quadrupolar residual and whether the inferred amplitude is compatible with the independent limits discussed below. This motivates the likelihood strategy described in Section~\ref{sec:data} and the falsification criteria discussed in Section~\ref{sec:diagnostics}.

\end{revision}
\subsection{Connection with BAO, standard sirens and parallax tests}

The quadrupolar distance correction has different observational manifestations in different probes, and this multi-probe structure is essential for assessing whether a signal is physical. \rev{In supernova data, it appears as an angular modulation of the Hubble-diagram residuals; in BAO data, it affects the mapping between observed angles, redshift intervals and comoving separations; in standard sirens, it enters the luminosity distance inferred from the gravitational-wave amplitude; and in cosmic-parallax and redshift-drift measurements, it appears as a direct kinematical signature of anisotropic expansion.} These observables are affected by different systematics, so consistency among them would be far more significant than a quadrupole detected in only one catalogue.

BAO deserve special care because most public BAO constraints are already compressed under FLRW assumptions. In an isotropic background the standard combinations are
\begin{equation}
 D_H(z)=\frac{c}{H(z)},\qquad D_M(z)=(1+z)D_A(z),\qquad
 D_V(z)=\left[zD_H(z)D_M^2(z)\right]^{1/3}.
 \label{eq:baoflrwdefs}
\end{equation}
\rev{In a Bianchi I spacetime, these quantities are not unique scalar functions of redshift alone. The radial scale probes a direction-dependent expansion rate, whereas the transverse scale depends on the orientation of the screen basis relative to the principal axes of the metric.} Schematically one should replace
\begin{equation}
 D_H(z)\rightarrow D_H(z,\hat n),\qquad
 D_A(z)\rightarrow D_A(z,\hat n,\rev{\varphi_{\rm scr}}),
 \label{eq:baodirectional}
\end{equation}
\rev{where $\varphi_{\rm scr}$ denotes the orientation of the transverse separation on the observer screen. The appearance of $\varphi_{\rm scr}$ is not a minor detail: in an anisotropic geometry the two transverse directions need not be equivalent, and the area distance measured by a beam is the determinant of a two-dimensional Jacobi map rather than a single scalar radius.}

A useful way to express this in the language of clustering analyses is through anisotropic dilation parameters. Relative to a fiducial FLRW model, one may define
\begin{equation}
 \alpha_\parallel(z,\hat n)=\frac{D_H^{\rm BI}(z,\hat n)/r_d^{\rm BI}}{D_H^{\rm fid}(z)/r_d^{\rm fid}},
 \qquad
 \alpha_{\perp A}(z,\hat n)=\frac{D_{A,A}^{\rm BI}(z,\hat n)/r_d^{\rm BI}}{D_A^{\rm fid}(z)/r_d^{\rm fid}},
 \label{eq:bao_alpha_bi}
\end{equation}
where $A=1,2$ labels the two screen directions. The FLRW BAO analysis effectively collapses this information into one radial dilation and one transverse dilation, or into the isotropized combination
\begin{equation}
 \alpha_{\rm iso}\simeq \left(\alpha_\parallel\alpha_{\perp1}\alpha_{\perp2}\right)^{1/3}.
 \label{eq:alpha_iso_bi}
\end{equation}
In a weakly anisotropic Bianchi I model the leading correction to these quantities may again be expanded in monopole and quadrupole pieces,
\begin{equation}
 \alpha_X(z,\hat n)=\alpha_{X0}(z)\left[1+B_X(z)Q(\hat n)+\cdots\right],
 \qquad X=\parallel,\perp1,\perp2,
 \label{eq:bao_quadrupole_alpha}
\end{equation}
\rev{Additional dependence on the screen orientation is also possible. Therefore, the usual isotropized quantity $D_V/r_d$ is a monopole compression of information that can contain quadrupolar angular dependence.}

\rev{This observation has a practical consequence. Published BAO likelihoods can be used conservatively as constraints on the mean expansion history and on the maximum allowed deviation from FLRW. They should not, however, be interpreted as direct likelihoods for a Bianchi I geometry unless the compression procedure is repeated in the anisotropic model. A complete BAO test would return to the clustering likelihood before isotropic compression, compute the anisotropic Alcock--Paczynski mapping in the Bianchi I metric, propagate it into the two-point correlation function or power spectrum, and test whether the monopole, quadrupole and higher multipoles of the clustering signal are consistent with the same preferred axis found in the supernova Hubble diagram \cite{Eisenstein2005BAO,Alam2021Eboss,DESI2024DR1BAO,DESI2025DR2BAO,DESI2025Cosmology}. Such an analysis is demanding but conceptually important: a supernova quadrupole with no compatible BAO imprint would be difficult to interpret as a property of the homogeneous background.}

\rev{Standard sirens provide a cleaner geometrical complement. In general relativity, gravitational and electromagnetic waves follow the same background null cones, while the waveform amplitude provides a luminosity distance without a supernova absolute-magnitude calibration. When an electromagnetic counterpart identifies the host galaxy, or when a statistical host association supplies a redshift, each event contributes a measurement of $D_L(z,\hat n)$ with a known sky direction \cite{Schutz1986StandardSirens,Abbott2017StandardSiren,Cai2018Sirens}. Individual uncertainties can be large because of inclination--distance degeneracy, weak lensing and detector antenna patterns, but these systematics differ from those of Type Ia supernovae. A future catalogue of standard sirens could therefore test whether a quadrupole in the electromagnetic Hubble diagram is reproduced by an independent distance ladder.}

\rev{The standard-siren version of the test is straightforward in principle. One may fit}
\begin{equation}
 D_L^{\rm GW}(z_i,\hat n_i)=D_{L,0}^{\rm GW}(z_i)
 \left\{1+A_{\rm GW}(z_i)\left[(\hat n_i\cdot\hat e_{\rm GW})^2-\frac{1}{3}\right]\right\}
 \label{eq:sirenquadrupole}
\end{equation}
\rev{The amplitude and axis can then be compared with the corresponding supernova values. In a purely geometrical interpretation one expects}
\begin{equation}
 \hat e_{\rm GW}\simeq \hat e_{\rm SN},
 \qquad
 A_{\rm GW}(z)\simeq A_{\rm SN}(z),
 \label{eq:siren_sn_consistency}
\end{equation}
within the different statistical errors and selection functions. Agreement would support a geometrical interpretation, but a disagreement would point instead to calibration, population, selection or local-environment effects. This comparison is especially valuable because a siren quadrupole is not degenerate with the supernova absolute magnitude $M$ in the same way as an electromagnetic distance-ladder measurement.

\rev{Strong-lensing time delays and cosmic chronometers provide additional, although less direct, checks. Time-delay lenses constrain combinations of angular-diameter distances between observer, lens and source. In a Bianchi I spacetime, these distances depend on the directions of the lens and source and on the orientation of the lens plane. An anisotropic background would therefore enter the time-delay distance as}
\begin{equation}
 D_{\Delta t}(z_l,z_s,\hat n)=(1+z_l)
 \frac{D_A(z_l,\hat n)D_A(z_s,\hat n)}{D_A(z_l,z_s,\hat n)}
 \left[1+\Delta_{\Delta t}(\hat n)\right],
 \label{eq:timedelay_aniso}
\end{equation}
\rev{The last factor summarizes the anisotropic correction. By contrast, cosmic chronometers are closer to probes of the mean expansion rate because they infer $H(z)$ from differential galaxy ages. They are therefore less sensitive to the preferred axis, but they help prevent a directional fit from compensating for an incorrect monopole expansion history.}

Cosmic parallax supplies an even more direct, although observationally more difficult, kinematical test. In an anisotropically expanding homogeneous universe, the angular separation between two distant sources is not exactly constant in time; the proper-motion field induced by the background shear is obtained by projecting the shear action onto the observer screen,
\begin{equation}
 \dot{\hat n}^{\,i}\simeq -\left(\delta^{ij}-\hat n^i\hat n^j\right)\sigma_{jk,0}\hat n^k,
 \label{eq:proper_motion_shear}
\end{equation}
up to sign conventions associated with the definition of the incoming direction. For the axisymmetric case this gives the characteristic angular dependence
\begin{equation}
 \dot\theta \sim \Delta H_{\rm ax,0}\,\sin\theta\cos\theta,
 \label{eq:parallaxschematic}
\end{equation}
where $\theta$ is the angle from the preferred axis and $\Delta H_{\rm ax,0}$ is the present directional expansion contrast. The signal is extremely small, but its angular structure is tied to the same shear tensor that enters Eq.~\eqref{eq:Hq_shear} \cite{Quercellini2009CosmicParallax,Fontanini2009CosmicParallax,Campanelli2011Parallax}. On the other hand, a detection of a distance quadrupole with no corresponding parallax or redshift-drift signature would require a careful explanation.

Redshift drift gives a related test on longer time baselines; in FLRW cosmology the Sandage--Loeb signal is controlled by $(1+z)H_0-H(z)$, in Bianchi I the same observable becomes direction dependent because both the observer expansion and the source expansion are projected along the photon direction. Schematically,
\begin{equation}
 \dot z(\hat n)=(1+z)H_{\rm loc,o}(\hat n)-H_{\rm loc,s}(\hat n_s)+{\cal I}_\sigma(z,\hat n),
 \label{eq:redshift_drift_bi}
\end{equation}
where ${\cal I}_\sigma$ denotes the integrated contribution from the evolution of the shear along the null ray. Although this measurement is not yet a competitive constraint on the Hubble tension, it is conceptually important because it tests the time derivative of the same anisotropic geometry that affects distances.

\rev{The observable imprint of Bianchi I geometry is not a single shifted value of $H_0$. It is a correlated set of monopole and quadrupole modifications in redshift, luminosity distance, angular-diameter distance, standard-ruler measurements, proper-motion fields and possibly redshift drift. A viable anisotropic interpretation must satisfy three consistency conditions. First, the preferred axis inferred from independent probes should be stable within the uncertainties. Second, the redshift dependence of the amplitudes should be compatible with a common shear or anisotropic-stress history. Third, the monopole expansion history should remain consistent with CMB, BAO, BBN, cosmic chronometers and local calibration data. Accordingly, the anisotropy must be fitted as a catalogue-level, direction-dependent signal and then confronted with BAO, CMB, standard-siren and cosmic-parallax consistency tests.}

\begin{revision}
\section{The Minimal Shear-Only Model}\label{sec:minimal}

The most economical Bianchi I extension of flat $\Lambda$CDM is obtained by keeping the matter sector isotropic and allowing only the freely decaying geometrical shear, in this case the mean expansion is governed by Eq.~\eqref{eq:Hsheronly}, and the new degree of freedom is the present-day dimensionless shear density $\Omega_{\sigma0}$. The parameter set may be written as
\begin{equation}
 \Theta_{\rm min}=\{H_0,\Omega_{m0},\Omega_{b0},\Omega_{r0},\Omega_{\Lambda0},\Omega_{\sigma0},M\},
 \label{eq:thetamin}
\end{equation}
where $M$ denotes the supernova absolute-magnitude nuisance parameter or, equivalently, the combination of absolute magnitude and $H_0$ normalization. In a strictly flat Bianchi I background these parameters are not all independent, because the closure condition in Eq.~\eqref{eq:closure} fixes one of the density parameters and a convenient implementation is therefore to sample
\begin{equation}
 \{H_0,\Omega_{m0},\Omega_{b0},\Omega_{r0},\Omega_{\sigma0},M\}
\end{equation}
with
\begin{equation}
 \Omega_{\Lambda0}=1-\Omega_{m0}-\Omega_{r0}-\Omega_{\sigma0},
 \label{eq:omegalambda_closure_minimal}
\end{equation}
\rev{while imposing $\Omega_{\sigma0}\geq0$. This non-negativity is not a phenomenological prior but a consequence of the definition $\Omega_{\sigma0}=\sigma_0^2/(3H_0^2)$. The minimal model therefore differs from many effective dark-radiation or early-dark-energy parameterizations: the shear term has a fixed sign and a fixed scaling, and its influence cannot be tuned independently at different epochs.}

The first aspect is scaling and early-Universe consistency. The defining feature of the shear-only model is the hierarchy produced by the $a^{-6}$ scaling,
\begin{equation}
 \frac{\rho_\sigma}{\rho_m}\propto a^{-3},\qquad
 \frac{\rho_\sigma}{\rho_r}\propto a^{-2},
 \label{eq:shearratios}
\end{equation}
equivalently,
\begin{equation}
 \Omega_\sigma(z)=\frac{\Omega_{\sigma0}(1+z)^6}{E^2(z)},
 \qquad E(z)=\frac{H(z)}{H_0}.
 \label{eq:omegasigmaz}
\end{equation}
and during radiation domination, $E^2(z)\simeq \Omega_{r0}(1+z)^4$, and therefore
\begin{equation}
 \Omega_\sigma(z)\simeq \frac{\Omega_{\sigma0}}{\Omega_{r0}}(1+z)^2.
 \label{eq:omegasigmarad}
\end{equation}
This expression is the simplest way to understand the severity of the early-Universe bound. Recombination occurs at $z\sim 10^3$, so even a present-day value that appears negligible at late times can be amplified by more than six orders of magnitude relative to radiation; Big-bang nucleosynthesis probes even earlier epochs, where the amplification is stronger; hence BBN and CMB limits suppress the allowed present shear density to a level at which its direct impact on the low-redshift distance scale is generally very small \cite{Fields2020BBN,Pitrou2018BBN,Planck2020Isotropy}.

Figure~\ref{fig:shearevolution} visualizes this point for three illustrative values of $\Omega_{\sigma0}$ in a fiducial flat background with matter, radiation and a cosmological constant. The figure should not be read as a fit to data; it is a diagnostic plot showing the unavoidable lever arm between the local Universe and the pre-recombination Universe.

\begin{figure}[H]
\centering
\includegraphics[width=0.68\textwidth]{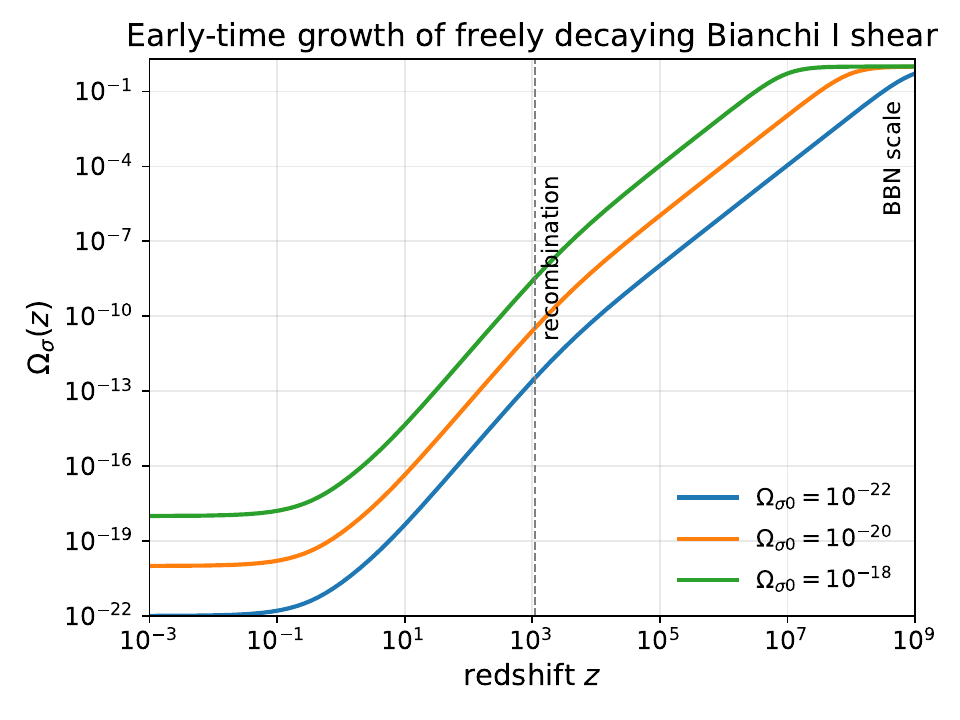}
\caption{\small Evolution of the fractional shear contribution $\Omega_\sigma(z)$ for freely decaying Bianchi I shear, computed from Eq.~\eqref{eq:omegasigmaz}. The curves use illustrative present-day values of $\Omega_{\sigma0}$ and a fiducial flat background. Even when the present-day shear is extremely small, the $a^{-6}$ scaling makes it grow rapidly toward recombination and BBN, explaining why early-Universe constraints are so restrictive.}
\label{fig:shearevolution}
\end{figure}

A second aspect is the impact on the mean distance scale. The reference model must be specified consistently. In the flat comparison adopted here, $H_0$, $\Omega_{m0}$ and $\Omega_{r0}$ are held fixed, while the closure relation changes the cosmological-constant density according to Eq.~\eqref{eq:omegalambda_closure_minimal}. Consequently,
\begin{equation}
 E_{\rm BI}^2(z)-E_{\Lambda{\rm CDM}}^2(z)
 =\Omega_{\sigma0}\left[(1+z)^6-1\right],
 \label{eq:deltaE2_flat}
\end{equation}
and, for small $\Omega_{\sigma0}$,
\begin{equation}
 \frac{\delta H(z)}{H_{\Lambda{\rm CDM}}(z)}\simeq
 \frac{\Omega_{\sigma0}\left[(1+z)^6-1\right]}
 {2E_{\Lambda{\rm CDM}}^2(z)}.
 \label{eq:deltaHminimal}
\end{equation}
The corresponding shift in the isotropic luminosity distance is
\begin{equation}
 \frac{\delta D_L^{\rm iso}(z)}{D_L^{\Lambda {\rm CDM}}(z)}
 \simeq -\Omega_{\sigma0}
 \frac{\displaystyle \int_0^z dz'\,
 \frac{(1+z')^6-1}{2E_{\Lambda{\rm CDM}}^3(z')}}
 {\displaystyle \int_0^z dz'\,\frac{1}{E_{\Lambda{\rm CDM}}(z')}}.
 \label{eq:deltaDLminimal}
\end{equation}
The subtraction of unity in Eqs.~\eqref{eq:deltaE2_flat}--\eqref{eq:deltaHminimal} enforces the same present-day normalization, $E_{\rm BI}(0)=E_{\Lambda{\rm CDM}}(0)=1$. A comparison at fixed $\Omega_{\Lambda0}$ would instead describe a different present-day normalization and is not used below. The sign in Eq.~\eqref{eq:deltaDLminimal} reflects the fact that a positive shear density increases the mean expansion rate at $z>0$ and decreases the distance to a fixed redshift.

\rev{The bounds used in the following comparison are external constraints adopted from the existing literature, principally Ref.~\cite{Akarsu2019Bianchi}; no new BBN, CMB or BAO parameter fit is performed in this paper. The purpose of the calculation below is different: it translates those established bounds into the distance-level and catalogue-level observables derived from the optical analysis of Section~\ref{sec:observables}.}

\sectionlead{Quantitative scale comparison}

The axisymmetric normalization gives a direct relation between the physical shear density and the directional Hubble contrast. From Eqs.~\eqref{eq:axisymsigmascalar} and \eqref{eq:omegasigma},
\begin{equation}
 \Omega_{\sigma0}=\frac{B_{H0}^2}{9},
 \qquad |B_{H0}|=3\sqrt{\Omega_{\sigma0}}.
 \label{eq:BH_omegasigma_relation}
\end{equation}
Because $\Omega_{\sigma0}$ is quadratic in the shear, it determines only the magnitude $|B_{H0}|$, not its sign. The sign distinguishes $H_{\parallel0}>H_{\perp0}$ from $H_{\parallel0}<H_{\perp0}$ and must be specified independently; all scale comparisons below are therefore reported in absolute value. The combined BAO+CMB constraint $\Omega_{\sigma0}\lesssim10^{-15}$ and the representative BBN requirement $\Omega_{\sigma0}\lesssim10^{-23}$ found for the minimal model in Ref.~\cite{Akarsu2019Bianchi} therefore imply
\begin{equation}
 |B_{H0}|\lesssim9.5\times10^{-8}\quad({\rm BAO+CMB}),
 \qquad
 |B_{H0}|\lesssim9.5\times10^{-12}\quad({\rm BBN}).
 \label{eq:BH_bounds_numeric}
\end{equation}
At $z=0.15$, Eq.~\eqref{eq:AD_lowz_shearonly} then gives $|A_D|\lesssim1.1\times10^{-11}$ and
\begin{equation}
 |A_\mu(0.15)|=\frac{5}{\ln10}|A_D(0.15)|
 \lesssim2.4\times10^{-11}\ {\rm mag}
 \label{eq:Amu_BBN_numeric}
\end{equation}
for the BBN-limited model. This is the explicit low-redshift optical output of the worked example.

The mean and directional effects should be compared separately. For $\Omega_{m0}=0.315$, $\Omega_{r0}=9\times10^{-5}$ and $z=0.15$, the one-dimensional integral in Eq.~\eqref{eq:deltaDLminimal} gives
\begin{equation}
 \frac{\delta D_L^{\rm iso}}{D_L^{\Lambda{\rm CDM}}}
 \simeq-0.2587\,\Omega_{\sigma0}.
 \label{eq:mean_distance_numeric}
\end{equation}
Thus the BBN limit produces a mean-distance change below $2.6\times10^{-24}$, whereas a one-percent mean-distance shift would require $\Omega_{\sigma0}\simeq3.9\times10^{-2}$. At low redshift, $D_L\propto H_0^{-1}$; therefore, to leading order, a fractional shift of the mean luminosity distance corresponds to an equal and opposite fractional bias in the value of $H_0$ recovered with an isotropic template. The linear estimate for a shift equal to the fractional \rev{Planck 2018--SH0ES 2022 benchmark separation},
\begin{equation}
 \epsilon_{H_0}\equiv\frac{73.04-67.36}{67.36}=0.0843,
 \label{eq:h0_fractional_gap}
\end{equation}
would require $\Omega_{\sigma0}\simeq0.33$ through the mean expansion and is already outside the perturbative regime.

\rev{Recent community-level combinations give a slightly larger early--late separation than the Planck 2018--SH0ES 2022 benchmark adopted in our numerical illustration; this update strengthens, rather than weakens, the conclusion that freely decaying shear is observationally negligible} \cite{H0DN2026}.

For the directional effect, the largest value of $|Q|$ is $2/3$, so the maximum local modulation is
\begin{equation}
 \left|\frac{\delta H}{H}\right|_{\rm dir,max}
 =\frac{2}{3}|B_{H0}|=2\sqrt{\Omega_{\sigma0}}.
 \label{eq:directional_max_shift}
\end{equation}
A one-percent maximally aligned modulation therefore requires $\Omega_{\sigma0}=2.5\times10^{-5}$, while a modulation equal to $\epsilon_{H_0}$ requires $\Omega_{\sigma0}=1.78\times10^{-3}$. Both values exceed the representative BBN limit by approximately eighteen to twenty orders of magnitude.

An analytic polar-cap toy window quantifies quadrupole-to-monopole leakage without being identified with any particular survey selection function. For a catalogue uniformly distributed inside a polar cap $0\leq\theta\leq\theta_c$ around the preferred axis,
\begin{equation}
 \langle Q\rangle_{\rm cap}
 =\frac{\mu_c+\mu_c^2}{3},
 \qquad \mu_c\equiv\cos\theta_c.
 \label{eq:Q_cap_average}
\end{equation}
For $\theta_c=60^\circ$, $\langle Q\rangle_{\rm cap}=0.25$. Taking the catalogue to be concentrated around the explicit worked-example redshift $z_{\rm eff}=0.15$, Eq.~\eqref{eq:AD_numeric_lowz} gives $|A_D|\simeq1.142|B_{H0}|$. Equation~\eqref{eq:H0windowbias_finitez} then yields
\begin{equation}
 \left|\frac{\delta H_0^{\rm fit}}{H_0}\right|
 \simeq0.25\,|A_D(0.15)|
 \simeq0.856\sqrt{\Omega_{\sigma0}}.
 \label{eq:cap_bias_finitez}
\end{equation}
A one-percent scalar leakage therefore requires $\Omega_{\sigma0}\simeq1.36\times10^{-4}$; reproducing the \rev{Planck 2018--SH0ES 2022 benchmark separation} would require $\Omega_{\sigma0}\simeq9.69\times10^{-3}$, for which the weak-anisotropy approximation is no longer reliable. This polar cap is an analytic toy window, not a mock generated from the sky and redshift distribution of a real survey. The hierarchy is summarized in Table~\ref{tab:numerical_hierarchy}.

\begin{table}[H]
\centering
\caption{Representative present-day shear densities required to generate specified low-redshift effects. The ``maximum directional'' row assumes a line of sight along the symmetry axis; the ``$60^\circ$ cap'' row is the analytic polar-cap toy-window calculation of Eq.~\eqref{eq:Q_cap_average}. \rev{The $8.43\%$ column corresponds to the Planck 2018--SH0ES 2022 benchmark separation.} Values in the last column that are not small should be read only as scale indicators because the linear expansion then breaks down.}
\label{tab:numerical_hierarchy}
\begin{tabular}{lcc}
\toprule
Effect & $1\%$ shift & \rev{$8.43\%$ benchmark shift}\\
\midrule
Mean distance at $z=0.15$ & $3.9\times10^{-2}$ & $3.3\times10^{-1}$\\
Maximum directional modulation & $2.5\times10^{-5}$ & $1.8\times10^{-3}$\\
Finite-$z$ monopole leakage, $60^\circ$ cap & $1.4\times10^{-4}$ & $9.7\times10^{-3}$\\
\bottomrule
\end{tabular}
\end{table}

This completed calculation shows quantitatively, rather than only qualitatively, why freely decaying shear cannot shift the inferred $H_0$ at an interesting level while satisfying early-Universe bounds.

\rev{Accordingly, the novelty of this comparison does not reside in re-establishing the BBN, CMB or BAO exclusion of a phenomenologically large freely decaying shear component. It resides in showing explicitly how an established bound on $\Omega_{\sigma0}$ propagates into the directional Hubble contrast $B_{H0}$, the fractional distance quadrupole $A_D(z)$, the magnitude quadrupole $A_\mu(z)$ and the monopole leakage generated by a finite catalogue window.}

This observation is central for the interpretation of the model. A shear-only Bianchi I component cannot be treated as a generic late-time deformation of $\Lambda$CDM; it is a constrained geometrical component with a built-in early-time consistency condition. If $\Omega_{\sigma0}$ is made large enough to create a substantial low-redshift shift in the distance ladder, Eq.~\eqref{eq:omegasigmarad} typically makes it unacceptable at recombination or nucleosynthesis. If it is made small enough to satisfy early-Universe constraints, Eqs.~\eqref{eq:deltaHminimal} and \eqref{eq:deltaDLminimal} imply that it is normally too small to produce a competitive resolution of the Hubble tension. This conclusion agrees with earlier phenomenological bounds on Bianchi I extensions of $\Lambda$CDM \cite{Akarsu2019Bianchi,Akarsu2023CurvatureAnisotropy,Gron2024Symmetry}.

The third aspect concerns the directional signal and the status of the minimal model. The minimal model is still valuable because it gives a sharply defined null hypothesis for anisotropic cosmology and its directional signal is not an arbitrary angular template: it is tied to the same shear tensor that enters the mean Friedmann equation. In the weakly anisotropic regime one expects a quadrupolar modulation of the Hubble diagram of the form described by Eq.~\eqref{eq:quadDL}, with a finite-redshift amplitude controlled by both the integrated shear and the Jacobi map. Beyond the explicit low-redshift result in Eq.~\eqref{eq:AD_lowz_shearonly}, one may write only schematically,
\begin{equation}
 A_D(z)\sim \int_0^z \frac{\sigma(z')}{H(z')}\frac{dz'}{1+z'},
 \label{eq:Aminimal_integrated}
\end{equation}
up to angular and convention-dependent factors. The proportionality coefficient in this schematic finite-redshift expression is not universal; it must be obtained from the redshift mapping and the optical equations. Nevertheless, the explicit calculation above shows that the same early-Universe constraint that limits the mean shear contribution also suppresses the direction-dependent distance residual.

This connection makes the minimal model useful in three ways. First, it is predictive: there is no freedom to choose an arbitrary redshift profile for the anisotropy once the shear dynamics are fixed. Second, it supplies a clean baseline for Bayesian comparison with flat $\Lambda$CDM, because it adds only a small number of physical parameters. Third, it separates two logically different possibilities. If the data show no statistically significant quadrupole, the FLRW compression of the Hubble diagram is supported. If the data instead show a quadrupole that is too large to be compatible with freely decaying shear, the result should not be advertised as evidence for the minimal model; it should instead be interpreted as evidence for systematics, local-structure effects, or a more general Bianchi I scenario with sustained anisotropic stress.

The practical conclusion is therefore conservative. The minimal shear-only model should not be presented as a stand-alone solution of the Hubble tension. Its role is to quantify how much room is left for homogeneous geometrical anisotropy once the early Universe is respected, and to provide a reference against which more flexible late-time anisotropic models can be judged.

\end{revision}
\begin{revision}
\section{Phenomenological Late-Time Anisotropy}\label{sec:phenomenology}

The restrictive nature of the minimal model motivates a broader phenomenological description. It is important to distinguish between Bianchi I as a spacetime geometry and the shear-only model as one particular dynamical realization: Bianchi I geometry only states that the Universe is homogeneous and admits different scale factors along different axes, while the shear-only model adds the much stronger assumption that the matter sector has no anisotropic stress. If this assumption is relaxed, late-time anisotropy can be sustained without being identical to an $a^{-6}$ stiff component.

A general stress-energy tensor compatible with the principal-axis Bianchi I background may be written as
\begin{equation}
 T^i{}_j=p\delta^i{}_j+\pi^i{}_j, \qquad \pi^i{}_i=0,
 \label{eq:anisstress}
\end{equation}
where $\pi^i{}_j$ is the trace-free anisotropic stress; the shear evolution equation is then schematically
\begin{equation}
 \dot\sigma^i{}_j+3H\sigma^i{}_j=8\pi G\pi^i{}_j.
 \label{eq:shearwithpi}
\end{equation}
\rev{The homogeneous shear no longer has to decay exactly as $a^{-3}$. If $\pi^i{}_j$ is associated with a late-time component, such as anisotropic dark energy, vector fields, elastic stresses or an effective modification of the gravitational sector, the anisotropic expansion can be concentrated in the redshift interval probed by supernovae, BAO and local distance-ladder measurements. Such models are therefore observationally different from the minimal shear-only scenario \cite{Koivisto2008AnisotropicDE,Rodrigues2008AnisotropicLambda,Appleby2013ProbingADE,Verma2025AnisotropicDE}.}

\par\medskip\noindent\textbf{Effective pressure-skewness description.}\par\nopagebreak\smallskip\nopagebreak

A compact way to parameterize the source is to introduce directional equations of state,
\begin{equation}
 p_i=w_i\rho_X,\qquad i=1,2,3,
 \label{eq:directionalpressures}
\end{equation}
with mean equation of state and skewness parameters
\begin{equation}
 w=\frac{1}{3}(w_1+w_2+w_3),\qquad
 \delta_i=w_i-w,\qquad \sum_i\delta_i=0,
 \label{eq:skewness}
\end{equation}
\rev{The anisotropic stress is then}
\begin{equation}
 \pi^i{}_j=\rho_X\,\mathrm{diag}(\delta_1,\delta_2,\delta_3).
 \label{eq:pi_skewness}
\end{equation}
In the axisymmetric case one sets $w_1=w_2=w_\perp$ and $w_3=w_\parallel$ and a single skewness parameter can then be defined as
\begin{equation}
 \Delta w=w_\parallel-w_\perp,
 \label{eq:deltaw_axis}
\end{equation}
so that $\Delta w=0$ restores isotropic pressure. The corresponding directional expansion contrast,
\begin{equation}
 \Delta H_{\rm ax}(t)=H_\parallel(t)-H_\perp(t),
 \label{eq:DeltaH_axis}
\end{equation}
obeys an equation of the schematic form
\begin{equation}
 \dot{\Delta H}_{\rm ax}+3H\Delta H_{\rm ax}\simeq 8\pi G\rho_X\Delta w,
 \label{eq:DeltaH_source}
\end{equation}
where the precise numerical coefficient depends on the normalization of the axisymmetric shear variable. Eq.~\eqref{eq:DeltaH_source} illustrates the qualitative point: a small pressure skewness can source a slowly varying anisotropic expansion if it is attached to a component that becomes dynamically relevant at late times.

This effective-fluid formulation is useful because it connects the present work to earlier Bianchi-I and anisotropic-dark-energy phenomenology \cite{Koivisto2008AnisotropicDE,Rodrigues2008AnisotropicLambda,Appleby2013ProbingADE,Tedesco2024CosmicShear,Verma2025AnisotropicDE} and it also clarifies what can and cannot be inferred from a purely geometrical fit. A measured quadrupole in the Hubble diagram would constrain a combination of shear history and anisotropic stress; it would not by itself identify the microscopic origin of the stress.

\par\medskip\noindent\textbf{Distance-level parameterization.}\par\nopagebreak\smallskip\nopagebreak

For a data-driven analysis it is useful to work directly at the level of the luminosity distance. In the axisymmetric case the leading correction can be written as
\begin{equation}
 D_L(z,\hat n)=D_L^{\rm FLRW}(z)\left\{1+A_D(z)
 \left[(\hat n\cdot\hat e)^2-\frac{1}{3}\right]\right\},
 \label{eq:DLaxisphen}
\end{equation}
\rev{The subtraction of $1/3$ ensures that the angular average of the correction vanishes. Thus $D_L^{\rm FLRW}(z)$ remains the monopole distance, while $A_D(z)$ measures the fractional quadrupolar deformation around the axis $\hat e$.}

\rev{Several phenomenological forms for $A_D(z)$ are useful, depending on the question being asked. A conservative saturating profile is}
\begin{equation}
 A_D(z)=A_{D0}\frac{z}{z+z_*},
 \label{eq:Azsat}
\end{equation}
which vanishes at the observer and approaches a constant at large redshift. A localized late-time profile is
\begin{equation}
 A_D(z)=A_{D0}\exp\left[-\frac{(z-z_*)^2}{2s_z^2}\right],
 \label{eq:Azgauss}
\end{equation}
which tests whether a possible signal is concentrated in the redshift range where the supernova Hubble diagram is most sensitive. A physically motivated profile relates the amplitude to the integrated shear,
\begin{equation}
 A_D(z)=\eta_D\int_0^z \frac{\sigma(z')}{H(z')}\frac{dz'}{1+z'},
 \label{eq:Azintegral}
\end{equation}
where $\eta_D$ is only a shorthand for the redshift and optical response. It must not be regarded as a universal constant: a dynamical Bianchi I calculation must determine it from the redshift mapping and the Jacobi map. Equation~\eqref{eq:AD_map_foc_decomposition} gives this determination explicitly at low redshift.

For small $A_D(z)$, the induced distance-modulus residual is
\begin{equation}
 \delta\mu(z,\hat n)\simeq \frac{5}{\ln 10}A_D(z)
 \left[(\hat n\cdot\hat e)^2-\frac{1}{3}\right],
 \label{eq:deltamu_axisphen}
\end{equation}
\rev{This expression shows why the anisotropy is observationally subtle. A quadrupole with $|A_D|\sim10^{-3}$ corresponds to a residual of only a few millimagnitudes, comparable to calibration and selection effects in modern supernova samples. A robust detection therefore requires both a good statistical fit and stability under changes of sky mask, redshift cuts, peculiar-velocity treatment and photometric calibration.}

\par\medskip\noindent\textbf{Relation to ellipsoidal-universe models.}\par\nopagebreak\smallskip\nopagebreak

The ellipsoidal-universe approach provides a more geometrical interpretation of the same class of ideas. Instead of starting from an arbitrary modulation of $D_L$, it introduces a small eccentricity of the large-scale spatial geometry and examines whether this eccentricity can affect the CMB quadrupole, large-angle anomalies, the Hubble tension and related late-time tensions \cite{Cea2010Polarization,Cea2014Planck,Cea2022EllipsoidalH0}. This approach is especially relevant because it attempts to connect low-redshift anisotropic expansion with independent high-redshift signatures.

In the present paper, the ellipsoidal scenario is used as motivation and comparison rather than as an assumed complete model. The phenomenological amplitude $A_D(z)$ should be interpreted as an observational compression of a possible anisotropic geometry. A physical model must eventually specify the evolution of the eccentricity or shear, the source of anisotropic stress, the compatibility with CMB isotropy, and the perturbation sector. This distinction is essential: a distance-level quadrupole is a test statistic, whereas an ellipsoidal or anisotropic-stress cosmology is a dynamical theory.

\par\medskip\noindent\textbf{Priors and degeneracies.}\par\nopagebreak\smallskip\nopagebreak

The phenomenological parameters must be assigned priors that do not artificially prefer anisotropy. A natural choice is to take the preferred axis isotropically distributed on the sphere,
\begin{equation}
 p(\ell,b)\,d\ell\,db\propto \cos b\,d\ell\,db,
 \label{eq:axis_prior}
\end{equation}
with $0\leq \ell<2\pi$ and $-\pi/2\leq b\leq \pi/2$. \rev{The amplitude prior should be symmetric around zero if $A_D(z)$ is treated as an effective distance residual, because the sign of the quadrupole depends on whether distances are larger along or perpendicular to the preferred axis. By contrast, $\Omega_{\sigma0}$ in the minimal model is positive definite. Confusing these two parameters would lead to an inconsistent analysis.}

There are also important degeneracies. A dipolar residual can arise from peculiar velocities, a bulk flow, calibration gradients or Solar-system-frame corrections; it should not be absorbed into a Bianchi I quadrupole. A quadrupole can be produced by survey footprint, hemispherical calibration offsets or uneven redshift coverage. The likelihood must therefore include nuisance terms or null tests able to distinguish a physical quadrupole from survey geometry. In practice, the phenomenological model should be fitted together with calibration offsets, redshift-bin amplitudes, and mock catalogues generated with the same angular selection function. Only a signal stable under these tests can be interpreted as evidence for a departure from FLRW isotropy.

\begin{table}[H]
\centering
\caption{Notation used to separate physical shear, distance-level phenomenology and fitted tension statistics.}
\label{tab:symbols}
\begin{tabularx}{0.94\textwidth}{@{}lX@{}}
\toprule
Symbol & Meaning\\
\midrule
$\Delta H_{\rm ax}=H_\parallel-H_\perp$ & Axisymmetric directional expansion contrast.\\
$B_H(z)=\Delta H_{\rm ax}(z)/H(z)$ & Dimensionless Hubble-rate quadrupole; $B_{H0}$ is its present value.\\
$Q(\hat n;\hat e)=(\hat n\!\cdot\!\hat e)^2-1/3$ & Trace-free axial quadrupole on the sky.\\
$q$, $j$ & Deceleration and jerk parameters of the mean expansion.\\
$q_{\hat n}$, $j_{\hat n}$ & Directional cosmographic parameters appearing in the low-redshift map.\\
$A_D(z)$ & Fractional luminosity-distance quadrupole in $D_L/D_L^{\rm FLRW}-1=A_DQ$.\\
$A_\mu(z)=(5/\ln10)A_D(z)$ & The same quadrupole expressed in distance-modulus units.\\
$\lambda_i$ & Normalized trace-free shear eigenvalues, with $\sum_i\lambda_i=0$ and $\sum_i\lambda_i^2=1$.\\
$\delta H_0^{\rm fit}=H_0^{\rm BI}-H_0^{\rm FLRW}$ & Shift of the scalar value recovered by an isotropic fit; it is not the directional contrast $\Delta H_{\rm ax,0}$.\\
$T_{H_0}$ & Significance of the residual early--late Hubble tension.\\
\bottomrule
\end{tabularx}
\end{table}

\end{revision}
\section{Data Sets and Likelihood Strategy}\label{sec:data}

\rev{A credible test of Bianchi I anisotropy must combine two requirements that are often separated in standard FLRW analyses. The first is the usual reconstruction of the monopole expansion history. The second is the reconstruction of angular residuals at fixed redshift. A useful strategy is therefore to keep the mean expansion parameters and the directional modulation explicit throughout the likelihood, rather than fitting an isotropic Hubble diagram first and searching for anisotropy only in post-fit residuals.}

\begin{revision}
\subsection{Supernova likelihood}

Type Ia supernovae are the primary low-redshift probe of a directional Hubble diagram because each object has a redshift, a distance modulus and a sky position. They are therefore not only distance indicators but also angular tracers of the expansion field, which makes them especially useful in the present work. A conventional FLRW analysis compresses the catalogue into a monopole distance-redshift relation, whereas a Bianchi-I-oriented analysis must retain the line-of-sight information from the beginning. The relevant data vector for each object is therefore
\begin{equation}
 {\cal D}_i=\{z_i,\mu_i^{\rm obs},\sigma_{\mu,i},\hat n_i,s_i\},
 \label{eq:sn_datavector_basic}
\end{equation}
where $\hat n_i$ is the unit vector pointing to the source and $s_i$ labels the survey, calibration subset, photometric system, host-mass bin or other sample class to which the object belongs. Modern supernova compilations such as DES-SN5YR and Union3 make precisely this kind of catalogue-level treatment important, because their cosmological constraining power comes from combining many subsamples with different selection functions, calibration histories and sky footprints \cite{Abbott2024DESY5,Popovic2025DESReanalysis,Rubin2025Union3}. Earlier supernova isotropy studies also show that a robust anisotropy claim must control sky coverage, redshift cuts, light-curve standardization and local-velocity corrections simultaneously \cite{Zhao2013Anisotropy,Lin2016JLAIsotropy,Javanmardi2015Probing,Colin2019Evidence,Soltis2019Percent,Aluri2023Isotropy,McConvilleColgain2023Pantheon,Sorrenti2023PantheonSH0ES,Cowell2023QuadrupolarPantheon,Hu2024PantheonCP,Bengaly2024PantheonSH0ES,Sah2025PantheonAnisotropy}.

For a catalogue of $N$ supernovae one may define the residual vector
\begin{equation}
 \Delta\mu_i=\mu_i^{\rm obs}-\mu^{\rm th}(z_i,\hat n_i;\Theta),
 \label{eq:residualvector}
\end{equation}
where $\Theta$ denotes the cosmological, anisotropic and nuisance parameters. With the full covariance matrix $C$, the Gaussian likelihood is
\begin{equation}
 -2\ln{\cal L}_{\rm SN}=\Delta\bm\mu^{T}C^{-1}\Delta\bm\mu+\ln |2\pi C|.
 \label{eq:SNlike}
\end{equation}
The use of the full covariance matrix is essential because, if only diagonal errors are retained, an angularly coherent calibration mode can be mistaken for a physical quadrupole. Conversely, if all systematics are compressed into an isotropic nuisance parameter, a genuine low-amplitude angular signal may be artificially removed.

The theoretical distance modulus should be evaluated as
\begin{equation}
 \mu^{\rm th}(z_i,\hat n_i;\Theta)=5\log_{10}\left[\frac{D_L(z_i,\hat n_i;\Theta)}{\mathrm{Mpc}}\right]+25+\Delta M_{s_i}+\Delta\mu_i^{\rm vel}+\Delta\mu_i^{\rm lens},
 \label{eq:mu_directional_with_cal}
\end{equation}
where $\Delta M_{s_i}$ denotes possible survey-, subsample- or calibrator-dependent magnitude offsets, while $\Delta\mu_i^{\rm vel}$ and $\Delta\mu_i^{\rm lens}$ denote the mean corrections associated with peculiar velocities and weak-lensing magnification. These terms may be treated either as explicit corrections with associated covariance or as nuisance contributions marginalized over in the likelihood. They are not optional in a precision anisotropy search: a small calibration gradient across the sky or an imperfect correction for coherent velocities can mimic a preferred axis even if the underlying cosmology is isotropic.

For the phenomenological Bianchi I model introduced in Section~\ref{sec:phenomenology}, the luminosity distance may be written in the form
\begin{equation}
 D_L(z_i,\hat n_i;\Theta)=D_L^{\rm mono}(z_i;\Theta_{\rm iso})
 \left[1+A_D(z_i;\Theta_A)Q_i(\hat e)\right],
 \qquad
 Q_i(\hat e)=(\hat n_i\cdot\hat e)^2-\frac{1}{3},
 \label{eq:sn_bianchi_distance_model}
\end{equation}
\rev{where $\hat e$ is the preferred axis and $A_D(z)$ is the fractional-distance quadrupole.} Defining $A_\mu(z)=(5/\ln10)A_D(z)$, the weak-anisotropy regime gives
\begin{equation}
 \mu^{\rm th}_i\simeq \mu^{\rm mono}(z_i)+\frac{5}{\ln 10}A_D(z_i)Q_i(\hat e)+\Delta M_{s_i}+\Delta\mu_i^{\rm vel}+\Delta\mu_i^{\rm lens}.
 \label{eq:sn_linearized_quad_model}
\end{equation}
This linearized expression is useful for fast tests, for constructing mock catalogues and for understanding degeneracies. However, it should be checked against the full distance expression whenever the posterior allows amplitudes that are not perturbatively small.

This requirement concerns not only the amplitude of the anisotropy but also the redshift range. Even for weak shear, the $z^2$ cosmographic expression derived in Section~\ref{sec:observables} is not a substitute for the exact finite-redshift optical solution when supernova or BAO measurements at $z\gtrsim\mathcal O(1)$ are included.

A useful diagnostic decomposition is
\begin{equation}
 \Delta\mu_i=\Delta\mu_i^{\rm mono}+\Delta\mu_i^{\rm dip}+\Delta\mu_i^{\rm quad}+\epsilon_i,
 \label{eq:sn_residual_components}
\end{equation}
where the monopole tests the mean expansion, the dipole absorbs velocity-like or calibration-gradient effects, and the quadrupole is the leading Bianchi I contribution. A practical model may write the angular part as
\begin{equation}
 \Delta\mu_i^{\rm ang}=D(z_i)\,\hat n_i\cdot\hat p
 +A_\mu(z_i)\left[(\hat n_i\cdot\hat e)^2-\frac{1}{3}\right]
 +\sum_{s} S_{is}\Delta M_s,
 \label{eq:sn_dipole_quadrupole_offsets}
\end{equation}
where $\hat p$ is a dipole direction, $\hat e$ is the quadrupole axis, $S_{is}=1$ if the $i$th object belongs to subset $s$ and zero otherwise, and $A_\mu(z)$ is the distance-modulus form of the quadrupole amplitude. The dipole term should be retained even if the scientific target is the quadrupole, because many local effects are dominantly dipolar and a detection of Bianchi-I-like anisotropy should not be claimed unless the quadrupole improves the fit after allowing for the dipole and for the relevant survey nuisance parameters.

The covariance matrix should include statistical errors, intrinsic scatter, light-curve standardization covariance, peculiar-velocity uncertainties, lensing scatter, calibration covariance and any survey-level systematic covariance supplied by the data release. To this end, it is useful to write schematically
\begin{equation}
 C=C_{\rm stat}+C_{\rm int}+C_{\rm cal}+C_{\rm vel}+C_{\rm lens}+C_{\rm model}.
 \label{eq:sn_covariance_decomposition}
\end{equation}
\rev{The velocity covariance is particularly important at low redshift. Peculiar velocities generate coherent angular correlations that are not produced by homogeneous Bianchi I expansion, but they can project onto similar low-order spherical harmonics if the sky coverage is incomplete. For this reason, the analysis should be repeated for several minimum-redshift cuts, for example $z_{\rm min}=0.01$, $0.02$ and $0.03$, and for independent treatments of the velocity field. A physical Bianchi I signal should remain stable under these changes.}

The absolute magnitude and $H_0$ normalization require special care. In a supernova-only Hubble diagram the combination of $M$ and $H_0$ is nearly degenerate. If the local distance ladder is included, calibrator supernovae and Cepheid/TRGB/JAGB information must be represented by an explicit calibration likelihood rather than by simply fixing $M$. In the anisotropic problem this point is even more important because the calibrator sample has its own sky distribution. A calibration subset concentrated in a limited angular region can correlate $M$ with the quadrupole amplitude; therefore, one can expose this degeneracy by sampling the absolute magnitude, the survey offsets and the quadrupole amplitude simultaneously, or by analytically marginalizing over linear nuisance parameters.

\rev{If the nuisance parameters are linear, the marginalization can be performed in closed form. Let the residual vector be written as}
\begin{equation}
 \Delta\bm\mu=\bm y-\bm X\bm a-\bm m(\Theta_{\rm nl}),
 \label{eq:sn_linear_nuisance_setup}
\end{equation}
where $\bm a$ contains linear nuisance amplitudes such as $M$, $\Delta M_s$, a constant dipole amplitude or a set of redshift-bin quadrupole amplitudes, while $\Theta_{\rm nl}$ contains nonlinear parameters such as $\Omega_{m0}$, $H_0$ and the preferred axis. With Gaussian priors on $\bm a$, the marginalized likelihood is obtained by replacing the inverse covariance with the corresponding Schur-complement expression. Even when this analytic step is not used, Eq.~\eqref{eq:sn_linear_nuisance_setup} is a useful diagnostic because it identifies which combinations of angular templates are nearly degenerate with calibration modes.

\rev{The preferred axis should be parameterized in a way that is invariant under $\hat e\to-\hat e$, since an axial quadrupole does not distinguish the two orientations. A convenient choice is to sample $(\ell,b)$ on one hemisphere with the isotropic prior $p(\ell,b)\propto\cos b$. Equivalently, one may sample a unit vector and identify antipodal points. The posterior for the axis should then be compared with the survey mask. If the best-fit axis is aligned with a survey boundary, a Galactic-extinction pattern, a calibration hemisphere or a known observing strategy, the physical interpretation becomes weak.}

\rev{Finally, the supernova likelihood should be accompanied by posterior-predictive and null tests. The same pipeline should be run on isotropic mock catalogues generated with the real sky mask, redshift distribution and covariance matrix. It should also be run after random rotations of the sky positions, after scrambling residuals within redshift bins, and after splitting the sample by survey, host mass, colour, stretch and calibrator class. A genuine homogeneous anisotropic expansion should produce a stable quadrupolar pattern tied to a single axis and a smooth redshift dependence. A signal that appears only in one survey, one redshift bin or one calibration subset is more naturally interpreted as a systematic effect.}

\end{revision}
\subsection{BAO, sound horizon and anisotropic distances}

BAO data constrain combinations of transverse and radial distances and therefore provide an important consistency test for any anisotropic interpretation of the Hubble tension. In an FLRW model the standard distance variables are
\begin{equation}
 D_H(z)=\frac{c}{H(z)}, \qquad D_M(z)=c\int_0^z\frac{dz'}{H(z')},
 \label{eq:BAOdistances}
\end{equation}
and
\begin{equation}
 D_V(z)=\left[zD_H(z)D_M^2(z)\right]^{1/3}.
 \label{eq:DV}
\end{equation}
Most published BAO constraints are given as measurements of $D_M(z)/r_d$, $D_H(z)/r_d$, $D_V(z)/r_d$, or equivalent Alcock-Paczynski combinations, where $r_d$ is the sound horizon at the baryon drag epoch. These measurements are among the most powerful late-time geometrical constraints on $\Lambda$CDM and on extensions of the background expansion \cite{Eisenstein2005BAO,Alam2021Eboss,DESI2024DR1BAO,DESI2025DR2BAO,DESI2025Cosmology}.

In a Bianchi I background the quantities in Eqs.~\eqref{eq:BAOdistances} and \eqref{eq:DV} are not sufficient to describe the full observable, because transverse distances and radial expansion rates depend on the orientation of the galaxy pair relative to the principal axes. The radial separation probes a direction-dependent expansion rate, while the transverse separation probes the screen-projected Jacobi map. Schematically one should replace
\begin{equation}
 D_H(z)\rightarrow D_H(z,\hat n),
 \qquad
 D_A(z)\rightarrow D_A(z,\hat n,\rev{\varphi_{\rm scr}}),
 \label{eq:bao_directional_replacement}
\end{equation}
\rev{Here $\hat n$ is the line of sight and $\varphi_{\rm scr}$ is the orientation of the transverse separation on the screen. The additional angle $\varphi_{\rm scr}$ has no analogue in an exactly isotropic FLRW model; it appears because an anisotropic background can stretch the two screen directions differently.}

For an axisymmetric weak-anisotropy model, the leading correction can be written in a form parallel to the supernova case,
\begin{equation}
 \frac{D_X(z,\hat n)}{r_d}=\left[\frac{D_X(z)}{r_d}\right]_{\rm mono}
 \left[1+A_X(z)\left((\hat n\cdot\hat e)^2-\frac{1}{3}\right)\right],
 \qquad X\in\{H,M,A,V\},
 \label{eq:bao_phenomenological_modulation}
\end{equation}
\rev{where $A_X(z)$ need not be identical for radial, transverse and volume-averaged observables. The distinction matters because BAO do not measure the same optical quantity as supernovae. Supernovae constrain luminosity distance along individual lines of sight, whereas BAO compare the observed angular and redshift separations of galaxy pairs with a standard ruler. In a true Bianchi I likelihood the functions $A_H(z)$, $A_M(z)$ and $A_A(z)$ should be derived from the same metric and Jacobi map rather than fitted independently.}

A conservative likelihood can nevertheless use public BAO summaries as monopole constraints; if a BAO data release provides a vector
\begin{equation}
 \bm d_{\rm BAO}=\left(D_M(z_1)/r_d,\,D_H(z_1)/r_d,\ldots,D_V(z_k)/r_d\right),
 \label{eq:bao_data_vector}
\end{equation}
with covariance $C_{\rm BAO}$, one may evaluate
\begin{equation}
 -2\ln{\cal L}_{\rm BAO}=
 \left[\bm d_{\rm BAO}-\bm t_{\rm BAO}(\Theta_{\rm mono})\right]^T
 C_{\rm BAO}^{-1}
 \left[\bm d_{\rm BAO}-\bm t_{\rm BAO}(\Theta_{\rm mono})\right],
 \label{eq:bao_monopole_likelihood}
\end{equation}
\rev{Here $\bm t_{\rm BAO}$ is computed from the mean expansion history. This use of BAO is intentionally conservative: it asks whether the mean Bianchi I expansion is already inconsistent with the isotropically compressed standard-ruler data; it does not claim that the published BAO vector is a complete Bianchi I observable.}

\rev{The sound horizon $r_d$ introduces an additional issue. If the analysis assumes standard early-Universe physics, $r_d$ can be constrained by CMB and BBN information. If the analysis is designed to be sound-horizon independent, $r_d$ should be sampled as a free parameter: these two choices answer different questions. Fixing or tightly constraining $r_d$ tests whether late-time anisotropy can reconcile low-redshift distances with the early-Universe calibration, marginalizing over $r_d$ tests whether a directional signal exists in the late-time geometry independently of the early-time standard ruler. In the second case the model is less powerful for resolving the numerical Hubble tension, but it becomes a cleaner test of isotropy.}

A more complete BAO treatment would return to the uncompressed clustering likelihood. The observed separations of two galaxies can be decomposed into a radial part and a transverse part, and an assumed fiducial cosmology is used to convert angles and redshifts into comoving coordinates. In FLRW analyses the mismatch between the fiducial and trial cosmology is encoded in the two Alcock-Paczynski scaling parameters
\begin{equation}
 \alpha_\parallel(z)=\frac{D_H(z)/r_d}{D_H^{\rm fid}(z)/r_d^{\rm fid}},
 \qquad
 \alpha_\perp(z)=\frac{D_M(z)/r_d}{D_M^{\rm fid}(z)/r_d^{\rm fid}}.
 \label{eq:alpha_parallel_perp}
\end{equation}
In a Bianchi I geometry these factors become functions of sky direction and, for the transverse part, of the orientation of the pair on the screen:
\begin{equation}
 \alpha_\parallel(z)\rightarrow \alpha_\parallel(z,\hat n),
 \qquad
 \alpha_\perp(z)\rightarrow \alpha_\perp(z,\hat n,\rev{\varphi_{\rm scr}}).
 \label{eq:anisotropic_ap_scalings}
\end{equation}
\rev{These direction-dependent AP scalings would modify the monopole, quadrupole and hexadecapole of the galaxy correlation function. A future analysis should therefore propagate the Bianchi I distance map into the two-point correlation function or power spectrum before fitting the BAO peak.}

\rev{This distinction is important for interpreting apparent tension relief. A model may improve a supernova-only Hubble diagram by introducing a quadrupole, yet fail when confronted with BAO if the same preferred axis and amplitude predict an anisotropic standard-ruler distortion that is not observed. Conversely, a weak BAO quadrupole consistent with the supernova axis would provide a much stronger geometrical case than either data set alone. The most useful intermediate test is to split BAO measurements by sky region or survey footprint, fit the monopole in each region and ask whether the regional shifts follow the same quadrupolar template inferred from supernovae. This test is not as complete as a raw-clustering likelihood, but it is less dependent on the detailed modelling of the full anisotropic correlation function.}

\rev{In the likelihood hierarchy adopted here, BAO therefore play three roles. First, the isotropic BAO monopole constrains the mean expansion history and prevents the anisotropic fit from moving too far away from the standard distance ladder. Second, the treatment of $r_d$ distinguishes early-calibrated and sound-horizon-free versions of the test. Third, future anisotropic BAO analyses can check whether the same preferred axis seen in standard candles is present in standard rulers. Only when these three levels are mutually consistent can a Bianchi I interpretation be considered physically credible.}

\begin{revision}
\subsection{CMB, BBN and chronometers}\label{sec:cmb_bbn_chronometers}

\rev{CMB information enters a Bianchi I analysis through two logically distinct channels that should not be mixed. The first channel is the usual isotropic, or monopole, information:} the angular acoustic scale, the physical baryon and cold-dark-matter densities, the damping scale and the distance to the surface of last scattering. The second channel is genuinely anisotropic: the absence of a large Bianchi-type quadrupolar distortion in the observed CMB temperature and polarization maps. A practical likelihood must therefore separate a compressed background prior from an explicit shear prior. Without this separation, one risks using an FLRW CMB constraint outside its domain of validity.

For the monopole part one may use, as a first approximation, compressed CMB quantities such as
\begin{equation}
 \bm q_{\rm CMB}=\left(\ell_A,\,R,\,\omega_b\right),
 \qquad
 \omega_b\equiv \Omega_{b0}h^2,
 \label{eq:cmbcompressedvector}
\end{equation}
where
\begin{equation}
 \ell_A=\pi\frac{D_M(z_*)}{r_s(z_*)},
 \qquad
 R=\frac{\sqrt{\Omega_{m0}}H_0D_M(z_*)}{c}.
 \label{eq:shiftparameters}
\end{equation}
Here $z_*$ is the redshift of photon decoupling, $D_M$ is the monopole comoving angular-diameter distance and $r_s(z_*)$ is the sound horizon at decoupling; the corresponding Gaussian prior may be written as
\begin{equation}
 -2\ln {\cal L}_{\rm CMB}^{\rm mono}
 =\left(\bm q_{\rm CMB}^{\rm th}-\bm q_{\rm CMB}^{\rm obs}\right)^T
 C_{\rm CMB}^{-1}
 \left(\bm q_{\rm CMB}^{\rm th}-\bm q_{\rm CMB}^{\rm obs}\right).
 \label{eq:cmbmonolike}
\end{equation}
This expression is appropriate only if the anisotropy is sufficiently small that the CMB distance prior can be interpreted as a constraint on the volume-averaged background. If the anisotropic sector produces order-one changes in the distance to last scattering or in the photon geodesics, the compressed prior is no longer adequate and the full anisotropic Boltzmann problem must be solved.

The anisotropic CMB part can be imposed through the dimensionless shear variables
\begin{equation}
 \Sigma_i(z)\equiv \frac{H_i(z)-H(z)}{H(z)},
 \qquad
 \sum_i\Sigma_i=0,
 \qquad
 \Sigma_H^2(z)\equiv \frac{1}{2}\sum_i \Sigma_i^2(z)
 =\frac{\sigma^2(z)}{H^2(z)}=3\Omega_\sigma(z),
 \label{eq:sigmadefdatasection}
\end{equation}
In the minimal shear-only model the shear energy fraction evolves as
\begin{equation}
 \Omega_\sigma(z)=\frac{\Omega_{\sigma0}(1+z)^6}{E^2(z)},
 \qquad
 E^2(z)=\Omega_{m0}(1+z)^3+\Omega_{r0}(1+z)^4+\Omega_{\Lambda0}
 +\Omega_{\sigma0}(1+z)^6,
 \label{eq:omegasigma_datasection}
\end{equation}
so that even a very small present-day value can become unacceptable at recombination. A conservative phenomenological implementation is to add a penalty term of the form
\begin{equation}
 -2\ln {\cal L}_{\rm CMB}^{\rm shear}
 =\left[\frac{\Omega_\sigma(z_*)}{\Omega_{\sigma,*}^{\rm max}}\right]^2
 +\left[\frac{\beta_{\rm rms}(z_*)}{\beta_*^{\rm max}}\right]^2,
 \label{eq:cmbshearpenalty}
\end{equation}
where $\beta_{\rm rms}^2\equiv (\beta_1^2+\beta_2^2+\beta_3^2)/3$ measures the accumulated anisotropic distortion of the spatial metric. The two terms represent related but not identical constraints: $\Omega_\sigma(z_*)$ controls the instantaneous anisotropic expansion, whereas $\beta_{\rm rms}(z_*)$ controls the integrated distortion of photon propagation and the large-angle pattern. The precise numerical bounds should be calibrated with a dedicated anisotropic CMB calculation, but Eq.~\eqref{eq:cmbshearpenalty} is a useful way of preventing the late-time likelihood from selecting models already excluded by the observed isotropy of the microwave background \cite{Aghanim2020PlanckParams,Planck2020Isotropy,Louis2025ACTDR6,Calabrese2025ACTExtended,Farren2025ACTUnWISE,Vitrier2025SPT3G}.

BBN supplies an independent early-time constraint because the shear term behaves like a stiff component. During radiation domination one has approximately
\begin{equation}
 \frac{\rho_\sigma}{\rho_r}(z)
 =\frac{\Omega_{\sigma0}}{\Omega_{r0}}(1+z)^2,
 \label{eq:shear_to_rad_ratio}
\end{equation}
and therefore
\begin{equation}
 \frac{\delta H_{\rm BBN}}{H}\simeq
 \frac{1}{2}\frac{\rho_\sigma}{\rho_r}
 =\frac{1}{2}\frac{\Omega_{\sigma0}}{\Omega_{r0}}(1+z)^2
 \qquad (\rho_\sigma\ll\rho_r).
 \label{eq:bbn_deltaH}
\end{equation}
Equivalently, one may express the shear contribution as an effective extra radiation component,
\begin{equation}
 \Delta N_{\rm eff}^{\sigma}(z)
 =\frac{8}{7}\left(\frac{11}{4}\right)^{4/3}
 \frac{\rho_\sigma(z)}{\rho_\gamma(z)},
 \label{eq:dneffsigma}
\end{equation}
with the important caveat that the redshift scaling is not the same as that of true free-streaming radiation. The BBN likelihood can then be represented schematically as
\begin{equation}
 -2\ln {\cal L}_{\rm BBN}
 =\left(\bm y_{\rm BBN}^{\rm th}-\bm y_{\rm BBN}^{\rm obs}\right)^T
 C_{\rm BBN}^{-1}
 \left(\bm y_{\rm BBN}^{\rm th}-\bm y_{\rm BBN}^{\rm obs}\right),
 \qquad
 \bm y_{\rm BBN}=\left(Y_p,\,{
m D/H},\,\omega_b\right),
 \label{eq:bbnlike}
\end{equation}
or, in a compressed analysis, as a prior on $\Delta N_{\rm eff}^{\sigma}$ or directly on $\Omega_\sigma(z_{\rm BBN})$ \cite{Fields2020BBN,Pitrou2018BBN}. This is one of the strongest reasons why a purely decaying shear component is not a natural mechanism for producing a large present-day shift in the inferred Hubble constant.

Cosmic chronometers provide a lower-redshift measurement of the expansion rate from differential galaxy ages. In the axisymmetric Bianchi I case, the local expansion rate along the observed line of sight can be written, to leading order, as
\begin{equation}
 H_{\hat n}(z)=H(z)+\sigma_{ij}(z)\hat n^i\hat n^j
 =H(z)\left[1+B_H(z)Q(\hat n)\right],
 \label{eq:hparallelchronometers}
\end{equation}
where $Q(\hat n)=(\hat n\cdot\hat e)^2-1/3$, $\hat e$ is the symmetry axis, and
$B_H(z)\equiv\Delta H_{\rm ax}(z)/H(z)$ with
$\Delta H_{\rm ax}=H_{\parallel}-H_{\perp}$ the difference between the two principal directional expansion rates. This notation distinguishes the arbitrary line-of-sight rate $H_{\hat n}$ from the principal-axis quantity $H_{\parallel}$ and avoids introducing an undefined scalar amplitude. Present chronometer catalogues, however, are not usually supplied with a directional response function precise enough to support a full anisotropic fit. The conservative choice is therefore to compare them with the mean expansion rate,
\begin{equation}
 H_{\rm CC}^{\rm th}(z_k)=H(z_k),
 \label{eq:ccmean}
\end{equation}
using
\begin{equation}
 -2\ln {\cal L}_{\rm CC}
 =\left(\bm H_{\rm CC}^{\rm th}-\bm H_{\rm CC}^{\rm obs}\right)^T
 C_{\rm CC}^{-1}
 \left(\bm H_{\rm CC}^{\rm th}-\bm H_{\rm CC}^{\rm obs}\right).
 \label{eq:cclike}
\end{equation}
Chronometers are therefore not expected to be the discovery channel for a small Bianchi I signal, but they are valuable because they constrain the monopole expansion without invoking the sound horizon and help reduce degeneracies among $H_0$, $\Omega_{m0}$ and late-time dark-energy parameters \cite{Jimenez2002Chronometers,Moresco2016Chronometers,Moresco2022Chronometers}.

\end{revision}
\begin{revision}
\subsection{Local \texorpdfstring{$H_0$}{H0} measurements and standard sirens}\label{sec:localh0sirens}

Local distance-ladder measurements can be included at two different levels of compression. The simplest implementation is a scalar Gaussian prior,
\begin{equation}
 -2\ln {\cal L}_{H_0}^{\rm scalar}
 =\frac{\left[H_0^{\rm fit}(\Theta)-H_0^{\rm loc}\right]^2}{\sigma_{H_0}^2},
 \label{eq:h0scalarprior}
\end{equation}
where $H_0^{\rm loc}$ and $\sigma_{H_0}$ are the published local value and uncertainty. In an anisotropic analysis, however, the quantity entering Eq.~\eqref{eq:h0scalarprior} should not automatically be identified with the mean expansion rate $H_0$. It should be the value that would be recovered by applying the same isotropic fitting procedure, angular mask and redshift cuts to the anisotropic model. Symbolically,
\begin{equation}
 H_0^{\rm fit}(\Theta)
 =\arg\min_{\mathcal H}
 \left[\bm\mu^{\rm th}(\Theta)-\bm\mu^{\rm FLRW}(\mathcal H)\right]^T
 C_{\rm loc}^{-1}
 \left[\bm\mu^{\rm th}(\Theta)-\bm\mu^{\rm FLRW}(\mathcal H)\right].
 \label{eq:h0fitoperator}
\end{equation}
This equation makes explicit that the observed scalar $H_0$ is an estimator rather than a fundamental observable. If the sky coverage is anisotropic, the estimator can be biased by the angular structure of $D_L(z,\hat n)$ even when the mean expansion is unchanged. Direct tests of anisotropic Hubble expansion based on directional variations of $H_0$ therefore provide an important complementary check of the supernova-based quadrupole analysis \cite{Boubel2025AnisotropicHubble}.

A more informative treatment keeps the ladder hierarchy and the angular information. In the supernova standardization notation one may write, schematically,
\begin{equation}
 m_{B,i}=M_B+\mu^{\rm th}(z_i,\hat n_i;\Theta)
 -\alpha x_{1,i}+\beta_c c_i+\Delta_{\rm host,i}+\Delta_{\rm cal,s(i)}+\epsilon_i,
 \label{eq:laddermodel}
\end{equation}
where $x_1$ and $c$ are light-curve parameters, $\Delta_{\rm host}$ denotes host-mass or population corrections and $\Delta_{\rm cal,s(i)}$ represents survey- or subsample-dependent calibration offsets. Cepheid, TRGB, JAGB or megamaser anchors enter through priors on the absolute distance scale and therefore on $M_B$. The corresponding hierarchical likelihood may be written in block form as
\begin{equation}
 \ln {\cal L}_{\rm ladder}
 =\ln {\cal L}_{\rm anchors}
 +\ln {\cal L}_{\rm calibrators}
 +\ln {\cal L}_{\rm Hubble\ flow}
 +\ln {\cal L}_{\rm nuisance}.
 \label{eq:ladderlike}
\end{equation}
This form is preferable to a single $H_0$ prior whenever the purpose is to test anisotropy, because it allows the preferred-axis amplitude to be distinguished from absolute calibration, survey zero-points and local peculiar-velocity corrections \cite{Riess2022SH0ES,Freedman2024CCHP,Lee2024JAGB,Hoyt2025TRGB,Riess2025PerfectHost,Li2025TRGBComplete,Scolnic2025Coma}.

At low redshift the leading anisotropic correction to the inferred local Hubble rate can be written as
\begin{equation}
 H_0^{\rm eff}(\hat n)
 =H_0\left[1+B_{H0} Q(\hat n;\hat e)+\bm d\cdot\hat n\right],
 \qquad
 Q(\hat n;\hat e)=(\hat n\cdot\hat e)^2-\frac{1}{3},
 \label{eq:directionalh0}
\end{equation}
where $\hat e$ is the Bianchi symmetry axis in the axisymmetric limit, $B_{H0}$ is the Hubble-rate quadrupole and $\bm d$ is an optional dipole term used to absorb coherent peculiar velocities or calibration gradients. Equation~\eqref{eq:directionalh0} gives a direct null test. A genuine Bianchi I signal should be mainly quadrupolar and should remain stable when dipole corrections, velocity-field models, redshift cuts and survey masks are varied. By contrast, a signal dominated by $\bm d\cdot\hat n$ is more naturally interpreted as local motion, structure or calibration rather than homogeneous anisotropic expansion.

The covariance of the local Hubble-flow sample should include the usual magnitude covariance and a velocity contribution. A simple form for the velocity part is
\begin{equation}
 C_{ij}^{v}=\left(\frac{5}{\ln 10}\right)^2
 \frac{\sigma_{v,i}\sigma_{v,j}}{cz_i\,cz_j}
 \left(\hat n_i\cdot\hat n_j\right)\,\Xi_{ij},
 \label{eq:velocitycovariance}
\end{equation}
\rev{Here $\Xi_{ij}$ parametrizes the coherence of the velocity field. In practice, this schematic expression may be replaced by the covariance supplied with the catalogue or by mock catalogues constructed with the same angular selection function. This step is essential because an underestimated velocity covariance can artificially enhance a low-redshift preferred axis.}

\rev{Standard sirens provide an independent distance ladder because gravitational waves measure luminosity distance without Cepheid or TRGB calibration. For an event with an electromagnetic counterpart and redshift $z_a$, a simplified siren likelihood is}
\begin{equation}
 -2\ln {\cal L}_{\rm siren}
 =\sum_a
 \frac{\left[D_{L,a}^{\rm GW}-D_L^{\rm th}(z_a,\hat n_a;\Theta)\right]^2}
 {\sigma_{D,a}^2+\left(\partial D_L/\partial z\right)^2\sigma_{z,a}^2}
 +\hbox{const.}
 \label{eq:sirenlike_simple}
\end{equation}
A more realistic analysis marginalizes over the inclination-distance degeneracy, weak-lensing scatter, detector selection function and host-identification uncertainty,
\begin{equation}
 {\cal L}_{\rm siren}
 =\prod_a\int dD_L\,dz\,d\Omega\,
 p_a^{\rm GW}(D_L,\Omega)\,p_a^{\rm host}(z,\Omega)
 \delta\!\big[D_L-D_L^{\rm th}(z,\Omega;\Theta)\big].
 \label{eq:sirenlike_general}
\end{equation}
The key point is that standard sirens test the same directional luminosity distance as supernovae but with different calibration and selection systematics \cite{Schutz1986StandardSirens,Abbott2017StandardSiren,Cai2018Sirens}. Agreement between a supernova quadrupole and a siren quadrupole would therefore be far more significant than either signal alone, whereas disagreement would point to survey systematics or local astrophysical effects.

\end{revision}
\begin{revision}
\subsection{Combined likelihood and model comparison}

The final stage of the analysis is the construction of a single likelihood in which the mean expansion history, the directional distance correction and the calibration nuisance parameters are varied simultaneously. This point is important because an anisotropic model cannot be tested by first fitting an isotropic cosmology and then searching for residuals in a second, independent step. Such a two-step procedure is useful as a diagnostic, but it does not propagate correlations among $H_0$, $M$, $r_d$, the preferred-axis direction and the quadrupolar amplitude. A catalogue-level likelihood is therefore required if the aim is to decide whether the anisotropic term genuinely changes the inferred Hubble scale or merely absorbs residual structure that is already present in the data.

A convenient way to define the combined problem is to collect all observables in a single data vector,
\begin{equation}
 {\bf d}=\left\{\bm \mu_{\rm SN},\, {\bf d}_{\rm BAO},\, {\bf d}_{\rm CMB},\,
 {\bf H}_{\rm CC},\, {\bf d}_{\rm loc},\, {\bf d}_{\rm BBN},\, {\bf d}_{\rm siren}\right\},
 \label{eq:combined_data_vector}
\end{equation}
where the entries denote, respectively, supernova distance moduli, BAO distance ratios, CMB compressed or full-likelihood information, cosmic-chronometer measurements, local distance-ladder constraints, BBN abundance information and, when available, standard-siren distances. The theory vector has the same structure,
\begin{equation}
 {\bf m}(\Theta)=\left\{\bm \mu_{\rm th},\, {\bf m}_{\rm BAO},\, {\bf m}_{\rm CMB},\,
 {\bf H}_{\rm th},\, {\bf m}_{\rm loc},\, {\bf m}_{\rm BBN},\, {\bf m}_{\rm siren}\right\},
 \label{eq:combined_theory_vector}
\end{equation}
\rev{This vector depends on both cosmological and nuisance parameters. If cross-correlations among the blocks are neglected, the total likelihood can be written schematically as}
\begin{equation}
 \ln{\cal L}_{\rm tot}=\ln{\cal L}_{\rm SN}+\ln{\cal L}_{\rm BAO}+\ln{\cal L}_{\rm CMB}
 +\ln{\cal L}_{H(z)}+\ln{\cal L}_{H_0}+\ln{\cal L}_{\rm BBN}
 +\ln{\cal L}_{\rm siren}+\ln{\cal L}_{\rm prior}.
 \label{eq:totlike}
\end{equation}
Equivalently, for a Gaussian block with covariance $C$, one has
\begin{equation}
 -2\ln {\cal L}_{X}=
 \left[ {\bf d}_{X}-{\bf m}_{X}(\Theta)\right]^{T}
 C_X^{-1}
 \left[ {\bf d}_{X}-{\bf m}_{X}(\Theta)\right]
 +\ln \det C_X+N_X\ln(2\pi),
 \label{eq:block_likelihood}
\end{equation}
\rev{Here $X$ denotes the probe. The determinant term is often irrelevant if the covariance is fixed, but it must be retained if the covariance depends on nuisance parameters, redshift cuts, velocity modelling or intrinsic scatter. In a realistic analysis, some blocks are not independent. Supernova calibration, local anchors and the Hubble-flow intercept share nuisance parameters; BAO and CMB share the sound-horizon calibration; and standard sirens may share large-scale-structure lensing covariance with electromagnetic distance probes. These correlations should be included whenever the published likelihood provides them. When they are unavailable, the analysis should state explicitly which compressed data products are being treated as independent approximations.}

The parameter vector for a simple axisymmetric phenomenological model may be written as
\begin{equation}
 \Theta=\{H_0,\Omega_{m0},\Omega_{b0},\Omega_{r0},\Omega_{\Lambda0},\Omega_{\sigma0},
 A_{D0},z_*,s_z,\ell,b,M,\Delta M_s,r_d,{\bf v}_{\rm bulk},{\bf \eta}_{\rm sys}\},
 \label{eq:paramvec}
\end{equation}
where $(\ell,b)$ specify the preferred axis in Galactic coordinates, $M$ is the supernova absolute-magnitude nuisance parameter, $\Delta M_s$ denotes calibration offsets, $r_d$ is included when BAO are used without imposing a fixed early-Universe calibration, ${\bf v}_{\rm bulk}$ describes possible coherent velocity contributions and ${\bf \eta}_{\rm sys}$ denotes additional survey-dependent nuisance parameters. The axis is equivalently represented by
\begin{equation}
 \hat e=(\cos b\cos \ell,\,\cos b\sin \ell,\,\sin b),
 \qquad 0\leq \ell<2\pi,\quad -\frac{\pi}{2}\leq b\leq \frac{\pi}{2}.
 \label{eq:axis_parametrization}
\end{equation}
Because $\hat e$ and $-\hat e$ define the same quadrupole, the prior must avoid double counting and one may either restrict the angular domain to a hemisphere or sample the full sphere and identify antipodal directions in the posterior. This detail is not cosmetic: an incorrect axis prior changes the Bayesian evidence and can make a weak directional preference appear more significant than it is.

Not all parameters in Eq.~\eqref{eq:paramvec} should be varied in every run. A transparent analysis should define a hierarchy of models, each one answering a different physical question,
\begin{align}
 {\cal M}_0 &: \hbox{flat }\Lambda{\rm CDM}, \nonumber\\
 {\cal M}_1 &: \hbox{minimal shear-only Bianchi I, }\Omega_{\sigma0}\neq 0,\ A_{D0} \hbox{ fixed by shear evolution}, \nonumber\\
 {\cal M}_2 &: \hbox{phenomenological quadrupole with the FLRW mean expansion fixed}, \nonumber\\
 {\cal M}_3 &: \hbox{phenomenological quadrupole with }H_0,\Omega_{m0},r_d\hbox{ and calibration parameters varied}, \nonumber\\
 {\cal M}_4 &: \hbox{late-time anisotropic-stress model with a specified redshift profile}. 
 \label{eq:model_hierarchy}
\end{align}
The sequence is useful because it separates physical shear from a purely residual-level quadrupole. Model ${\cal M}_1$ is predictive but tightly constrained by early-Universe data. Model ${\cal M}_2$ is a null test of the Hubble diagram at fixed background. Model ${\cal M}_3$ tests whether the quadrupole is degenerate with the scalar value of $H_0$ or with calibration parameters. Model ${\cal M}_4$ is needed only if the data prefer a redshift dependence that cannot be reproduced by freely decaying shear. Without this hierarchy, an apparent improvement of the fit may be wrongly attributed to Bianchi I geometry when it is actually produced by extra phenomenological freedom.

The prior structure should also be made explicit. A schematic prior factorization is
\begin{equation}
 \pi(\Theta)=\pi(H_0,\Omega_{m0},\Omega_{b0},\Omega_{r0},r_d)
 \pi(\Omega_{\sigma0})\pi(A_{D0},z_*,s_z)
 \pi(\hat e)\pi(M,\Delta M_s)
 \pi({\bf v}_{\rm bulk})\pi({\bf \eta}_{\rm sys}).
 \label{eq:prior_factorization}
\end{equation}
The prior on $\Omega_{\sigma0}$ should respect $\Omega_{\sigma0}\geq 0$ and should be reported both in linear and logarithmic form when upper bounds are quoted. The prior on $A_{D0}$ may be symmetric around zero for a phenomenological residual model, but it is not equivalent to the prior on $\Omega_{\sigma0}$ because a distance quadrupole can have either sign depending on the convention and on the underlying anisotropic-stress history. The axis prior should be isotropic on the sphere, $\pi(\hat e)d\Omega=d\Omega/(4\pi)$ before antipodal identification and calibration priors should be chosen from the corresponding survey or distance-ladder analysis rather than tuned to suppress the quadrupole. These choices must be published with the posterior constraints, because the evidence for a weak anisotropic signal can be prior-volume dominated.

Model comparison should not rely only on the best-fit $\chi^2$. A small improvement can be produced by fitting survey geometry, redshift-dependent calibration residuals or local-velocity structure. The Bayesian evidence for model ${\cal M}_i$ is
\begin{equation}
 Z_i=p({\bf d}|{\cal M}_i)=\int d\Theta_i\,{\cal L}({\bf d}|\Theta_i,{\cal M}_i)\,
 \pi(\Theta_i|{\cal M}_i),
 \label{eq:evidence}
\end{equation}
and the Bayes factor $B_{ij}=Z_i/Z_j$ compares two models after penalizing the prior volume. In practice, one should report $\Delta\chi^2$, the Akaike and Bayesian information criteria, the Bayesian evidence when feasible, and posterior predictive diagnostics \cite{Trotta2008Bayes,Liddle2007Information}. For a data set with $N$ effective points and $k$ fitted parameters,
\begin{equation}
 {\rm AIC}=\chi^2_{\rm min}+2k,
 \qquad
 {\rm BIC}=\chi^2_{\rm min}+k\ln N,
 \label{eq:aicbic}
\end{equation}
while the difference in BIC between a Bianchi I extension and the FLRW baseline is
\begin{equation}
 \Delta {\rm BIC}=\chi^2_{{\rm BI},\min}-\chi^2_{{\rm FLRW},\min}
 +(k_{\rm BI}-k_{\rm FLRW})\ln N.
 \label{eq:bic_difference}
\end{equation}
These criteria are not substitutes for the full evidence, but they provide useful safeguards against overinterpreting a modest reduction in $\chi^2$ obtained by adding an axis and one or more amplitude parameters.

Mock catalogues are indispensable. The same sky mask, redshift distribution, covariance matrix, calibration segmentation and peculiar-velocity prescription used for the real data should be applied to isotropic mock catalogues. For each mock, one should repeat the full axis scan and record the maximum apparent quadrupole. This procedure estimates the look-elsewhere effect associated with searching over directions. A useful posterior-predictive statistic is
\begin{equation}
 p_{\rm pp}=P\left[T({\bf d}^{\rm mock})\geq T({\bf d}^{\rm obs})\,|\,{\cal M}_0\right],
 \label{eq:posterior_predictive}
\end{equation}
where $T$ may be the likelihood improvement, the best-fit quadrupole amplitude, the concentration of the preferred-axis posterior, or a combined statistic. A very small value of $p_{\rm pp}$ would indicate that the isotropic model rarely generates a residual pattern as coherent as the observed one. A value of order unity would mean that the apparent axis is a common consequence of survey geometry and noise.

\rev{A second diagnostic is the impact on the Hubble tension itself. For each model, one should quote both the best-fit value of $H_0$ and the posterior shift relative to the isotropic analysis:}
\begin{equation}
 \delta H_0^{\rm fit}=H_0^{\rm BI}-H_0^{\rm FLRW},
 \qquad
 T_{H_0}=\frac{|H_0^{\rm loc}-H_0^{\rm early}|}
 {\sqrt{\sigma^2_{\rm loc}+\sigma^2_{\rm early}}}.
 \label{eq:h0_shift_tension}
\end{equation}
A model that fits a quadrupolar residual but leaves $T_{H_0}$ essentially unchanged has not solved the Hubble tension; it has only described an angular systematic or a subdominant geometrical effect. Conversely, a model that shifts $H_0$ significantly but violates CMB, BBN or BAO constraints is not a viable Bianchi I resolution. The relevant outcome is therefore a joint statement involving the monopole shift, the quadrupole amplitude, the preferred axis and the early-Universe consistency bounds.

The output of the analysis should include at least five quantities. First, one should report the posterior of $\Omega_{\sigma0}$ in the minimal model because this is the physical shear-only parameter. Second, one should report the posterior of the phenomenological fractional-distance amplitude $A_{D0}$ or the redshift-dependent function $A_D(z)$. Third, one should give the posterior distribution of the preferred axis, including credible regions on the sky and the antipodal degeneracy. Fourth, one should provide the inferred shift in $H_0$ relative to the isotropic fit. Fifth, one should show residual maps and mock-calibrated significance estimates. These quantities answer different questions. A bound on $\Omega_{\sigma0}$ tests the physical shear-only model. A bound on $A_{D0}$ tests the directional Hubble diagram. The preferred-axis posterior tests whether the direction is meaningful or prior dominated, the $H_0$ shift tests whether the anisotropic extension changes the tension, and mock-calibrated residual maps test robustness against survey geometry. Only the simultaneous satisfaction of these criteria would make the anisotropic interpretation cosmologically compelling.

\end{revision}
\section{Expected Results and Diagnostic Tests}\label{sec:diagnostics}

The purpose of this section is not to anticipate a preferred numerical outcome, but to define a set of decision criteria by which a Bianchi I interpretation of the Hubble tension can be accepted, rejected or downgraded to a phenomenological description of residual systematics: this distinction is essential. The model contains a physical degree of freedom, the shear of the homogeneous expansion, but any practical fit also contains nuisance directions, calibration parameters and sky-dependent selection effects and a statistically better fit is therefore not sufficient. A credible anisotropic interpretation requires coherence among four levels of evidence: improvement of the likelihood, stability of the preferred direction, compatibility with early-Universe bounds and persistence under alternative descriptions of local structure.

\sectionlead{Statistical decision criteria}

The baseline comparison must be made against a well-defined FLRW reference model with the same data vector, covariance matrix and calibration treatment. For a parameter vector $\Theta$, the relevant quantities are not only the minimum value of $\chi^2$, but also the marginalized posterior volume and the predictive performance of the model. In practice one should report
\begin{equation}
 \Delta\chi^2 = \chi^2_{\rm BI,min}-\chi^2_{\rm FLRW,min},
 \qquad
 \Delta {\rm BIC}=\Delta\chi^2+(k_{\rm BI}-k_{\rm FLRW})\ln N,
 \label{eq:diagnostic_bic}
\end{equation}
where $k$ is the number of fitted parameters and $N$ is the effective number of data points. The Bayesian evidence, or at least an information-criterion proxy, is necessary because an anisotropic model introduces an axis on the sphere and one or more amplitudes; a modest decrease of $\chi^2$ can easily be produced by fitting survey geometry rather than cosmology \cite{Trotta2008Bayes,Liddle2007Information}. For the same reason, any quoted improvement should be accompanied by posterior predictive residual maps, leave-one-survey-out tests and an explicit assessment of the look-elsewhere effect associated with scanning over directions.

A useful diagnostic for the Hubble tension itself is the change in the tension metric
\begin{equation}
 T_{H_0}=\frac{|H_0^{\rm loc}-H_0^{\rm early}|}
 {\sqrt{\sigma^2_{\rm loc}+\sigma^2_{\rm early}}}.
 \label{eq:tension_metric}
\end{equation}
The anisotropic model should not be considered successful merely because it broadens the posterior and therefore lowers $T_{H_0}$; a physically meaningful reduction requires the local and early-Universe determinations to move toward a common value in a way that is supported by the directional data. This criterion is important because a weakly constrained axis or a broad shear prior may create an apparent reduction in the numerical tension without providing physical evidence for anisotropy.

\sectionlead{FLRW limit and stability of the cosmological parameters}

\rev{The first admissible outcome is a null result. In that case, the posterior for the shear density $\Omega_{\sigma0}$ is compatible with zero, the late-time modulation amplitude $A_{D0}$ is consistent with zero, and the axis posterior is statistically isotropic on the sphere. Such a null result would be scientifically informative: it would strengthen the empirical domain of validity of the FLRW compression and provide a direct comparison with previous Bianchi I constraints and with ellipsoidal-Universe analyses of shear and eccentricity \cite{Akarsu2019Bianchi,Tedesco2018ShearJerk,Tedesco2024CosmicShear}.}

\rev{The null limit should be formulated in terms of parameter stability, not only in terms of the anisotropic amplitude. Let $p$ denote one of the scalar parameters $H_0$, $\Omega_{m0}$ or $r_d$. A useful shift statistic is}
\begin{equation}
 S_p=\frac{|p_{\rm BI}-p_{\rm FLRW}|}{\sqrt{\sigma^2(p_{\rm BI})+\sigma^2(p_{\rm FLRW})}},
 \label{eq:shift_statistic}
\end{equation}
\rev{If $S_p\ll1$ after marginalization over $\Omega_{\sigma0}$, $A_{D0}$ and the preferred axis, then the FLRW parameter inference is robust. If $T_{H_0}$ is non-negligible while the anisotropic amplitude remains statistically insignificant, the result should be interpreted as a parameter-volume effect or prior sensitivity rather than as evidence for anisotropic expansion. Conversely, if a small but nonzero anisotropic amplitude is correlated with a reproducible shift in $H_0$, the correlation matrix should be reported explicitly; otherwise the physical interpretation of the shift remains ambiguous.}

\begin{revision}
\sectionlead{Directional Hubble diagram}

The principal late-time observable is the angular structure of the Hubble-diagram residuals. For a supernova in the direction $\hat n_i$, a minimal quadrupolar residual can be written schematically as
\begin{equation}
 \Delta\mu_i = \mu_i-\mu_{\rm FLRW}(z_i)
 \simeq A_\mu(z_i)\left[(\hat n_i\cdot\hat e)^2-\frac{1}{3}\right]
 +\Delta\mu_{\rm pv}(z_i,\hat n_i)+\Delta\mu_{\rm cal}+\epsilon_i,
 \label{eq:residual_decomposition}
\end{equation}
where $A_\mu(z)=(5/\ln10)A_D(z)$ is the quadrupole amplitude in magnitudes, $\hat e$ is the preferred quadrupole axis, and $\Delta\mu_{\rm pv}$ denotes the peculiar-velocity contribution, $\Delta\mu_{\rm cal}$ denotes calibration and survey-dependent offsets, and $\epsilon_i$ contains statistical noise and intrinsic scatter. Eq.~(\ref{eq:residual_decomposition}) makes clear why a preferred axis is not by itself a cosmological detection. The quadrupolar term must be separated from low-redshift velocity fields, photometric zero-point errors, Malmquist-like selection effects and the survey mask.

A professional analysis should therefore test the axis stability under redshift cuts, survey cuts, alternative peculiar-velocity reconstructions, covariance inflation and independent light-curve standardizations. The axis posterior should be presented as a probability distribution on the sphere, not as a single best-fit direction; hemispherical fits are useful as exploratory tools, but they should be supplemented by spherical-harmonic decompositions and posterior predictive maps. Existing directional-supernova studies are relevant here because they show both the potential sensitivity of the Hubble diagram to anisotropy and the danger of confusing anisotropic cosmology with sample geometry or local structure \cite{Zhao2013Anisotropy,Lin2016JLAIsotropy,Javanmardi2015Probing,Colin2019Evidence,Soltis2019Percent}.

The ellipsoidal-Universe literature provides a useful theoretical bridge between shear, eccentricity and angular residuals. The analyses of the CMB quadrupole and of ellipsoidal cosmic shear show how a small geometrical eccentricity can project onto quadrupolar observables \cite{Campanelli2006Ellipsoidal,Campanelli2007CMBQuadrupole,Tedesco2024CosmicShear}. The existence of large-angle CMB anomalies motivates such tests of statistical isotropy, but it does not by itself establish Bianchi I expansion as their cause \cite{Schwarz2016Anomalies,Aluri2023Isotropy}. In the present work, the quadrupolar template in Fig.~\ref{fig:quadrupoleresidual} should be regarded as a diagnostic template: it is a way to search for a coherent angular pattern, not by itself a complete dynamical solution of the Hubble tension.

\sectionlead{Consistency with CMB, BBN and anisotropic stress}

\rev{A late-time anisotropic signal has to satisfy a much stronger test than a supernova-only fit. In the minimal shear-only model, the shear contribution scales as a stiff component, $\rho_\sigma\propto a^{-6}$. A value large enough to alter low-redshift distances therefore becomes unacceptable at recombination or during primordial nucleosynthesis unless the present-day amplitude is extremely small. This early-Universe lever arm is the central physical obstruction to a minimal Bianchi I solution of the Hubble tension. The model must consequently be tested against CMB statistical-isotropy limits and BBN bounds before any claim of physical relevance is made \cite{Fields2020BBN,Pitrou2018BBN,Planck2020Isotropy}.}

A useful way to state the constraint is to compare the shear density with the radiation density,
\begin{equation}
 R_\sigma(z)\equiv \frac{\rho_\sigma(z)}{\rho_r(z)}
 =\frac{\Omega_{\sigma0}}{\Omega_{r0}}(1+z)^2.
 \label{eq:Rsigma_diag}
\end{equation}
The quadratic growth with redshift is already enough to show why the local and early-Universe requirements are difficult to reconcile. If one demands $R_\sigma(z_{\rm BBN})\ll 1$ at nucleosynthesis, then $\Omega_{\sigma0}$ must be many orders of magnitude below the value that would produce a percent-level shift in the low-redshift Hubble diagram through the mean expansion. Equivalently, one can express the shear contribution as an effective extra relativistic component,
\begin{equation}
 \Delta N_{\rm eff}^{(\sigma)}
 \simeq \frac{\rho_\sigma}{\rho_{\nu,1}}
 =\frac{8}{7}\left(\frac{11}{4}\right)^{4/3}
 \frac{\rho_\sigma}{\rho_\gamma},
 \label{eq:deltaNeff_shear_diag}
\end{equation}
where $\rho_{\nu,1}$ is the energy density of one standard neutrino species after $e^+e^-$ annihilation. Equation~\eqref{eq:deltaNeff_shear_diag} should be understood as a diagnostic translation rather than as a claim that shear is literally a free-streaming neutrino species. It is useful because it puts the Bianchi I shear contribution on the same scale as the BBN and CMB limits usually quoted for additional radiation-like energy density.

The CMB imposes a second, independent consistency condition. A homogeneous shear produces an integrated anisotropic redshift and can contribute to the large-angle temperature pattern. Schematically, one may write the Bianchi contribution to a quadrupolar temperature anisotropy as
\begin{equation}
 \left(\frac{\Delta T}{T}\right)_{\rm BI}(\hat n)
 \sim \int_{t_*}^{t_0}\sigma_{ij}(t)\hat n^i\hat n^j\,dt,
 \label{eq:CMBshearintegral_diag}
\end{equation}
where $t_*$ denotes last scattering. A practical likelihood can therefore include a penalty of the form
\begin{equation}
 -2\ln {\cal L}_{\rm CMB}^{\rm iso}
 =\left({\bm q}_{\rm BI}-{\bm q}_{\rm obs}\right)^T
 C_q^{-1}
 \left({\bm q}_{\rm BI}-{\bm q}_{\rm obs}\right),
 \label{eq:CMBquad_penalty_diag}
\end{equation}
where ${\bm q}$ denotes the five quadrupole coefficients or a compressed set of isotropy estimators. The exact construction of ${\bm q}_{\rm BI}$ is model dependent, but the principle is not: a late-time preferred axis cannot be accepted if the same history predicts an unacceptable CMB quadrupole, polarization pattern or violation of Planck statistical-isotropy constraints.

There are two logically distinct ways to proceed: the conservative strategy is to impose the early-Universe bounds directly on $\Omega_{\sigma0}$ and ask whether any observable late-time anisotropy remains, and the constructive strategy is to replace freely decaying shear by a sourced anisotropy, for example through anisotropic stress in the dark sector, vector-field dynamics, primordial magnetic fields or another component whose redshift dependence differs from $a^{-6}$. In terms of e-fold time $N=\ln a$, the shear equation may be written schematically as
\begin{equation}
 \frac{d}{dN}\left(\frac{\sigma^i{}_j}{H}\right)
 +\left(3+\frac{d\ln H}{dN}\right)\frac{\sigma^i{}_j}{H}
 =\frac{8\pi G}{H^2}\pi^i{}_j,
 \label{eq:sourced_shear_diag}
\end{equation}
which makes explicit that the time dependence of the anisotropy is determined by the anisotropic stress $\pi^i{}_j$. This is the sense in which anisotropic-dark-energy and ellipsoidal-Universe models are relevant: they do not merely add a preferred axis, but attempt to control the time dependence of the anisotropic contribution \cite{Koivisto2008AnisotropicDE,Rodrigues2008AnisotropicLambda,Appleby2013ProbingADE,Verma2025AnisotropicDE,Campanelli2011AnisotropicDE}.

\rev{A detected late-time amplitude should be reported together with the reconstructed redshift dependence $A_D(z)$. A useful consistency variable is}
\begin{equation}
 A_D^{\rm shear}(z)=\eta_D\int_0^z
 \frac{\sigma(z')}{H(z')}\frac{dz'}{1+z'},
 \label{eq:Ashear_diag}
\end{equation}
where $\eta_D$ denotes the response fixed by the redshift and optical calculation; it is not a universal angular constant. If the measured $A_D(z)$ is consistent with Eq.~\eqref{eq:Ashear_diag} and with the $a^{-6}$ scaling, the early-Universe constraints are expected to dominate. If instead $A_D(z)$ is localized at low redshift or tracks the dark-energy density, then the minimal shear-only model is not the correct interpretation and the result should be described as evidence for an effective anisotropic stress or for residual systematics. This distinction avoids the common mistake of using a phenomenological anisotropic template and then interpreting it as freely evolving Bianchi I shear without checking the dynamics.

The diagnostic outcome should therefore be classified explicitly. A bound on $\Omega_{\sigma0}$ constrains a physical shear-only model. A bound on $A_{D0}$ constrains a distance-level quadrupole. A reconstruction of $\pi^i{}_j$ or of a pressure-skewness parameter concerns a physical source of anisotropy. These three statements are not interchangeable. The presentation should report them separately; otherwise, the reader cannot know whether the analysis has constrained geometry, phenomenology or the microphysics of the anisotropic stress.

\end{revision}
\sectionlead{Degeneracy with radial inhomogeneity and local structure}

A further source of degeneracy is the distinction between anisotropy and inhomogeneity. Bianchi I is spatially homogeneous and direction dependent; Lema\^{\i}tre--Tolman--Bondi models are radially inhomogeneous and need not contain a global shear axis. Both classes can modify the distance-redshift relation, but they do so with different angular and redshift signatures. Comparative analyses are therefore important, especially because inhomogeneous and anisotropic models have both been discussed in connection with apparent acceleration, fractal-like matter distributions and distance-ladder observables \cite{Fanizza2015InhomAniso,Cosmai2019FractalLTB}.

The most dangerous contaminant for a low-redshift anisotropy search is the peculiar-velocity field. To leading order, a line-of-sight velocity perturbation changes the distance modulus by
\begin{equation}
 \Delta\mu_v(z,\hat n)\simeq
 -\frac{5}{\ln 10}\frac{\bm v\cdot\hat n}{cz},
 \label{eq:velocity_mu_diag}
\end{equation}
\rev{up to observer-frame and convention-dependent terms. This contribution is dipolar at the simplest level, but realistic velocity flows, survey masks and redshift-dependent selection effects can project power into quadrupolar templates. Therefore, a Bianchi I quadrupole should never be fitted without simultaneously modelling dipole terms, velocity covariance and possible calibration gradients.}

Radial inhomogeneity produces a different signature, in an LTB-like description the local expansion may be written schematically as
\begin{equation}
 H_{\rm loc}(r)=\bar H(z)+\delta H(r),
 \label{eq:LTB_H_diag}
\end{equation}
\rev{where the correction depends primarily on distance from the observer or from a preferred centre. If the observer is displaced from the centre, the radial profile can also generate an apparent dipole and higher multipoles. This differs from the Bianchi I prediction, in which the leading homogeneous correction is a global quadrupole,}
\begin{equation}
 \Delta\mu_{\rm BI}(z,\hat n)\propto A_D(z)
 \left[(\hat n\cdot\hat e)^2-\frac{1}{3}\right].
 \label{eq:BI_quad_diag}
\end{equation}
The practical distinction is therefore not simply ``anisotropic'' versus ``isotropic'', but whether the residual is better described by a fixed axis with a coherent redshift evolution, by a radial density profile, by a local bulk flow, or by survey-dependent calibration terms.

A robust pipeline should compare at least four hypotheses: a standard FLRW model with conventional peculiar-velocity corrections; a Bianchi I model with a global axis and no radial density profile; a local-structure or LTB-inspired model with radial dependence but no global anisotropic shear; and a hybrid nuisance model containing dipole, quadrupole and survey-zero-point terms without assigning them a cosmological origin. The comparison can be organized through the residual model
\begin{equation}
 \Delta\mu_i=m_0(z_i)+\bm d(z_i)\cdot\hat n_i
 +A_D(z_i)Q(\hat n_i;\hat e)
 +\Delta\mu_{\rm LSS}(z_i,\hat n_i)
 +\Delta\mu_{{\rm cal},s(i)}+\epsilon_i,
 \label{eq:full_residual_diag}
\end{equation}
where $Q=(\hat n\cdot\hat e)^2-1/3$. The role of Eq.~\eqref{eq:full_residual_diag} is not to introduce an over-flexible final model but to perform null tests. If the preferred anisotropic signal disappears when a local-density or bulk-flow term is included, the evidence favors local structure rather than global geometry. If the signal survives at redshifts where peculiar velocities and nearby density contrasts are subdominant, and if the same axis is recovered in independent data sets, the case for a genuine anisotropic background becomes substantially stronger.

This distinction also matters for the interpretation of the Hubble tension. A local void or radial density fluctuation can alter the locally inferred intercept of the Hubble diagram without changing the global CMB-inferred expansion history. By contrast, a homogeneous Bianchi I signal changes the angular structure of distances and should correlate with other observables such as BAO anisotropy, standard-siren distances and possibly cosmic parallax. The absence of these correlated signatures would not automatically exclude every phenomenological quadrupole, but it would make a global Bianchi I interpretation much less convincing.

\sectionlead{Falsification and possible positive evidence}

\rev{The model should be regarded as disfavored if the preferred axis is driven by one survey, by the lowest-redshift supernovae, by a known footprint asymmetry or by a calibration subset. It should also be disfavored if the inferred amplitude violates CMB or BBN bounds under the shear-only scaling, or if the reduction of the Hubble tension is achieved mainly by widening the posterior rather than by moving independent determinations of $H_0$ toward a common value. These are not secondary checks; they are part of the definition of a viable anisotropic interpretation.}

A useful falsification test is to compare the improvement produced by the quadrupole with the improvement produced by less physical nuisance templates, for example,
\begin{equation}
 \Delta\chi^2_{Q|D}=\chi^2_{\rm dipole}-\chi^2_{\rm dipole+quadrupole}
 \label{eq:quad_improvement_diag}
\end{equation}
\rev{measures the value of the quadrupole after a dipole has already been allowed. A Bianchi I interpretation requires $\Delta\chi^2_{Q|D}$ to remain significant after survey offsets, redshift cuts and velocity covariance are varied. If the quadrupole is significant only before these nuisance terms are included, it should be interpreted as a catalogue-level residual, not as evidence for homogeneous anisotropic expansion.}

\rev{The preferred direction must also be reproducible. If $\hat e_{\rm SN}$, $\hat e_{\rm BAO}$, $\hat e_{\rm siren}$ and $\hat e_{\rm CMB}$ denote axes inferred from different probes, their angular consistency can be summarized by}
\begin{equation}
 \cos\Delta\theta_{XY}=|\hat e_X\cdot\hat e_Y|.
 \label{eq:axis_consistency_diag}
\end{equation}
\rev{Let us observe that the absolute value reflects the fact that an axis has no orientation: $\hat e$ and $-\hat e$ describe the same quadrupolar geometry and a positive result would require the posterior distributions for these axes to overlap within their uncertainties, not merely to have best-fit values that appear visually close on a sky map. Conversely, a supernova-only axis that is inconsistent with BAO, CMB or standard-siren constraints would be strong evidence against a global Bianchi I interpretation.}

A second positive-evidence condition is consistency of the redshift evolution. In a coherent model the amplitudes reconstructed from different probes should satisfy, within calibration and kernel differences,
\begin{equation}
 A_{\rm SN}(z)\simeq A_{\rm BAO}(z)\simeq A_{\rm siren}(z)
 \simeq A_{\rm model}(z),
 \label{eq:amplitude_consistency_diag}
\end{equation}
\rev{where $A_{\rm model}$ is either the shear integral of Eq.~\eqref{eq:Ashear_diag} or a sourced-anisotropy prediction. If the signal is present only in one redshift bin or one survey but not in independent tracers of the same light cone, the conservative interpretation is survey systematics or local structure. If the same redshift trend appears in supernovae, BAO and standard sirens, the result becomes much harder to dismiss as a calibration artefact.}

\rev{A positive result would therefore require a restrictive pattern. The same quadrupolar residual should appear in independent distance indicators; the axis should be stable under covariance, calibration and redshift-cut variations; the amplitude should have a plausible redshift dependence; and the inferred geometry should remain compatible with CMB, BAO and BBN constraints. In that case, the result would not merely state that Bianchi I gives a better fit. It would identify a falsifiable failure mode of the scalar FLRW compression of the data: the possibility that the measured Hubble tension contains a small but coherent angular component. This would be the main physical value of the analysis, whether or not the final amplitude is large enough to resolve the full numerical tension.}

\rev{The final decision should be stated in deliberately conservative language. The hierarchy of outcomes is: no evidence for anisotropy; a phenomenological quadrupole with no robust physical interpretation; a late-time anisotropic-stress signal requiring a model beyond freely decaying shear; or, most restrictively, a coherent Bianchi I background signal compatible with all early- and late-time data. Present data may well favor the first or second outcome. The framework is useful precisely because it makes the distinction testable.}

\begin{revision}
\section{Discussion}\label{sec:discussion}

\rev{The strongest point in favor of the Bianchi I approach is conceptual clarity. It asks whether the scalar FLRW compression of cosmological data is adequate at the precision demanded by the Hubble tension; this is a geometrical question that can be tested. The framework is also conservative because it preserves spatial homogeneity and does not introduce a preferred centre. It therefore differs from Lema\^{\i}tre--Tolman--Bondi void models, which relax homogeneity and can place the observer near a special position. The approach is also connected to the broader observational programme of testing the cosmological principle through dipoles, quadrupoles and large-scale rest-frame comparisons \cite{Secrest2025CosmicDipole}.}

The strongest limitation is also clear. Freely decaying shear behaves like a stiff component and is strongly constrained by early-Universe data. Therefore, the minimal shear-only model cannot be advertised as a simple solution of the Hubble tension. Any claim of a large late-time anisotropic effect must either identify a physical source of anisotropic stress, demonstrate that the signal is a phenomenological residual not captured by the minimal model, or show that the data analysis itself is sensitive to an angular selection effect.

\rev{This early-Universe obstruction is inherited from the existing Bianchi I literature and is not presented here as a new constraint. The contribution of the present work is to connect that established dynamical limitation to the observable chain $\Omega_{\sigma0}\rightarrow B_{H0}\rightarrow A_D(z)\rightarrow A_\mu(z)\rightarrow\delta H_0^{\rm fit}$, including the leading local Jacobi contribution and the finite-window quadrupole-to-monopole leakage.}

The literature contains different perspectives on this problem. Exact and phenomenological Bianchi I studies often emphasize that shear is tightly constrained and that a robust solution of the Hubble tension is difficult \cite{Akarsu2019Bianchi,Gron2024Symmetry,Deliyergiyev2025MNRAS}. The ellipsoidal-universe literature emphasizes that small anisotropies of spatial geometry may help with large-angle CMB anomalies and may soften the Hubble tension under specific assumptions about the dark ages \cite{Campanelli2006Ellipsoidal,Cea2014Planck,Cea2022EllipsoidalH0}. Directional supernova analyses provide empirical tests of whether the magnitude--redshift relation is isotropic \cite{Zhao2013Anisotropy,Lin2016JLAIsotropy,Javanmardi2015Probing,Colin2019Evidence,Soltis2019Percent}. These strands should not be treated as mutually exclusive and they should be combined in a single inference framework that makes clear which assumptions are physical, which are phenomenological and which are purely diagnostic.

A useful way to frame the novelty of the present paper is the following. The Hubble tension is usually treated as a conflict between early and late determinations of a scalar parameter. In a Bianchi I framework, the scalar parameter $H_0$ is replaced by a mean expansion rate plus a direction-dependent expansion pattern. This replacement does not automatically solve the tension; it changes the question. One asks whether the scalar compression of the data into a single FLRW value is hiding a small angular dependence. This is a falsifiable question, especially with all-sky supernova samples, DESI-like large-scale-structure measurements and future gravitational-wave standard sirens.

\end{revision}
\begin{revision}
\section{Conclusions}\label{sec:conclusions}

We have developed a quantitative framework for studying the Hubble tension in a Bianchi type I cosmological background. The framework separates the mean expansion rate, the directional Hubble quadrupole, the fractional luminosity-distance quadrupole, and the scalar value recovered by an isotropic catalogue fit. This separation is essential because these quantities coincide in FLRW but are physically and statistically distinct once exact isotropy is relaxed.

The principal new result is a completed weak-shear axisymmetric calculation through relative order $z^2$. The calculation separates the redshift--affine-parameter contribution from the Jacobi-focusing contribution. For freely decaying shear with isotropic pressure we find
\begin{equation}
 A_D(z)=-B_{H0}+\frac{2q_0-1}{2}B_{H0}z
 +\frac{5-q_0-18q_0^2+6j_0}{12}B_{H0}z^2
 +{\cal O}(z^3,B_{H0}^2).
\end{equation}
The direct quadrupolar Ricci-focusing coefficient vanishes in the minimal model, while isotropic Ricci focusing contributes through the direction-dependent normalization of the distance series. This result replaces the purely schematic use of $A_D(z)$ in the original manuscript.

This expression is a local low-redshift benchmark rather than a global fitting formula. Its use at $z_{\rm eff}=0.15$ is consistent with the intended cosmographic regime, whereas an application to the full redshift extent of current supernova and BAO samples requires the simultaneous numerical integration of the exact null-geodesic and Jacobi equations. The Jacobi term retained in the analytic result is the leading observer-centred coefficient of the cumulative optical-focusing expansion. It is complete at the stated local cosmographic order, but it does not replace integration of the evolving Ricci and Weyl tidal fields along finite-redshift null geodesics.

\rev{The strong early-Universe exclusion of a phenomenologically large freely decaying shear component is a known result and is not claimed here as a new cosmological constraint. The original result of the present analysis is its explicit optical and observational translation: a specified shear history is mapped into the redshift-dependent luminosity-distance quadrupole, including the leading local Jacobi contribution, and subsequently into the scalar bias generated by a finite catalogue window.}

\rev{Using previously established early-Universe bounds, our optical calculation makes the observational limitation of the minimal model explicit at the level of luminosity distances, distance-modulus residuals and isotropic $H_0$ fits.} A representative BBN bound $\Omega_{\sigma0}\lesssim10^{-23}$ implies $|B_{H0}|\lesssim9.5\times10^{-12}$ and $|A_\mu(0.15)|\lesssim2.4\times10^{-11}$ mag. A one-percent maximally aligned directional shift would instead require $\Omega_{\sigma0}=2.5\times10^{-5}$, and a shift comparable with the \rev{$8.43\%$ Planck 2018--SH0ES 2022 benchmark separation} would require $\Omega_{\sigma0}=1.78\times10^{-3}$. An analytic $60^\circ$ polar-cap toy window evaluated at $z_{\rm eff}=0.15$ requires $1.36\times10^{-4}$ and $9.69\times10^{-3}$, respectively. The mean-expansion channel is even less efficient at low redshift. These scales differ from the BBN limit by many orders of magnitude, so freely decaying shear cannot provide a viable resolution of the Hubble tension.

Bianchi I nevertheless remains useful as a geometrical null test. A viable late-time anisotropic signal would have to appear as a stable, redshift-dependent quadrupole whose axis and amplitude are consistent among supernovae, BAO, local distance-ladder measurements and future standard sirens. If a measured quadrupole is too large or has the wrong redshift dependence to be produced by freely decaying shear, it should be interpreted as evidence for survey systematics, local structure, or a specified source of late-time anisotropic stress rather than as evidence for the minimal model.

The paper should therefore be read as a framework with a worked quantitative baseline, not as a claim that Bianchi I anisotropy already resolves the Hubble tension. Its falsifiable question is whether the scalar FLRW compression of precision distance data hides a coherent angular component that survives catalogue-window, calibration, peculiar-velocity, BAO, CMB and BBN consistency tests.

\appendix

\end{revision}
\section{Expanded Bianchi I Kinematics and Shear Evolution}\label{app:shear}

This appendix gives the derivation behind the main formulae used in the text. The aim is to make explicit which results depend only on the Bianchi I geometry and which depend on the additional assumption that the matter sector has no anisotropic stress.

\subsection{Metric variables and trace-free anisotropy}

Start from the diagonal Bianchi I line element
\begin{equation}
 ds^2=-dt^2+\sum_{i=1}^{3}a_i^2(t)(dx^i)^2.
 \label{eq:Ametric}
\end{equation}
The average scale factor is chosen as the geometric mean,
\begin{equation}
 a(t)=\left(a_1a_2a_3\right)^{1/3}.
 \label{eq:Aavscale}
\end{equation}
It is then natural to define anisotropy variables $\beta_i$ through
\begin{equation}
 a_i(t)=a(t)e^{\beta_i(t)},\qquad \sum_i\beta_i(t)=0.
 \label{eq:Abeta}
\end{equation}
The trace-free condition in Eq.~\eqref{eq:Abeta} is not an extra dynamical constraint. It is a definition that assigns the isotropic volume expansion to $a(t)$ and the anisotropic shape distortion to the two independent combinations of $\beta_i$. A constant shift of a given $\beta_i$ can be absorbed into a rescaling of the corresponding spatial coordinate, but the time dependence of $\beta_i$ is physical because it changes the directional Hubble rates and photon propagation.

The directional Hubble rates are
\begin{equation}
 H_i=\frac{\dot a_i}{a_i}=H+\dot\beta_i,
 \qquad H=\frac{\dot a}{a},
 \label{eq:AHi}
\end{equation}
and the condition $\sum_i\dot\beta_i=0$ gives
\begin{equation}
 H=\frac{1}{3}\sum_i H_i.
 \label{eq:AHmean}
\end{equation}
In the comoving orthonormal frame, the shear eigenvalues are simply
\begin{equation}
 \sigma_i=H_i-H=\dot\beta_i,
 \qquad \sum_i\sigma_i=0.
 \label{eq:Asigmai}
\end{equation}
The shear scalar is therefore
\begin{equation}
 \sigma^2=\frac{1}{2}\sigma_{ij}\sigma^{ij}
 =\frac{1}{2}\sum_i(H_i-H)^2
 =\frac{1}{2}\sum_i\dot\beta_i^2.
 \label{eq:Ashearscalar}
\end{equation}
This is the expression used in the main text.

\subsection{Shear propagation without anisotropic stress}

For a homogeneous irrotational spacetime with isotropic pressure and vanishing anisotropic stress, the trace-free part of the Einstein equations gives
\begin{equation}
 \dot\sigma^i{}_j+3H\sigma^i{}_j=0.
 \label{eq:Ashearprop}
\end{equation}
Multiplying by $a^3$ gives
\begin{equation}
 \frac{d}{dt}\left(a^3\sigma^i{}_j\right)=0,
 \label{eq:Ashearconservation}
\end{equation}
so that
\begin{equation}
 \sigma^i{}_j=\frac{\Sigma^i{}_j}{a^3},
 \label{eq:Asigmasol}
\end{equation}
where $\Sigma^i{}_j$ is a constant trace-free matrix. In the principal-axis frame this becomes
\begin{equation}
 \dot\beta_i=\frac{\Sigma_i}{a^3},\qquad \sum_i\Sigma_i=0.
 \label{eq:Abetadot}
\end{equation}
The scalar amplitude is then
\begin{equation}
 \sigma^2=\frac{1}{2a^6}\sum_i\Sigma_i^2\equiv \frac{\Sigma^2}{a^6}.
 \label{eq:Asigmaa6}
\end{equation}
Eq.~\eqref{eq:Asigmaa6} is the origin of the effective stiff scaling in the mean Friedmann equation.

The anisotropy variables themselves are obtained by integration,
\begin{equation}
 \beta_i(a)-\beta_i(a_0)=\Sigma_i\int_{a_0}^{a}\frac{d\tilde a}{\tilde a^4 H(\tilde a)}.
 \label{eq:Abetaint}
\end{equation}
This formula is useful because it shows the difference between the instantaneous shear, which scales as $a^{-3}$, and the accumulated metric anisotropy, which depends on the entire expansion history. In a late-time de Sitter-like phase, $H\simeq \mathrm{constant}$ and the integral converges; the shear decays while the metric anisotropy approaches a constant that can partly be absorbed into spatial coordinates.

\subsection{Mean Einstein equations and effective stiff behavior}

The $00$ Einstein equation for Bianchi I can be written as
\begin{equation}
 3H^2=8\pi G\rho+\sigma^2.
 \label{eq:Aconstraint}
\end{equation}
The mean Raychaudhuri equation is
\begin{equation}
 \dot H=-4\pi G(\rho+p)-\sigma^2.
 \label{eq:AdotH}
\end{equation}
Together with Eq.~\eqref{eq:Asigmaa6}, these equations show that the shear contribution behaves, at the level of the mean background, as a component with effective equation of state $w=1$. Defining
\begin{equation}
 \Omega_{\sigma0}=\frac{\sigma_0^2}{3H_0^2},
 \label{eq:Aomegasigma0}
\end{equation}
the shear-only mean expansion rate is
\begin{equation}
 E^2(a)=\Omega_{r0}a^{-4}+\Omega_{m0}a^{-3}+\Omega_{\Lambda0}+\Omega_{\sigma0}a^{-6}.
 \label{eq:AE2}
\end{equation}
The corresponding fractional shear contribution at redshift $z$ is
\begin{equation}
 \Omega_\sigma(z)=\frac{\Omega_{\sigma0}(1+z)^6}{E^2(z)}.
 \label{eq:AOmegasigmaz}
\end{equation}
This relation makes clear why early-Universe constraints are severe. During radiation domination,
\begin{equation}
 \Omega_\sigma(z)\simeq\frac{\Omega_{\sigma0}}{\Omega_{r0}}(1+z)^2,
 \label{eq:AOmegasigmaRad}
\end{equation}
while during matter domination,
\begin{equation}
 \Omega_\sigma(z)\simeq\frac{\Omega_{\sigma0}}{\Omega_{m0}}(1+z)^3.
 \label{eq:AOmegasigmaMatter}
\end{equation}
Thus a percent-level shear today would be catastrophically large at recombination or BBN. A viable shear-only model must therefore have an extremely small present value, or else the anisotropy must be generated or sustained only at late times by additional physics.

\subsection{Axisymmetric limit and relation to an ellipsoidal universe}

Many phenomenological applications use a locally rotationally symmetric, or axisymmetric, limit. Let
\begin{equation}
 \beta_1=\beta_2=-\beta,\qquad \beta_3=2\beta.
 \label{eq:ALRSbeta}
\end{equation}
Then
\begin{equation}
 a_{\perp}=ae^{-\beta},\qquad a_{\parallel}=ae^{2\beta},
 \label{eq:ALRSscale}
\end{equation}
and
\begin{equation}
 H_{\perp}=H-\dot\beta,
 \qquad H_{\parallel}=H+2\dot\beta.
 \label{eq:ALRSH}
\end{equation}
The shear scalar becomes
\begin{equation}
 \sigma^2=\frac{1}{2}\left[2\dot\beta^2+(2\dot\beta)^2\right]
 =3\dot\beta^2.
 \label{eq:ALRSshear}
\end{equation}
The local directional expansion rate for a line of sight making an angle $\theta$ with the symmetry axis is
\begin{equation}
 H_{\hat n}=H+\sigma_{ij}n^i n^j
 =H+\dot\beta\left(3\cos^2\theta-1\right).
 \label{eq:ALRSHn}
\end{equation}
Equivalently, after subtracting the angular average,
\begin{equation}
 \frac{\delta H_{\hat n}}{H}=B(t)\left(\cos^2\theta-\frac{1}{3}\right),
 \label{eq:ALRSquad}
\end{equation}
with $B(t)=3\dot\beta/H$ in this convention. This is the mathematical origin of the quadrupolar form used in the main phenomenological model.

If the spatial geometry is described in terms of an ellipsoidal eccentricity, one may relate the eccentricity to the ratio of directional scale factors. For small anisotropy,
\begin{equation}
 \frac{a_{\parallel}}{a_{\perp}}=e^{3\beta}\simeq 1+3\beta.
 \label{eq:Aellipsoidratio}
\end{equation}
Different authors define the eccentricity with slightly different conventions, but in all cases the leading observable effect is controlled by a small trace-free quadrupolar deformation. This explains why Bianchi I, ellipsoidal-universe and directional-Hubble-law parameterizations often lead to similar angular structures even when their physical interpretations differ.

\subsection{Including anisotropic stress}

When the source has anisotropic stress, Eq.~\eqref{eq:Ashearprop} is replaced by
\begin{equation}
 \dot\sigma^i{}_j+3H\sigma^i{}_j=8\pi G\pi^i{}_j.
 \label{eq:Awithpi}
\end{equation}
The formal solution is
\begin{equation}
 \sigma^i{}_j(t)=\frac{1}{a^3(t)}\left[\Sigma^i{}_j+8\pi G\int^{t}a^3(t')\pi^i{}_j(t')dt'\right].
 \label{eq:Aformalpi}
\end{equation}
This equation is important for interpretation. A measured late-time quadrupolar signal cannot automatically be identified with freely decaying primordial shear. It may instead point to a source term active at low redshift, such as anisotropic dark energy, vector fields or another effective anisotropic stress. Conversely, if no such source is specified, the conservative expectation is the rapidly decaying shear-only behavior.

\begin{revision}
\subsection{Practical initial conditions}

For numerical work, choose a present-day shear amplitude $\Omega_{\sigma0}$, one continuous eigenvalue-shape parameter $\psi$, and a rotation matrix $R$ that maps the principal-axis frame to the sky frame. A non-redundant continuous parameterization of the normalized trace-free eigenvalues is
\begin{align}
 \lambda_1&=\sqrt{\frac{2}{3}}\cos\psi,\nonumber\\
 \lambda_2&=\sqrt{\frac{2}{3}}\cos\left(\psi-\frac{2\pi}{3}\right),\nonumber\\
 \lambda_3&=\sqrt{\frac{2}{3}}\cos\left(\psi+\frac{2\pi}{3}\right),
 \label{eq:qpsi_parameterization}
\end{align}
which automatically satisfies
\begin{equation}
 \lambda_1+\lambda_2+\lambda_3=0,
 \qquad \lambda_1^2+\lambda_2^2+\lambda_3^2=1.
 \label{eq:q_constraints}
\end{equation}
Thus, after the amplitude has been fixed, the eigenvalue shape has one rather than two continuous degrees of freedom. With labelled principal axes one may use $0\leq\psi<2\pi$. If permutations are physically identified, impose the ordering $\lambda_1\geq\lambda_2\geq\lambda_3$ and restrict the non-redundant domain to $0\leq\psi\leq\pi/3$; boundary points correspond to axisymmetric degeneracies. The dimensional present-day shear eigenvalues are then
\begin{equation}
 \sigma_{i0}=\sqrt{6\Omega_{\sigma0}}\,H_0\lambda_i,
 \label{eq:initialshear}
\end{equation}
and, for freely decaying shear, $\sigma_i(a)=\sigma_{i0}a^{-3}$. The shear tensor in sky coordinates is
\begin{equation}
 \sigma_{ab}=R_a{}^i\,\sigma_i\,R_b{}^i.
 \label{eq:sheartensor_rotation}
\end{equation}
The model therefore contains one amplitude, one continuous eigenvalue-shape parameter, and three orientation angles in the fully triaxial case. In an axisymmetric analysis the eigenvalue shape is fixed up to its sign and the rotation reduces to the two angles specifying the preferred axis.

\end{revision}
\section{Expanded Practical Likelihood Pipeline}\label{app:pipeline}

This appendix gives a more complete likelihood recipe than the schematic outline in the main text. It is intended as a direct guide for converting the theoretical framework into a numerical analysis.

\subsection{Data vector}

For a directional supernova analysis, the minimal data vector for each object is
\begin{equation}
 d_i=\{z_i,\mu_i,\sigma_{\mu_i},\hat n_i,\mathrm{survey}_i,\mathrm{calibration\ group}_i\}.
 \label{eq:Bdatavector}
\end{equation}
In a modern analysis the diagonal uncertainty $\sigma_{\mu_i}$ is not sufficient; one must use the full covariance matrix,
\begin{equation}
 C=C_{\rm stat}+C_{\rm sys}+C_{\rm calib}+C_{\rm pv}+C_{\rm lens},
 \label{eq:Bcov}
\end{equation}
where the terms denote statistical errors, systematic covariance, calibration covariance, peculiar-velocity covariance and lensing scatter. If the covariance matrix is supplied by the survey, it should be used rather than reconstructed from simplified errors.

The sky position should be converted into a unit vector,
\begin{equation}
 \hat n_i=(\cos b_i\cos\ell_i,\cos b_i\sin\ell_i,\sin b_i),
 \label{eq:Bunitvec}
\end{equation}
where $(\ell_i,b_i)$ are Galactic longitude and latitude. For equatorial coordinates, one first converts $(\alpha_i,\delta_i)$ into Cartesian unit vectors and then rotates into the desired coordinate system. The preferred axis is similarly written as
\begin{equation}
 \hat e=(\cos b_e\cos\ell_e,\cos b_e\sin\ell_e,\sin b_e).
 \label{eq:Baxisvec}
\end{equation}

\subsection{Supernova likelihood}

For the axisymmetric phenomenological model,
\begin{equation}
 \mu_i^{\rm th}=5\log_{10}\left[\frac{D_L^{\rm FLRW}(z_i;\Theta_{\rm bg})}{\mathrm{Mpc}}
 \left(1+A_D(z_i)Q_i\right)\right]+25+M_{\rm off},
 \label{eq:Bmutheory}
\end{equation}
where
\begin{equation}
 Q_i=(\hat n_i\cdot\hat e)^2-\frac{1}{3}.
 \label{eq:BQi}
\end{equation}
For $|A_D(z_i)Q_i|\ll 1$, this can be linearized as
\begin{equation}
 \mu_i^{\rm th}\simeq \mu_i^{\rm FLRW}+\frac{5}{\ln 10}A_D(z_i)Q_i+M_{\rm off}.
 \label{eq:Bmutheorylin}
\end{equation}
The Gaussian likelihood is
\begin{equation}
 -2\ln{\cal L}_{\rm SN}=\left(\bm\mu^{\rm obs}-\bm\mu^{\rm th}\right)^T C^{-1}
 \left(\bm\mu^{\rm obs}-\bm\mu^{\rm th}\right)+\ln|2\pi C|.
 \label{eq:BSNlike}
\end{equation}
If the absolute calibration is not included, the nuisance offset $M_{\rm off}$ should be analytically marginalized or sampled. A useful analytic marginalization over a linear offset with a flat prior is obtained by defining
\begin{equation}
 \chi^2(M)=\left(\bm r-M\bm 1\right)^TC^{-1}\left(\bm r-M\bm 1\right),
 \label{eq:BchiM}
\end{equation}
where $\bm r=\bm\mu^{\rm obs}-\bm\mu^{\rm th}(M=0)$. The best-fit offset is
\begin{equation}
 M_{\rm bf}=\frac{\bm 1^TC^{-1}\bm r}{\bm 1^TC^{-1}\bm 1}.
 \label{eq:BMbest}
\end{equation}
This prevents a spurious monopole shift from being confused with a quadrupole.

\subsection{BAO, chronometers and CMB priors}

For compressed BAO measurements, define the theoretical vector
\begin{equation}
 \bm v_{\rm BAO}^{\rm th}=\left\{\frac{D_M(z_k)}{r_d},\frac{D_H(z_k)}{r_d},\frac{D_V(z_k)}{r_d}\right\}
 \label{eq:BBAOvec}
\end{equation}
for the redshifts and observables supplied by the BAO release. The likelihood is
\begin{equation}
 -2\ln{\cal L}_{\rm BAO}=\left(\bm v^{\rm obs}-\bm v^{\rm th}\right)^TC_{\rm BAO}^{-1}
 \left(\bm v^{\rm obs}-\bm v^{\rm th}\right).
 \label{eq:BBAOlike}
\end{equation}
In a conservative first analysis $D_M$, $D_H$ and $D_V$ are computed from the mean expansion rate. A fully anisotropic BAO analysis would require the survey window function, the line-of-sight convention, the anisotropic Alcock--Paczynski mapping and the directional dependence of the standard ruler.

For cosmic chronometers,
\begin{equation}
 -2\ln{\cal L}_{H(z)}=\sum_{k,l}\left[H^{\rm obs}(z_k)-H^{\rm th}(z_k)\right]
 C_{kl}^{-1}
 \left[H^{\rm obs}(z_l)-H^{\rm th}(z_l)\right].
 \label{eq:Bchronolike}
\end{equation}
For CMB information one may either use a compressed parameter vector, such as the acoustic scale and physical densities, or impose a conservative prior on the shear. A schematic shear prior may be written as
\begin{equation}
 -2\ln{\cal L}_{\rm shear}=\frac{\Omega_{\sigma}(z_*)^2}{s_{\sigma,*}^2},
 \label{eq:Bshearprior}
\end{equation}
where $z_*$ is close to recombination and $s_{\sigma,*}$ is chosen to represent the allowed level of anisotropic expansion. This is only an effective compressed proxy for a real CMB likelihood, but it is preferable to ignoring early-Universe anisotropy altogether.

\subsection{Local \texorpdfstring{$H_0$}{H0} likelihood}

If a local determination is used as a scalar prior, one may write
\begin{equation}
 -2\ln{\cal L}_{H_0}=\frac{\left[H_0^{\rm model}-H_0^{\rm loc}\right]^2}{\sigma_{H_0}^2}.
 \label{eq:BLH0scalar}
\end{equation}
In a directional analysis, however, this should be regarded as a compressed approximation. A more informative approach is to use the calibrator and Hubble-flow supernova directions directly. In a low-redshift approximation,
\begin{equation}
 H_{0,i}^{\rm eff}=H_0\left[1+B_{H0}Q_i\right]+\delta H_{\rm pv}(z_i,\hat n_i),
 \label{eq:BH0eff}
\end{equation}
where $\delta H_{\rm pv}$ represents peculiar-velocity and local-flow corrections. This formulation tests whether a scalar local $H_0$ value is biased by an anisotropic sky distribution.

\subsection{Priors and parameter sampling}

A conservative prior set for the axisymmetric model is
\begin{align}
 H_0 &\in [50,90]~\mathrm{km\,s^{-1}\,Mpc^{-1}},\\
 \Omega_{m0} &\in [0.05,0.6],\\
 \Omega_{\sigma0} &\in [0,\Omega_{\sigma0}^{\rm max}],\\
 A_{D0} &\in [-A_{D,\rm max},A_{D,\rm max}],\\
 \ell_e &\in [0,2\pi),\qquad \sin b_e\in[-1,1].
 \label{eq:Bpriors}
\end{align}
The prior should be uniform on the sphere, which is why $\sin b_e$ rather than $b_e$ is sampled uniformly. For amplitudes that may span many orders of magnitude, a logarithmic prior may also be tested, but the result must be reported because anisotropy constraints can be prior-sensitive.

The posterior is
\begin{equation}
 P(\Theta|d)\propto {\cal L}_{\rm tot}(d|\Theta)\Pi(\Theta),
 \label{eq:Bposterior}
\end{equation}
where $\Pi(\Theta)$ is the prior. Sampling may be performed with Markov-chain Monte Carlo for parameter constraints and with nested sampling for Bayesian evidence. The evidence ratio between the anisotropic model and $\Lambda$CDM is useful because the anisotropic model has extra parameters. A model that reduces $\chi^2$ slightly but pays a large Occam penalty should not be interpreted as evidence for anisotropy.

\section*{Author Contributions}
The author is solely responsible for the conceptualization, methodology, formal analysis, writing, review and editing of the manuscript.

\section*{Funding}
This research received no external grant funding. The author acknowledges institutional support from INFN under the program TAsP, ``Theoretical Astroparticle Physics''.

\begin{revision}
\section*{Data Availability Statement}
No new observational data were generated or analysed in this work. All equations and parameter values required to reproduce the analytic calculations are given in the manuscript. The short numerical scripts used for the one-dimensional quadrature and for generating the figures are available from the author upon reasonable request.

\end{revision}
\section*{Acknowledgments}
The author acknowledges the hospitality of the CERN Department of Theoretical Physics.

\section*{Conflicts of Interest}
The author declares no conflict of interest.

\end{document}